\documentclass[aps,prd,groupedaddress,twocolumn,superscriptaddress]{revtex4}
\usepackage{graphicx}
\begin{document}
%\linenumbers

\title{Anisotropies in the diffuse gamma-ray background measured by the Fermi LAT}


%\author{AuthorList}
%\thanks{}
%\affiliation{AffiliationList}

\author{M.~Ackermann}
\affiliation{Deutsches Elektronen Synchrotron DESY, D-15738 Zeuthen, Germany}
\author{M.~Ajello}
\affiliation{W. W. Hansen Experimental Physics Laboratory, Kavli Institute for Particle Astrophysics and Cosmology, Department of Physics and SLAC National Accelerator Laboratory, Stanford University, Stanford, CA 94305, USA}
\author{A.~Albert}
\affiliation{Department of Physics, Center for Cosmology and Astro-Particle Physics, The Ohio State University, Columbus, OH 43210, USA}
\author{L.~Baldini}
\affiliation{Istituto Nazionale di Fisica Nucleare, Sezione di Pisa, I-56127 Pisa, Italy}
\author{J.~Ballet}
\affiliation{Laboratoire AIM, CEA-IRFU/CNRS/Universit\'e Paris Diderot, Service d'Astrophysique, CEA Saclay, 91191 Gif sur Yvette, France}
\author{G.~Barbiellini}
\affiliation{Istituto Nazionale di Fisica Nucleare, Sezione di Trieste, I-34127 Trieste, Italy}
\affiliation{Dipartimento di Fisica, Universit\`a di Trieste, I-34127 Trieste, Italy}
\author{D.~Bastieri}
\affiliation{Istituto Nazionale di Fisica Nucleare, Sezione di Padova, I-35131 Padova, Italy}
\affiliation{Dipartimento di Fisica ``G. Galilei", Universit\`a di Padova, I-35131 Padova, Italy}
\author{K.~Bechtol}
\affiliation{W. W. Hansen Experimental Physics Laboratory, Kavli Institute for Particle Astrophysics and Cosmology, Department of Physics and SLAC National Accelerator Laboratory, Stanford University, Stanford, CA 94305, USA}
\author{R.~Bellazzini}
\affiliation{Istituto Nazionale di Fisica Nucleare, Sezione di Pisa, I-56127 Pisa, Italy}
\author{E.~D.~Bloom}
\affiliation{W. W. Hansen Experimental Physics Laboratory, Kavli Institute for Particle Astrophysics and Cosmology, Department of Physics and SLAC National Accelerator Laboratory, Stanford University, Stanford, CA 94305, USA}
% ...

\collaboration{Fermi LAT Collaboration}
\noaffiliation

\author{E.~Komatsu}
\email{komatsu@astro.as.utexas.edu}
\affiliation{Texas Cosmology Center and Department of Astronomy, The University of
Texas at Austin, Austin, Texas 78712}

\date{\today}

\begin{abstract}
The contribution of unresolved sources to the diffuse gamma-ray background could induce anisotropies in this emission on small angular scales.  We analyze the angular power spectrum of the diffuse emission measured by the Fermi LAT at Galactic latitudes $|b| > 30^{\circ}$ in four energy bins spanning 1 to 50~GeV.  At multipoles $\ell \ge 155$, corresponding to angular scales $\lesssim 2^{\circ}$,  angular power above the photon noise level is detected at $>99.99$\% CL in the 1--2~GeV, 2--5~GeV, and 5--10~GeV energy bins, and at $>99$\% CL at 10--50~GeV.  Within each energy bin the measured angular power takes approximately the same value at all multipoles $\ell \ge 155$, suggesting that it originates from the contribution of one or more unclustered source populations.  The amplitude of the angular power normalized to the mean intensity in each energy bin is consistent with a constant value at all energies, $C_{\rm P}/\langle I \rangle^{2} = 9.05 \pm 0.84 \times 10^{-6}$ sr, while the energy dependence of $C_{\rm P}$ is consistent with the anisotropy arising from one or more source populations with power-law photon spectra with spectral index $\Gamma_{\rm s} = 2.40 \pm 0.07$.  We discuss the implications of the measured angular power for gamma-ray source populations that may provide a contribution to the diffuse gamma-ray background.
\end{abstract}


\maketitle

\section{Introduction}
\label{sec:intro}

The origin of the all-sky diffuse gamma-ray emission remains one of the outstanding questions in high-energy astrophysics.  First detected by OSO-3~\citep{Kraushaar:1972}, the isotropic gamma-ray background (IGRB) was subsequently measured by SAS-2~\citep{Fichtel:1975}, EGRET~
\citep{Sreekumar:1997un,Strong:2004ry}, and most recently by the Large Area Telescope (LAT) onboard the Fermi Gamma-ray Space Telescope (Fermi)~\citep{Abdo:2010nz:igrb}.  The term IGRB is used to refer to the observed diffuse gamma-ray emission which appears isotropic on large angular scales but may contain anisotropies on small angular scales.  The IGRB describes the collective emission of unresolved members of extragalactic source classes and Galactic source classes that contribute to the observed emission at high latitudes, and gamma-ray photons resulting from the interactions of ultra-high energy cosmic rays with intergalactic photon fields~\citep{Kalashev:2007sn}.

Confirmed gamma-ray source populations with resolved members are guaranteed to contribute to the IGRB\@ at some level via the emission from fainter, unresolved members of those source classes.  In the EGRET era the possibility that blazars are the dominant contributor to the IGRB intensity was extensively studied (e.g., \citep{Stecker:1996ma, Narumoto:2006qg, Inoue:2008pk}), however the level of the blazar contribution remains uncertain, with recent results suggesting different energy-dependent contributions from blazars, which amount to as little as $\sim 15$\% or as much as $\sim 100$\% of the Fermi-measured IGRB intensity, depending on the energy~\citep{Collaboration:2010gqa:srccounts,Abazajian:2010pc,Stecker:2010di,Malyshev:2011zi}.  Star-forming galaxies~\citep{Fields:2010bw} and gamma-ray millisecond pulsars~\citep{FaucherGiguere:2009df} may also provide a significant contribution to the IGRB at some energies.  However, substantial uncertainties in the properties of even confirmed source populations present a challenge to estimating the amount of emission attributable to each source class, and currently the possibility that the IGRB includes an appreciable contribution from unknown or unconfirmed gamma-ray sources, such as dark matter annihilation or decay (e.g., \citep{Ullio:2002pj, Elsaesser:2004ap, Oda:2005nv, Kuhlen:2008aw, Springel:2008zz, Zavala:2009zr, Abdo:2010dk:cosmowimp}), cannot be excluded.

The Fermi-measured IGRB energy spectrum is relatively featureless, following a simple power law to good approximation over a large energy range ($\sim 200$~MeV to $\sim 100$~GeV)~\citep{Abdo:2010nz:igrb}.  As a result, identifying the contributions from individual components based on spectral information alone is difficult.  However, in addition to the energy spectrum and average intensity, the IGRB contains angular information in the form of fluctuations on small angular scales~\citep{Ando:2005xg}.  The statistical properties of these small-scale anisotropies may be used to infer the presence of emission from unresolved source populations.  

If some component of the IGRB emission originates from an unresolved source population, rather than from a perfectly isotropic, smooth source distribution, the diffuse emission will contain fluctuations on small angular scales due to the varying number density of sources in different sky directions.  Unlike the Poisson fluctuations between pixels in a map of a truly isotropic source distribution (which we shall call ``photon noise''), which are due to finite event statistics, the fluctuations from an unresolved source population are inherent in the source distribution and will not decrease in amplitude even in the limit of infinite statistics.  Hence, with sufficient statistics, these fluctuations could be detected above those expected from the photon noise, and could be used to understand the origin of the diffuse emission.  

The angular power spectrum of the emission provides a metric for characterizing the intensity fluctuations.  For a source population modeled with a specific spatial and luminosity distribution, the angular power spectrum can be predicted and compared to the measured angular power spectrum; in this way an anisotropy measurement has the potential to constrain the properties of source populations.  Other approaches to using anisotropy information in the IGRB have also been considered.  For example, the 1-point probability distribution function (PDF), i.e. the distribution of the number of counts per pixel, is an alternative metric to characterize the fluctuations~\citep{Lee:2008fm,Dodelson:2009ih,Malyshev:2011zi}.  In addition, cross-correlating the gamma-ray sky with galaxy catalogs or the cosmic microwave background can be used to constrain the origin of the emission~\citep{Xia:2011ax}.

In recent years theoretical studies have predicted the angular power spectrum of the gamma-ray emission from several known and proposed source classes.  Established astrophysical source populations such as blazars~\citep{Ando:2006cr, Ando:2006mt, Miniati:2007ke}, star-forming galaxies~\citep{Ando:2009nk}, and Galactic millisecond pulsars~\citep{SiegalGaskins:2010mp} have been considered as possible contributors to the anisotropy of the IGRB\@.  In addition, it has been shown that the annihilation or decay of dark matter in Galactic subhalos~\citep{SiegalGaskins:2008ge, Ando:2009fp, Fornasa:2009qh} and extragalactic structures~\citep{Ando:2005xg, Ando:2006cr, Miniati:2007ke, Cuoco:2007sh, Zhang:2008rs, Taoso:2008qz, Fornasa:2009qh, Ibarra:2009nw, Cuoco:2010jb}, may generate an anisotropy signal in diffuse gamma-ray emission.  Interestingly, the predicted angular power spectra of these gamma-ray source classes in the multipole range of $\ell \sim 100$--$500$ are in most cases fairly constant in multipole (except for dark matter annihilation and decay signals, e.g.~\citep{Ando:2005xg, Ando:2006cr, Ibarra:2009nw}), although the amplitude of the predicted anisotropy varies between source classes.  This multipole-independent signal arises from the Poisson term in the angular power spectrum, which describes the anisotropy from an unclustered collection of point sources.  The multipole-independence of the predicted angular power spectra therefore indicates that the expected degree of intrinsic clustering of these gamma-ray source populations has a subdominant effect on the angular power spectra in this multipole range.  The angular power spectra of dark matter annihilation and decay signals are predicted to be smooth and relatively featureless, with the angular power generally falling off more quickly with multipole than Poisson angular power.

In this work we present a measurement of the angular power spectrum of the high-latitude emission detected by the Fermi LAT\@, using $\sim$ 22 months of data.  The data were processed with the Fermi Science Tools \footnote{{\tt http://fermi.gsfc.nasa.gov/ssc}}, and binned into maps covering several energy ranges.  Regions of the sky heavily contaminated by Galactic diffuse emission and known point sources were masked, and then angular power spectra were calculated on the masked sky for each energy bin using the HEALPix package~\footnote{{\tt  http://healpix.jpl.nasa.gov}}, described in~\citep{Gorski:2004by}.

To understand the impact of the instrument response on the measured angular power spectrum, several tailored validation studies were performed for this analysis.  The robustness of the anisotropy analysis pipeline was tested using a source model with known anisotropy properties that was simulated to include the effects of the instrument response and processed with the same analysis pipeline as the data.  The data processing was cross-checked to exclude the presence of anisotropies created by systematics in the instrument exposure calculation by using an event-shuffling technique (as used in \citep{Ackermann:2010ip:CREaniso}) that does not rely on the Monte-Carlo--based exposure calculation implemented in the Science Tools.  In addition, validation studies were performed to characterize the impact of foreground contamination, masking, and inaccuracies in the assumed point spread function (PSF).
 
We use a set of simulated models of the gamma-ray sky as a reference, and compare the angular power spectrum measured for the data to that of the models to identify any significant differences in anisotropy properties.  Finally, we compare the predicted anisotropy for several confirmed and proposed gamma-ray source populations to the measured angular power spectrum of the data.  

The data selection and map-making procedure are described in~\S\ref{sec:dataselect}, and the angular power spectrum calculation is outlined in \S\ref{sec:aps}.  The event-shuffling technique is presented in~\S\ref{sec:shuffle}, and the details of the models simulated to compare with the data are given in~\S\ref{sec:sims}. The results of the angular power spectrum measurement and the validation studies are presented in~\S\ref{sec:results}.  The energy dependence of the anisotropy is discussed in \S\ref{sec:energydep}, and the implications of the results for specific source populations are examined in~\S\ref{sec:discussion}.  The conclusions are summarized in \S\ref{sec:conclusions}.

\section{Data selection and processing}
\label{sec:dataselect}

The Fermi LAT is designed to operate primarily as a survey instrument, featuring both a wide field of view ($\sim$2.4~sr) and a large effective area ($\gtrsim 7000$~cm$^2$ for normally-incident photons above 1~GeV). The telescope is equipped with a 4$\times$4 array of modules, each consisting of a precision tracker and calorimeter, covered by an anti-coincidence detector that allows for rejection of charged particle events. Full details of the instrument, including technical information about the detector, onboard and ground data processing, and mission-oriented support, are given in~\citep{Atwood:2009ez}. 

We selected data taken from the beginning of scientific operations in early-August 2008 through early-June 2010, encompassing over 56.6 Ms of live time \footnote{MET Range: 239557414 - 298180536}.  We selected only ``diffuse'' class~\citep{Atwood:2009ez} events to ensure that the events are photons with high probability, and restricted our analysis to the energy range 1--50~GeV where the PSF of the LAT is small enough to allow for sufficient sensitivity to anisotropies at small angular scales.  The upper limit of 50~GeV was chosen because the small photon statistics above this energy severely limit the sensitivity of the analysis at the high multipoles of interest.  The data and simulations were analyzed with the LAT analysis software Science Tools version v9r15p4 using the standard P6\_V3 LAT instrument response functions (IRFs).  Detailed documentation of the Science Tools is given in~\footnote{{\tt http://fermi.gsfc.nasa.gov/ssc/data/analysis/\\documentation/}}.

In order to both promote near uniform sky exposure and to limit contamination from gamma rays originating in Earth's atmosphere, the tool \emph{gtmktime} was used to remove data taken during any time period when the LAT rocked to an angle exceeding 52$^\circ$ with respect to the zenith, and during any time period when the LAT was not in survey mode.  Beginning in its second year of operation (September~2009), Fermi has been operating in survey mode with a large rocking angle of 50$^{\circ}$, in contrast to the 35$^{\circ}$ rocking angle used during the first year of operation.  The rocking-angle cut is used to limit the amount of contamination from gamma rays produced in cosmic-ray interactions in the upper atmosphere by using only data taken when the Earth's limb was outside of the field of view (the Earth's limb has zenith angle $\sim 113^{\circ}$).  However, due to the LAT's large field of view, some Earth-limb gamma rays may be observed even when the rocking angle constraint is not exceeded, thus the \emph{gtselect} tool was also used to remove each individual event with a zenith angle exceeding 105$^\circ$.  We note that all events in the data set were detected while the Fermi spacecraft was outside of the South Atlantic Anomaly region in which the cosmic-ray fluxes at the altitude of Fermi are significantly enhanced.

In order to balance the need for a large effective area with the need for high angular resolution, the LAT uses a combination of thin tracker regions near the front of the instrument and thicker tracker regions in the back of the detector.  While the effective area of each region is comparable, the width of the PSF for events detected in the front trackers is approximately half that of events detected in the back of the instrument. For a measurement of the angular power at high multipoles, it is thus necessary to differentiate between photons observed in the front and back trackers of the Fermi LAT\@. In this study, we processed front- and back-converting events separately, using the \emph{gtselect} tool to isolate each set of events and calculating the exposure maps independently.  The P6\_V3\_DIFFUSE:FRONT and P6\_V3\_DIFFUSE:BACK IRFs were used to analyze the corresponding sets of events.

Taking the selection cuts into account, the integrated live time was calculated using \emph{gtltcube}.  We chose a pixel size of 0.125$^\circ$, which produces a HEALPix map with resolution parameter $N_{\rm side}=512$.  At this resolution, the suppression of angular power from the pixel window function is subdominant with respect to the suppression from the LAT PSF\@.  We adopted an angular step size $\cos(\theta)$ = 0.025 in order to finely grid the exposure map for different gamma-ray arrival directions in instrument coordinates. The exposure was then calculated using \emph{gtexpcube} with the same pixel size, for 42 logarithmic energy bins spanning 1.04~GeV -- 50.0~GeV.  These finely-gridded energy bins were then summed to build maps covering four larger energy bins, as described in \S\ref{sec:apsebins}.  Using the GaRDiAn package~\citep{Ackermann:2009zz}, both the photon counts and exposure maps were converted into HEALPix-format maps with $N_{\rm side}=512$.

\section{Angular power spectrum calculation}
\label{sec:aps}

We consider the angular power spectrum $C_{\ell}$ of an intensity map $I(\psi)$ where $\psi$ denotes the sky direction.  The angular power spectrum is given by the coefficients $C_{\ell} = \langle |a_{\ell m}|^{2} \rangle$, with the $a_{\ell m}$ determined by expanding the map in spherical harmonics, 
\begin{equation}
I(\psi)=\sum_{\ell m} a_{\ell m} Y_{\ell m}(\psi).
\end{equation}
The intensity angular power spectrum indicates the \emph{dimensionful} size of intensity fluctuations and can be compared with predictions for source classes whose collective intensity is known or assumed (as in, e.g.,~\citep{SiegalGaskins:2010mp}).  The intensity angular power spectrum of a single source class is not in general independent of energy due to the energy-dependence of the mean map intensity $\langle I \rangle$.  

We can also construct the \emph{dimensionless} fluctuation angular power spectrum by dividing the intensity angular power spectrum $C_{\ell}$ of a map by the mean sky intensity (outside of the mask, for a masked sky map) squared $\langle I \rangle^{2}$.  The fluctuation angular power spectrum characterizes the angular distribution of the emission independent of the intensity normalization.  Its amplitude for a single source class is the same in all energy bins if all members of the source class share the same observed energy spectrum, since this results in the angular distribution of the collective emission being independent of energy.  
Energy dependence in the fluctuation angular power due to variation of the energy spectra between individual members of the population is discussed in \S\ref{sec:energydep}.

\subsection{Energy dependence}
\label{sec:apsebins}
We calculate the angular power spectrum of the data and simulated models in four energy bins.  Using multiple energy bins increases sensitivity to source populations that contribute significantly to the anisotropy in a limited energy range, and may also aid in the interpretation of a measurement in terms of a detection of or constraints on specific source populations~\citep{Hensley:2009gh, Cuoco:2010jb}.  In addition, the detection of an energy dependence in the fluctuation angular power spectrum of the total emission (the anisotropy energy spectrum) may be used to infer the presence of multiple contributing source classes~\citep{SiegalGaskins:2009ux}.  In the case that a single source population dominates the anisotropy over a given energy range, the energy dependence of the intensity angular power spectrum can indicate the energy spectrum of that contributor.  

Since the LAT's angular resolution and the photon statistics depend strongly on energy, the sensitivity of the analysis is also energy-dependent: at low energies the LAT's PSF broadens, resulting in reduced sensitivity to small-scale anisotropies, while at high energies the measurement uncertainties are dominated by low statistics.  We calculate angular power spectra in the energy bins 1.04--1.99~GeV, 1.99--5.00~GeV, 5.00--10.4~GeV, and 10.4--50.0~GeV.  The map for each energy bin for the angular power spectrum analysis was created by summing the corresponding maps produced in finely-gridded energy bins, as described in \S\ref{sec:dataselect}.  

\subsection{Angular power spectrum of a masked sky}
\begin{figure*}
\includegraphics[width=0.45\textwidth]{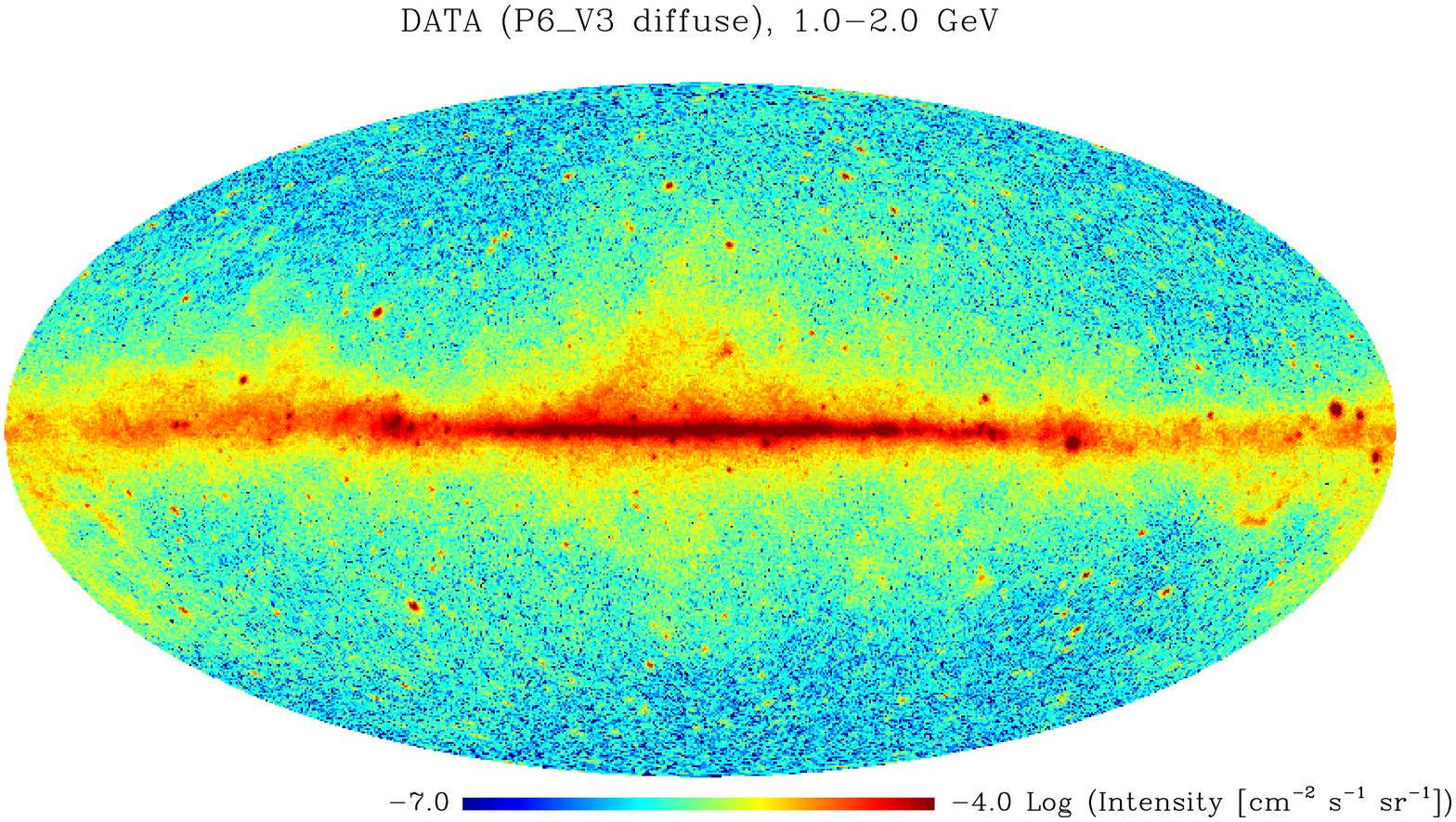}
\includegraphics[width=0.45\textwidth]{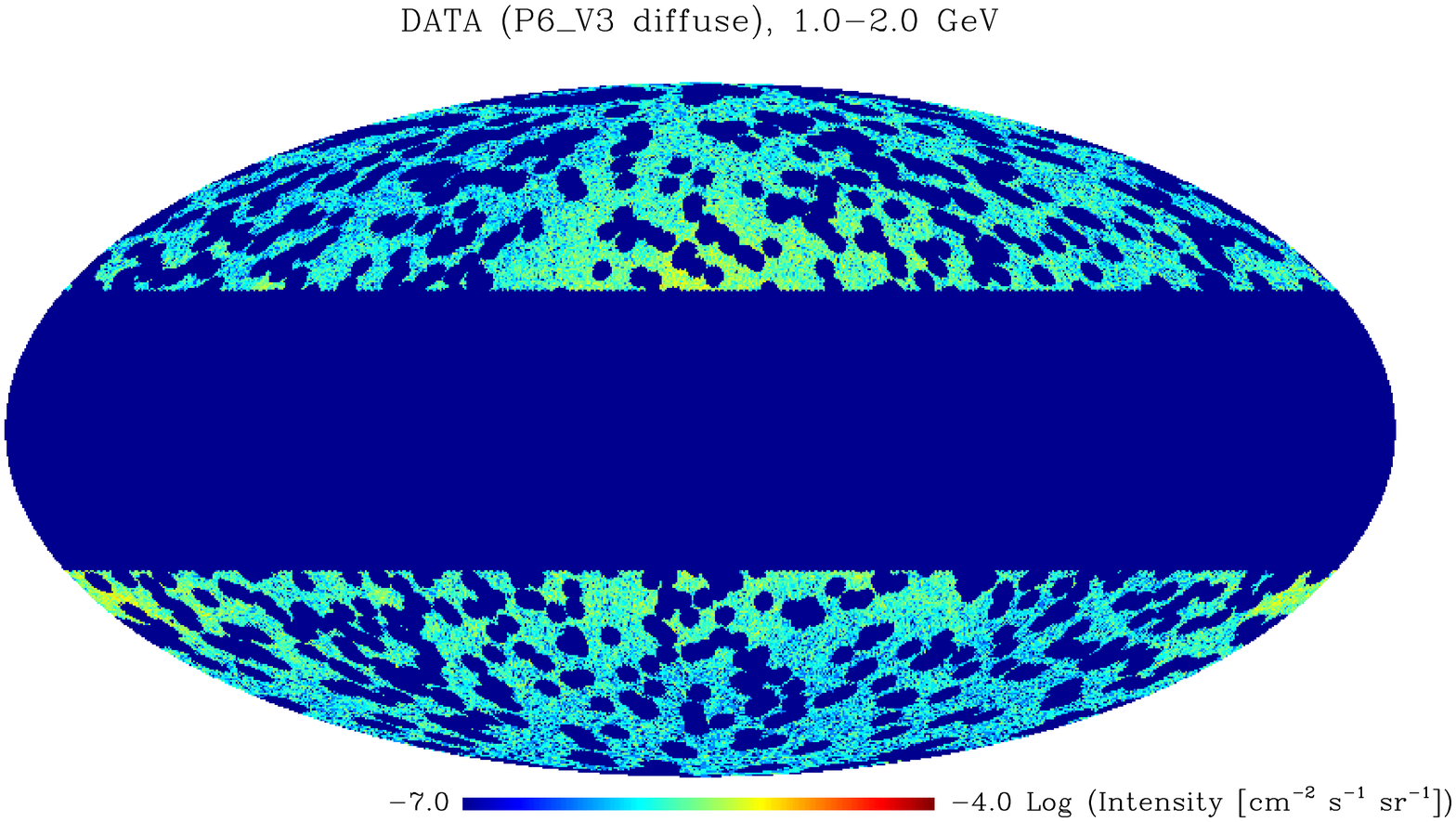}
\vspace{0.5cm}

\includegraphics[width=0.45\textwidth]{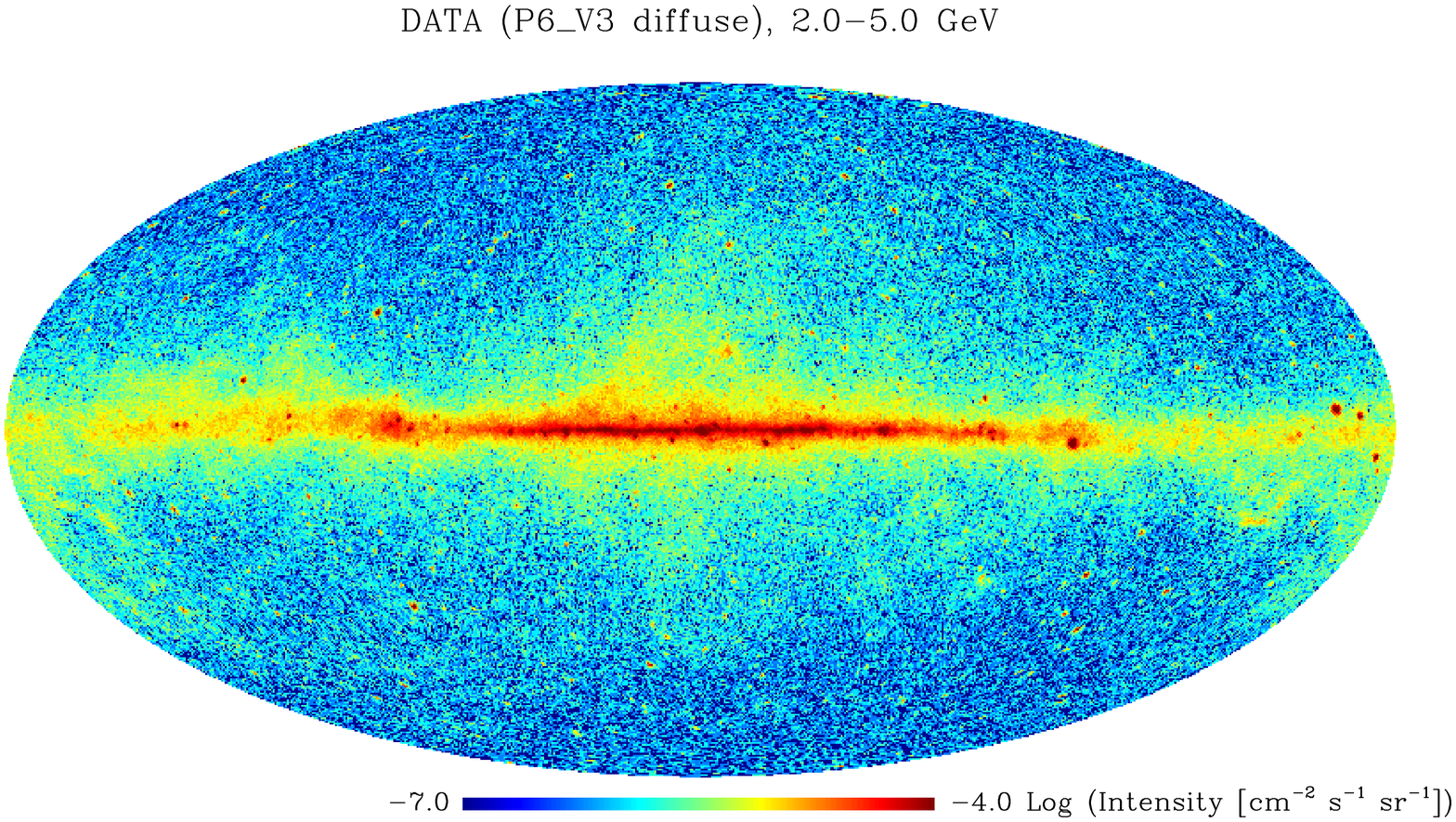}
\includegraphics[width=0.45\textwidth]{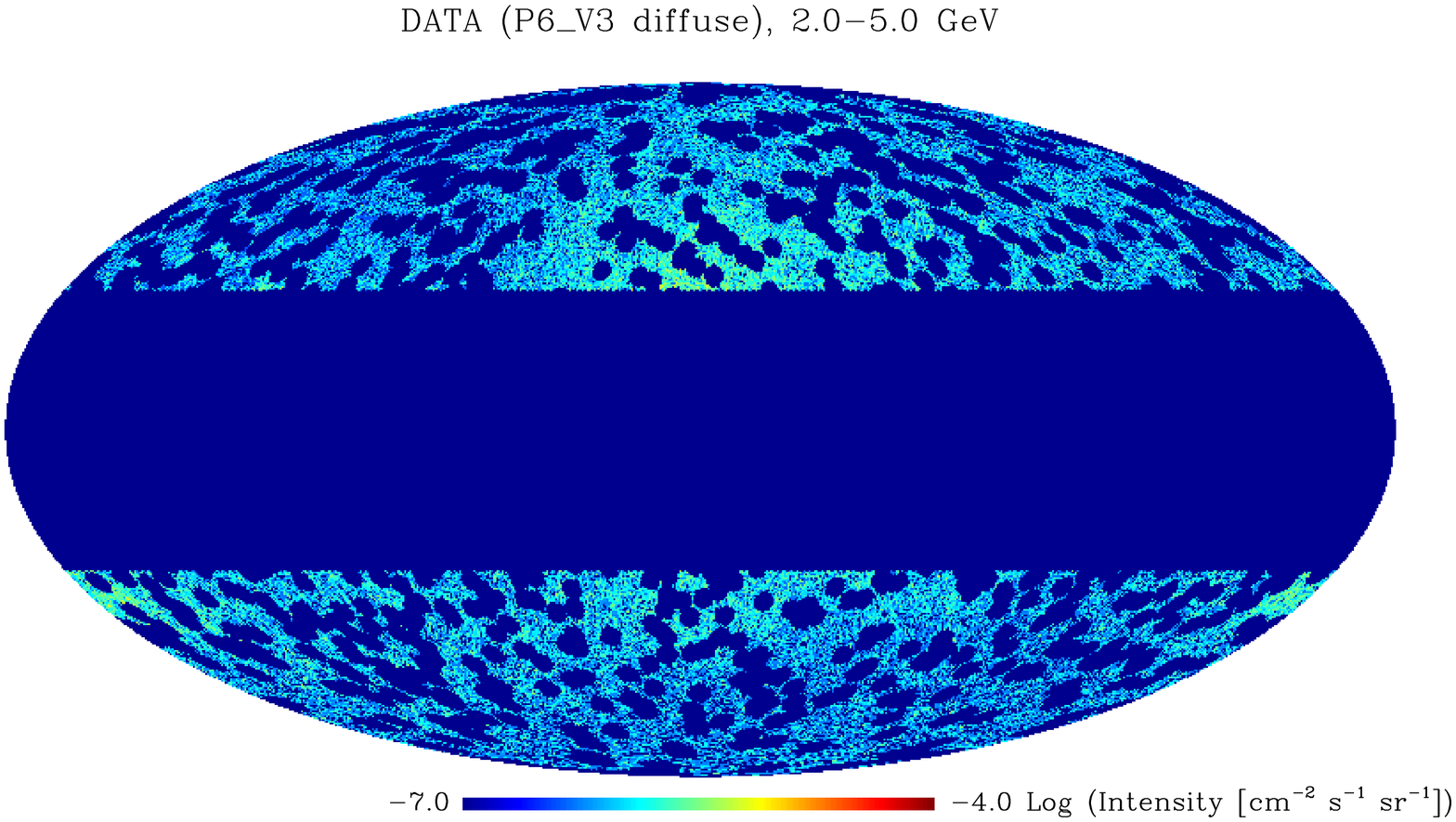}
\vspace{0.5cm}

\includegraphics[width=0.45\textwidth]{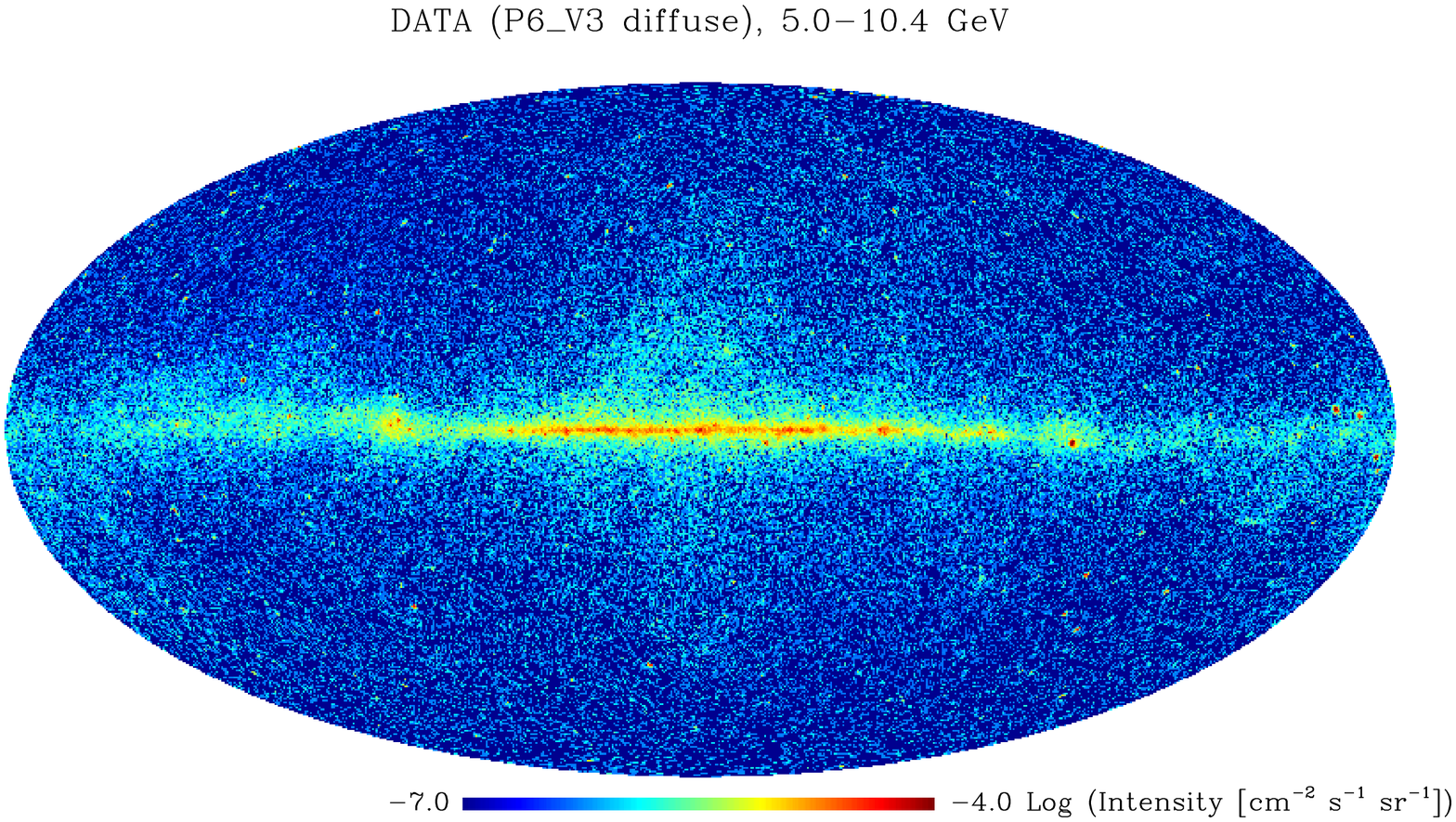}
\includegraphics[width=0.45\textwidth]{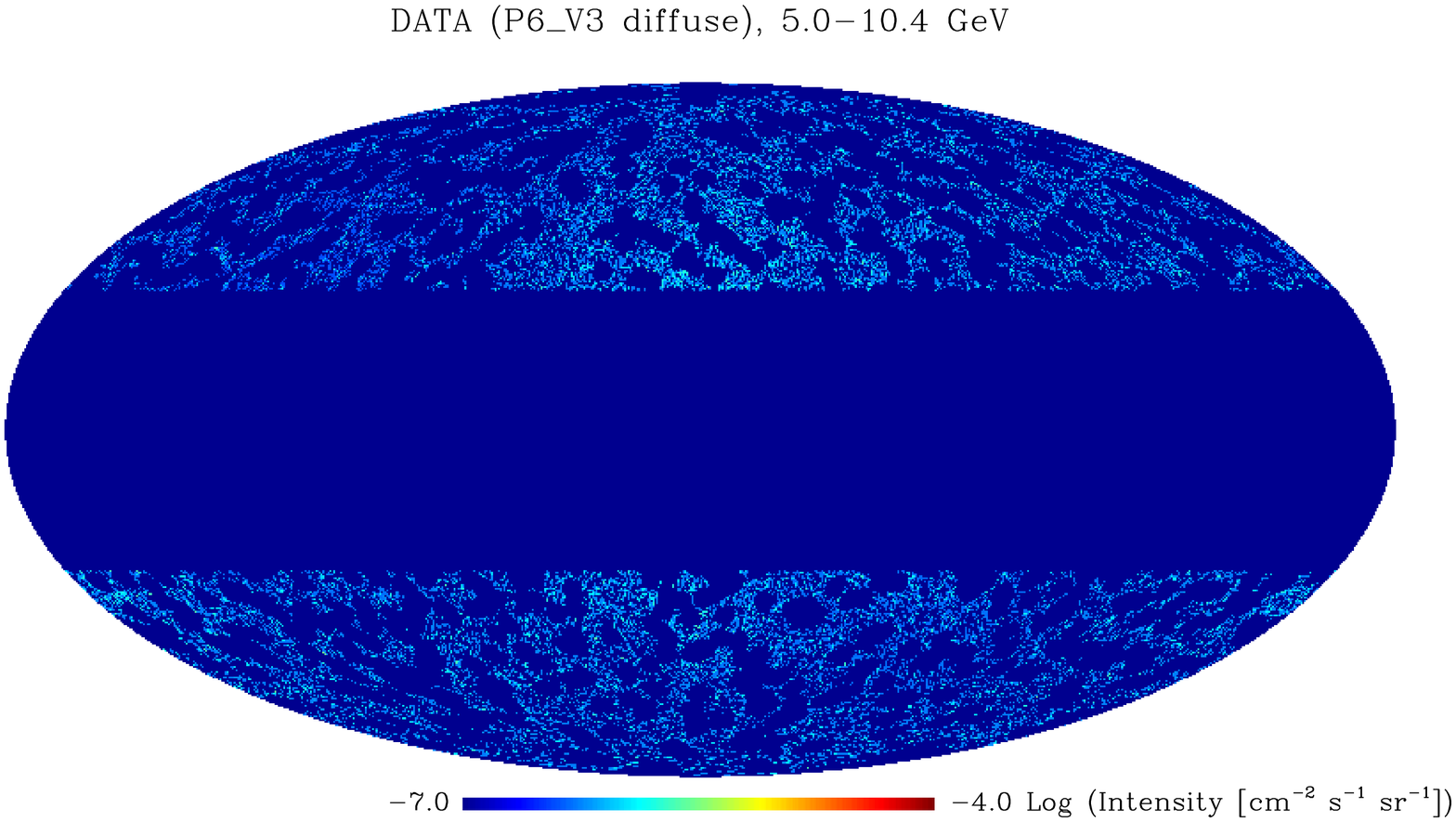}
\vspace{0.5cm}

\includegraphics[width=0.45\textwidth]{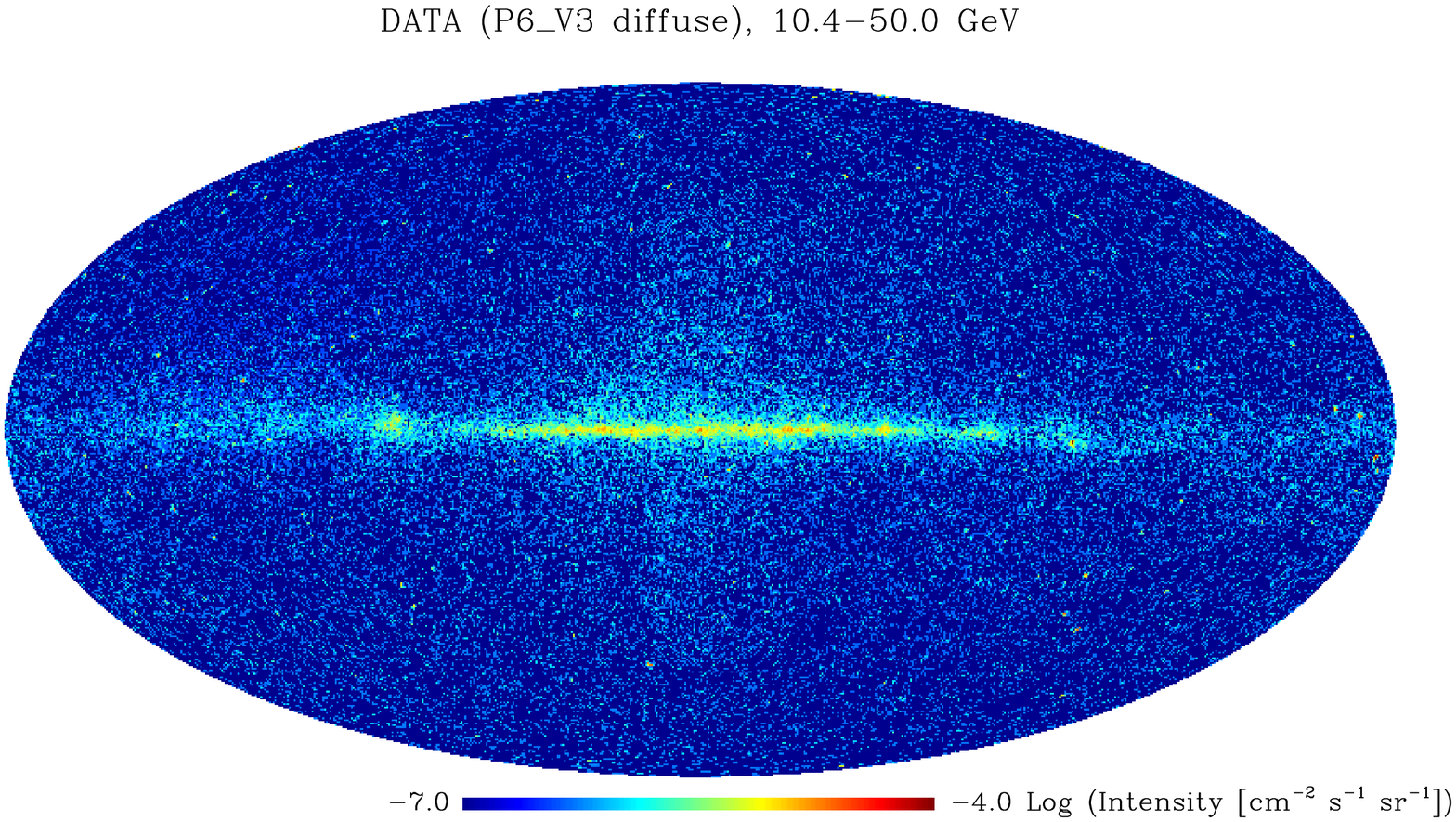}
\includegraphics[width=0.45\textwidth]{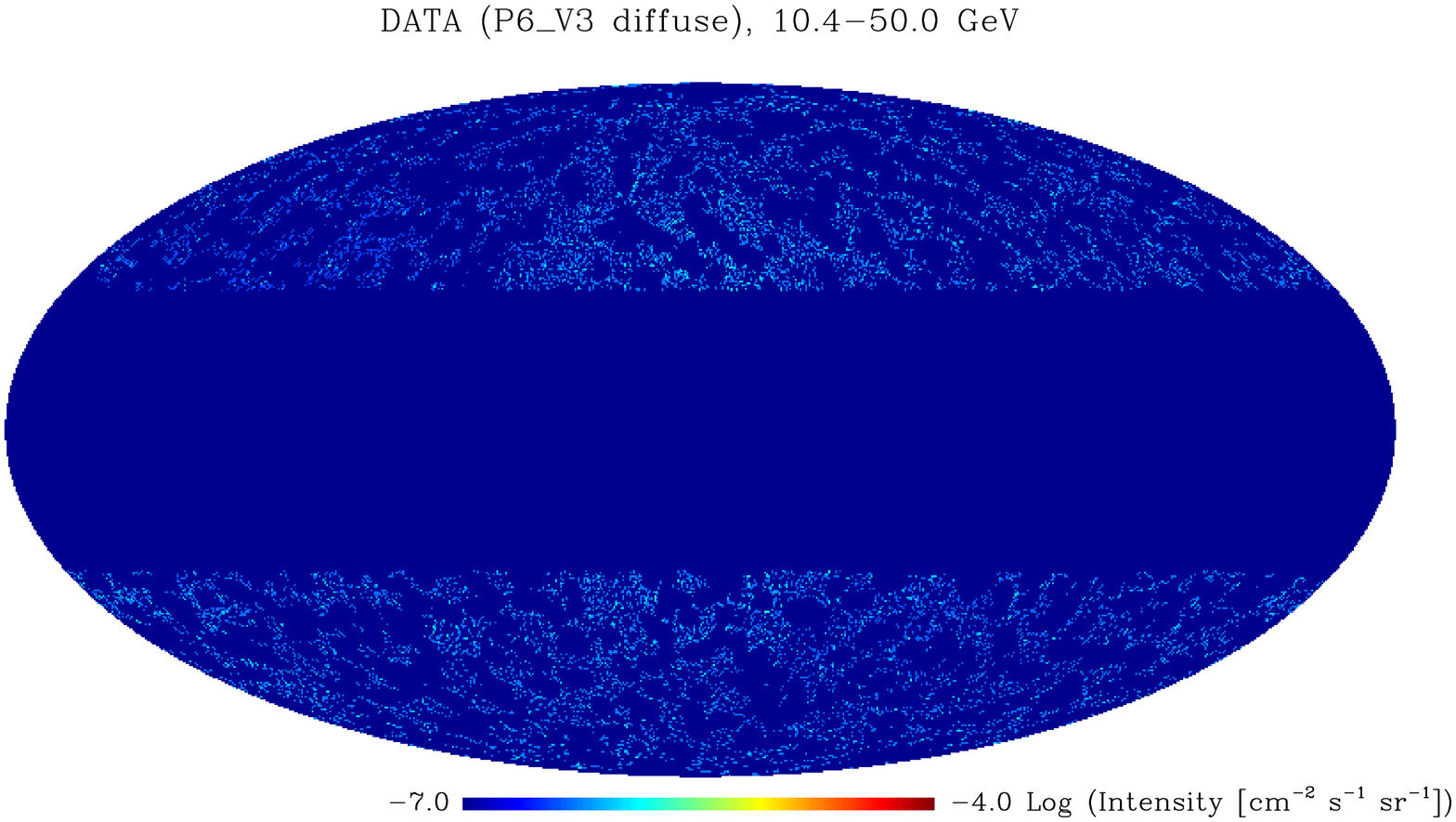}
\caption{All-sky intensity maps of the data in the four energy bins used in this analysis, in Galactic coordinates; the map projection is Mollweide.  The data shown are the average of the maps of the front- and back-converting events, and are shown unmasked (\emph{left panels}) and with the default mask applied (\emph{right panels}).  The mask excludes Galactic latitudes $|b|<30^{\circ}$ and a $2^{\circ}$ angular radius around each source in the 1FGL catalog.  The map images shown have been downgraded in resolution to $N_{\rm side}=128$ to improve the visual quality of the images; however, the analysis was performed on  the higher resolution maps as described in the text.\label{fig:intensmaps} }
\end{figure*}

The focus of this work is to search for anisotropies on small angular scales from unresolved source populations, hence the regions of the sky used in this analysis were selected to minimize the contribution of the Galactic diffuse emission from cosmic-ray interactions and the emission from known sources.  A mask excluding Galactic latitudes $|b|<30^{\circ}$ and a $2^{\circ}$ angular radius around each source in the 11-month Fermi LAT catalog~(1FGL)~\citep{Collaboration:2010ru:1fgl} was applied prior to performing the angular power spectrum calculations in all energy bins.  The fraction of the sky outside of this mask is $f_{\rm sky}= 0.325$.  The $2^{\circ}$ angular radius for the source masking approximately corresponds to the 95\% containment angle for events at normal incidence at 1~GeV (front/back average for P6\_V3 IRFs); the containment angle decreases with increasing energy.  The effect of the mask on the angular power spectra is discussed below and in \S\ref{sec:maskeffects}, and the impact of variations in the latitude cut is assessed in \S\ref{sec:latcompare}.  An all-sky intensity map of the data in each energy bin is shown in Fig.~\ref{fig:intensmaps}, both with and without applying the default mask.

The angular power spectra of the masked maps were calculated using HEALPix, after first removing the monopole and dipole terms.  To approximately correct for the power suppression due to masking, the raw angular power spectra output by HEALPix were divided by the fraction of the sky outside the mask, $f_{\rm sky}$.  This correction is valid at multipoles greater than $\sim 10$, where the power spectrum of the signal varies much more slowly than the window function, as detailed below.

When a fraction of the sky is masked, the measured spherical harmonics coefficients are related to the true, underlying spherical harmonics coefficients, $a_{\ell m}^{\rm true}$, via a matrix multiplication, $a_{\ell m}=\sum_{\ell'm'}a_{\ell'm'}^{\rm true}W_{\ell\ell'mm'}$, where $W_{\ell\ell'mm'}$ is the so-called coupling matrix given by $W_{\ell\ell'mm'}\equiv \int_{\Omega_{\rm obs}} d^2\hat{{\mathbf n}}Y_{\ell m}^*(\hat{\mathbf n})Y_{\ell'm'}(\hat{\mathbf n})$, where the integral is done only on the unmasked sky whose solid angle is $\Omega_{\rm obs}$.  Then, HEALPix returns a raw angular power spectrum, $C_{\ell}^{\rm raw}=(2\ell+1)^{-1}\sum_m |a_{\ell m}|^2$, whose ensemble average is related to the true power spectrum, $C_{\ell}^{\rm true}$, as
\begin{eqnarray}
 \langle C_{\ell}^{\rm raw}\rangle=\frac1{2\ell+1}\sum_{\ell'}C_{\ell'}^{\rm
  true}\sum_{mm'}|W_{\ell\ell'mm'}|^2.
\end{eqnarray}
Now, for a given mask, one may calculate $\sum_{mm'}|W_{\ell\ell'mm'}|^2$ and estimate $C_{\ell}^{\rm true}$ by inverting this equation. This approach is called the MASTER algorithm~\citep{Hivon:2001jp}, and has been shown to yield unbiased estimates of $C_{\ell}^{\rm true}$. However, while this is an unbiased estimator, it is not necessarily a minimum-variance one. In particular, when the coupling matrix is nearly singular because of, e.g., an excessive amount of mask or a complex morphology of mask, this estimator amplifies noise. We observed this amplification of noise when applying the MASTER algorithm to our data set. Therefore, we decided to use an approximate, but less noisy alternative.  It is easy to show (see, e.g., Eq.~A3 of~\citep{Komatsu:2001wu}) that, when $\sum_{mm'}|W_{\ell\ell'mm'}|^2$ peaks sharply at $\ell=\ell'$ and $C_{\ell}^{\rm true}$ varies much more slowly than the width of this peak, the above equation can be approximated as
\begin{eqnarray}
 \langle C_{\ell}^{\rm raw}\rangle\approx C_{\ell}^{\rm true}\frac{\Omega_{\rm
  obs}}{4\pi}=C_\ell^{\rm true}f_{\rm sky}.
\end{eqnarray}
This approximation eliminates the need for a matrix inversion.  We have verified that this method yields an unbiased result with substantially smaller noise than the MASTER algorithm at $\ell>10$. We adopt this method throughout this paper.

\subsection{Window functions}
The angular power spectrum calculated from a map is affected by the PSF of the instrument and the pixelization of the map, encoded in the beam window function $W_{\ell}^{\rm beam}$ and the pixel window function $W_{\ell}^{\rm pix}$ respectively, both of which can lead to a multipole-dependent suppression of angular power that becomes stronger at larger multipoles.  Depending upon whether the power spectrum originates from signal or noise, corrections for the beam and pixel window functions must be applied to the measurement differently. For our application, we must {\it not} apply any corrections to the photon shot noise (Poisson noise) term, while we must apply both the beam and pixel window function corrections to the signal term from, e.g., unresolved sources. While it is obvious why one must not apply the beam window correction to the photon noise term, it may not be so obvious why one must also not apply the pixel window correction to that same term. In fact, this statement is correct only for the shot noise, if the data are pixelized by the nearest-grid assignment (which we have adopted for our pixelizing scheme). This has been shown by Ref.~\citep{Jing:2004fq} (see Eq.~20 of that work) for a 3-dimensional density field, but the same is true for a 2-dimensional field, as we are dealing with here.  We have verified this using numerical simulations. 

In this paper, although we use maps at $N_{\rm side}=512$ (for
which the maximum multipole is $\ell_{\rm max}=1024$), we restrict
the analysis to $C_\ell$ up to $\ell_{\rm max} \sim 500$ where we have a reasonable
signal-to-noise ratio. For these multipoles the effect of the pixel window
function is negligible, and thus we shall simplify our analysis pipeline
by not applying the pixel window
correction to the observed power spectrum~\footnote{The precise form of the
pixel window function depends on the expected signal power spectrum
itself. We have not 
computed the form of the pixel window function appropriate for 
unresolved sources, and thus we have decided not to apply
pixel window corrections. (The pixel window functions provided by
HEALPix may not be 
adequate for our application.)
However, this approximation has a negligible impact on $C_{\ell}$ up to $\ell\sim500$.}. 
Therefore, our signal power spectrum estimator is given by
\begin{equation}
\label{eq:clest}
 C_\ell^{\rm signal} = \frac{C_{\ell}^{\rm raw}/f_{\rm sky} - C_{\rm N}}{(W_\ell^{\rm beam})^2},
\end{equation}
where $C_{\rm N} = \langle N_{\gamma,\rm pix} \rangle  \langle 1/A_{\rm pix}^{2} \rangle /\Omega_{\rm pix}$ is the photon noise term, with $N_{\gamma,\rm pix}$, $A_{\rm pix}$, and $\Omega_{\rm pix}$ the number of observed events, the exposure, and the solid angle, respectively, of each pixel, and the averaging is done over the unmasked pixels.  We approximate the photon noise term by $C_{\rm N} = \langle I \rangle^{2} 4\pi f_{\rm sky} / N_{\gamma}$, with $N_{\gamma}$ denoting the total number of observed events outside the mask.  This approximation is accurate at the percent level.
Note that while $C_{\ell}^{\rm raw}$ is always non-negative, it is possible for our estimator for the signal power spectrum $C_\ell^{\rm signal}$ to be negative due to the subtraction of the noise term.

The beam window function in multipole space associated with the full non-Gaussian PSF is given by 
\begin{equation}
\label{eq:beamwindow}
W_{\ell}^{\rm beam}(E)=
2\pi \int_{-1}^1 d\cos\theta P_{\ell}(\cos(\theta)) \rm{PSF}(\theta; E),
\end{equation}
where $P_{\ell}(\cos(\theta))$ are the Legendre polynomials and $\rm{PSF}(\theta; E)$ is the energy-dependent PSF for a given set of IRFs, with $\theta$ denoting the angular distance in the distribution function.  The PSF used corresponds to the average for the actual pointing and live time history of the LAT and over the off-axis angle, as given by the \emph{gtpsf} tool.  We calculate the beam window functions for both the front- and back-converting events.

The PSF of the LAT, and consequently the beam window function, varies substantially over the energy range used in this analysis, and also non-negligibly within each energy bin.  We treated this energy dependence by calculating an average window function $\langle W_{\ell}^{\rm beam}(E_{i}) \rangle$ for each energy bin $E_{i}$, weighted by the intensity spectrum of the events in each bin,  
\begin{equation}
\langle W_{\ell}^{\rm beam}(E_{i}) \rangle = \frac{1}{I_{\rm bin}} \int_{E_{{\rm min},i}}^{E_{{\rm max},i}}\! \! {\rm d}{E} \; W_{\ell}^{\rm beam} (E)\, \frac{{\rm d}N}{{\rm d}E},
\end{equation}
where $I_{\rm bin} \equiv \int_{E_{{\rm min},i}}^{E_{{\rm max},i}} {\rm d}E \, ({\rm d}N/{\rm d}E)$ and $E_{{\rm min},i}$ and $E_{{\rm max},i}$ are the lower and upper edges of each energy bin.  The differential intensity ${\rm d}N/{\rm d}E$ outside the mask in each map for the finely-gridded energy bins described in \S\ref{sec:dataselect} was used to approximate the energy spectrum for this calculation.

\subsection{Measurement uncertainties}
The $1\sigma$ statistical uncertainty $\sigma_{C_{\ell}}$ on the measured angular power spectrum coefficients $C_{\ell}^{\rm signal}$ is given by~\citep{Knox:1995dq}
\begin{equation}
\label{eq:sigma}
\sigma_{C_{\ell}}=\sqrt{\frac{2}{(2\ell+1) f_{\rm sky}\, \Delta \ell}} \left(C_{\ell}^{\rm signal} + \frac{C_{N}}{(W_{\ell}^{\rm beam})^{2}}\right),
\end{equation}
where $\Delta \ell$ is the width of the multipole bin (for binned data).

After implementing the corrections for masking and for the beam window function to estimate the signal angular power spectrum via Eq.~\ref{eq:clest}, the coefficients $C_{\ell}^{\rm signal}$ were binned in multipole with $\Delta \ell=50$ and averaged in each multipole bin, weighted by the measurement uncertainties,
\begin{equation}
\label{eq:weightedavg}
\langle C_{\ell} \rangle = \frac{\sum_{\ell} C_{\ell}/\sigma_{C_\ell}^{2}} {\sum_{\ell} 1/\sigma_{C_\ell}^{2}}
\end{equation}
with $C_{\ell}=C_{\ell}^{\rm signal}$ as calculated by Eq.~\ref{eq:clest} and $\sigma_{C_\ell}$ given by Eq.~\ref{eq:sigma} with $\Delta \ell = 1$ and $W_{\ell}^{\rm beam}=\langle W_{\ell}^{\rm beam}(E_{i})\rangle$ for the corresponding energy bin $E_{i}$.  As expected, we find that the statistical measurement uncertainties calculated at the linear center of each multipole bin via Eq.~\ref{eq:sigma} with $\Delta \ell = 50$ agree well with the scatter within each multipole bin.  The value of $C_{\ell}$ at multipoles $2 \le \ell \le 4$ was found in most cases to be anomalously large~\footnote{The coefficients $C_{0}$ and $C_{1}$ are negligibly small by construction since the monopole and dipole contributions were removed prior to calculating the angular power spectrum.}, indicating the presence of strong correlations on very large angular scales, such as those that could be induced by the shape of the mask and by contamination from Galactic diffuse emission.   To avoid biasing the value of the average $C_{\ell}^{\rm signal}$ in the first bin by the values at these low multipoles, the multipole bins begin at $\ell=5$.

Finally, the angular power spectra of the front- and back-converting events were combined by weighted averaging, weighting by the measurement uncertainty on each data point.  Due to the larger PSF associated with back-converting events, the measurement errors on the angular power spectra of the back-converting data set tend to be larger than those of the front-converting data set, particularly at low energies and high multipoles where the suppression of the raw angular power due to the beam window function is much stronger for the back-converting data set.  The difference between the measurement uncertainties associated with the front and back data sets is less prominent at higher energies.

\section{Event-shuffling technique}
\label{sec:shuffle}

One way to search for anisotropies is to first calculate the flux of particles from each direction in the sky (equal to the number of detected events from some direction divided by the exposure in the same direction), and then examine its directional distribution.  The flux calculation, which requires knowledge of the exposure, depends on the effective area of the detector and the accumulated observation live time. 

The effective area, calculated from a Monte Carlo simulation of the instrument, could suffer from systematic errors, such as miscalculations of the dependence of the effective area on the instrument coordinates (off-axis angle and azimuthal angle). Naturally, any systematic errors involved in the calculation of the exposure will propagate to the flux, possibly affecting its directional distribution.  If the magnitude of these systematic errors is comparable to or larger than the statistical power of the available data set, their effects on the angular distribution of the flux might masquerade as a real detectable anisotropy. For this reason, we cross-check our results using an alternative method to construct an exposure map that does not rely on the Monte-Carlo--based calculation of the exposure implemented in the Science Tools.

\begin{figure}[ht!]
\includegraphics[width=0.95\columnwidth,keepaspectratio,clip]{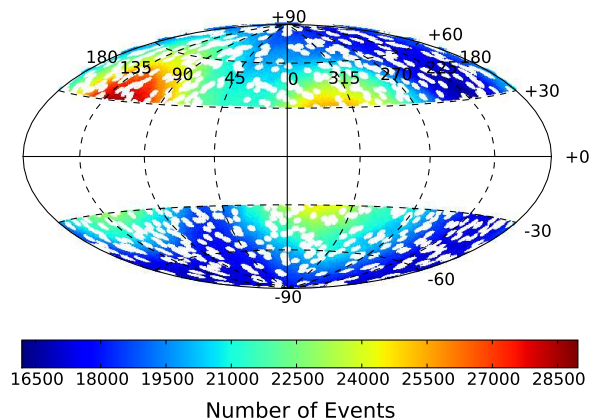}
\caption{No-anisotropy sky map created by summing 20k shuffled maps using front- and back-converting events with $E>1$~GeV, binned into a HEALPix map with $N_{\rm side}=256$. The map projection is Hammer-Aitoff.  The features in the no-anisotropy sky map result from the fact that the sky was not observed with uniform exposure.\label{noAniMap1GeV}}
\end{figure}

The starting point of this method is the construction of a sky map that shows how an isotropic sky would look as seen by the Fermi LAT.  This sky map, hereafter called the ``no-anisotropy sky map'', is directly proportional to the exposure map.  

One method of generating a no-anisotropy map is to randomize the reconstructed directions of the detected events (as in~\citep{Ackermann:2010ip:CREaniso}). In the case that the angular distribution of the flux is perfectly isotropic, a time-independent intensity should be detected when looking in any given detector direction. Possible time variation of the intensity would be due only to changes in the operating conditions of the instrument.  A set of isotropic events can be built by randomly coupling the times and the directions of real events in local instrument coordinates. The randomization in this analysis was performed by exchanging the direction of a given real event in the LAT frame with the direction of another event selected randomly from the data set with uniform probability. Using this information, the sky direction is re-evaluated for the two events.  By construction, the randomized data set preserves the exposure, the energy and angular (with respect to the LAT reference frame) distributions, and also accounts for the detector dead times.

As already discussed in \S\ref{sec:dataselect}, for this analysis a cut of 52$^\circ$ on the rocking angle was applied to limit possible photon contamination from the Earth's albedo.  For the shuffling technique, the analysis was performed with a reduced field of view of the instrument, namely, the events used were selected to have an off-axis angle less than 50$^\circ$. In this way, events with zenith angle exceeding 102$^\circ$ were removed.  This selection cut avoids introducing asymmetries in the exposure across the field of view due to cutting events based on zenith angle.

The randomization was performed using the masked sky map described in~\S\ref{sec:aps}, so that only real events with sky coordinates outside the masks were used, and the re-evaluated sky direction for each event was required to be in the unmasked region of the sky.  This randomization process was repeated 20k times, separately for the front- and back-converting events, each time producing a shuffled sky map that is compatible with an isotropic source distribution. The final no-anisotropy sky map for each energy bin was produced by taking the average of these 20k shuffled sky maps.  For the available event statistics, averaging 20k shuffled maps was reasonably effective at reducing the Poisson noise associated with the average number of events per pixel.  To reduce the number of required shuffled maps by increasing the average number of events per pixel, the shuffled maps were constructed at slightly lower resolution ($N_{\rm side}=256$) than was used in the default analysis.  When analyzing the anisotropy with these exposure maps from the shuffling technique, count maps at $N_{\rm side}=256$ were used to construct the intensity maps.  A no-anisotropy sky map is shown in Fig.~\ref{noAniMap1GeV}.  This sky map does not appear entirely uniform because the sky was not observed with uniform exposure. 

Although the no-anisotropy sky map is directly proportional to the exposure map, this method does not allow us to determine the absolute level of the exposure.  We therefore constructed intensity maps (with arbitrary normalization) by dividing the real data maps in each energy bin by the no-anisotropy map for that energy bin, after first smoothing the no-anisotropy map with a Gaussian beam with $\sigma=1^{\circ}$ to reduce the pixel-to-pixel fluctuations due to the finite number of events available to use in the randomization.  This smoothing beam size removes noise in the no-anisotropy sky map above $\ell \sim 200$, and was chosen because we focus our search for anisotropies in that multipole range.  Angular power spectra were then calculated from these intensity maps as in \S\ref{sec:aps}.  Due to the arbitrary normalization of these intensity maps, we calculate only fluctuation angular power spectra of the data when using the exposure map produced by this shuffling technique.

\section{Simulated models}
\label{sec:sims}

Detailed Monte Carlo simulations of Fermi LAT all-sky observations were performed to provide a reference against which to compare the results obtained for the real data set.  The simulations were produced using the \emph{gtobssim} tool, which simulates observations with the LAT of an input source model.  The \emph{gtobssim} tool generates simulated photon events for an assumed spacecraft pointing and live-time history, and a given set of IRFs.  The P6\_V3\_DIFFUSE IRFs and the actual spacecraft pointing and live-time history matching the observational time interval of the data were used to generate the simulated data sets.  

Two models of the gamma-ray sky were simulated.  Each model is the sum of three components: 
\begin{enumerate}
\item GAL -- a model of the Galactic diffuse emission 
\item CAT -- the sources in the 11-month catalog~(1FGL)~\citep{Collaboration:2010ru:1fgl}
\item ISO -- an isotropic background
\end{enumerate}

Both models include the same CAT and ISO components, and differ only in the choice of the model for the GAL component.  GAL describes both the spatial distribution and the energy spectrum of the Galactic diffuse emission.  The GAL component for the reference sky model used in this analysis (hereafter, MODEL) is the recommended Galactic diffuse model for LAT data analysis, {\tt gll\_iem\_v02.fit}~\footnote{The default diffuse model used in this analysis as well as the isotropic spectral template are available from {\tt http://fermi.gsfc.nasa.gov/ssc/data/access/lat/\\BackgroundModels.html}}, which has an angular resolution of 0.5$^{\circ}$.  This model was used to obtain the 1FGL catalog; a detailed description can be found in Ref.~\footnote{{\tt http://fermi.gsfc.nasa.gov/ssc/data/access/lat/\\ring\_for\_FSSC\_final4.pdf}}.

An alternate sky model (ALT MODEL) was simulated for comparison, in order to test the possible impact of variations in the Galactic diffuse model.  This model is internal to the LAT collaboration, and was built using the same method as {\tt gll\_iem\_v02.fit}, but differs primarily in the following ways:
(i) this model was constructed using 21 months of Fermi LAT observations, while {\tt gll\_iem\_v02.fit} was based on 9 months of data; and
(ii) additional large-scale structures, such as the Fermi bubbles~\citep{Su:2010qj}, are included in the model through the use of simple templates.

The sources in CAT were simulated with energy spectra approximated by single power laws, and with the locations, average integral fluxes, and photon spectral indices as reported in the 1FGL catalog.  All 1451 sources were included in the simulation.  ISO represents the sum of the Fermi-measured IGRB and an additional isotropic component presumably due to unrejected charged particles; for this component the spectrum template {\tt isotropic\_iem\_v02.txt} was used.
 
For both the MODEL and the ALT MODEL, the sum of the three simulated components results in a description of the gamma-ray sky that closely approximates the angular-dependent intensity and energy spectrum of the all-sky emission measured by the Fermi LAT\@.  Although the simulated models may not accurately reproduce some large-scale structures, e.g., Loop I~\citep{Casandjian:2009wq} and the Fermi bubbles, these features are not expected to induce anisotropies on the small angular scales on which we focus in this work.

\section{Results}
\label{sec:results}

In this section we present the measured angular power spectra of the data, followed by the results of validation studies which examine the effect of variations in the default analysis parameters, and by a comparison of the results for the data with those for simulated models.  We summarize the main results of the angular power spectrum measurements of the data and of key validation studies in Table~\ref{tab:cpfits}.

Unless otherwise noted, the results are shown for data and models with the angular power spectra calculated after applying the default source mask which excludes sources in the 1FGL catalog and Galactic latitudes $|b|<30^{\circ}$.  Due to the arbitrary normalization of the intensity maps calculated using the exposure map from shuffling, we show fluctuation angular power spectra for this data set.  Intensity angular power spectra are presented for all other data sets.

In the figures we show our signal angular power spectrum estimator $C_{\ell}^{\rm signal}$ (Eq.~\ref{eq:clest}), which represents the signal after correcting for the power suppression due to masking, subtracting the photon noise, and correcting for the beam window function.  A measurement that is inconsistent with zero thus indicates the presence of signal angular power.  The $C_{\ell}^{\rm signal}$ shown is the weighted average of this quantity for the maps of front and back events.  The fluctuation angular power spectra $C_{\ell}^{\rm signal}/\langle I \rangle^{2}$ were calculated by dividing $C_{\ell}^{\rm signal}$ of the front and back events by their respective $\langle I \rangle^{2}$, and then averaging the angular power spectra.
For conciseness, in the figure labels $C_{\ell}=C_{\ell}^{\rm raw}/f_{\rm sky}$ is the raw angular power spectrum output by HEALPix corrected for the effects of masking.  The error bars on points indicate the 1$\sigma$ statistical uncertainty in the measurement in each multipole bin as calculated by Eq.~\ref{eq:sigma} with $\Delta \ell = 50$ and with the bins beginning at $\ell=5$.  The binned data points are located at the linear center of each multipole bin.

\subsection{Angular power spectrum of the data}
\label{sec:cleaning}

We now present the results of the angular power spectrum analysis of the data.  We measure the angular power spectrum of the data after applying the default latitude cut and source mask, and refer to this as our default data analysis (DATA).  We also measure the angular power spectrum of the data using the same masking and analysis pipeline after performing Galactic foreground cleaning, described below, and refer to this as the cleaned data analysis (DATA:CLEANED).  These two measurements constitute our main results for the data, and so we discuss the energy dependence of the measured angular power (\S\ref{sec:energydep}) and present constraints on specific source populations (\S\ref{sec:discussion}) for the results of both the default and cleaned data analyses.

To minimize the impact of Galactic foregrounds we have employed a large latitude cut.  However, Galactic diffuse emission extends to very high latitudes and may not exhibit a strong gradient with latitude, and it is thus important to investigate to what extent our data set may be contaminated by a residual Galactic contribution.  For this purpose we attempt to reduce the Galactic diffuse contribution to the high-latitude emission by subtracting a model of the Galactic foregrounds from the data, and then calculating the angular power spectra of the residual maps.  For the angular power spectrum analysis of the residual maps (cleaned data) we note that the noise term $C_{\rm N}$ is calculated from the original (uncleaned) map, since subtracting the model from the data does not reduce the photon noise level.

In the following we use the recommended Galactic diffuse model  {\tt gll\_iem\_v02.fit}, which is also the default GAL model that we simulate, as described in \S\ref{sec:sims}.  To tailor the model to the high-latitude sky regions considered in this work, the normalization of the model was adjusted by refitting the model to the data only in the regions outside the latitude mask.  For the fit we used GaRDiAn which convolves the model with the instrument response (effective area and PSF\@). The normalization obtained in this way is, however, very close to the nominal one, within a few percent.

\begin{figure*}
\includegraphics[width=0.45\textwidth]{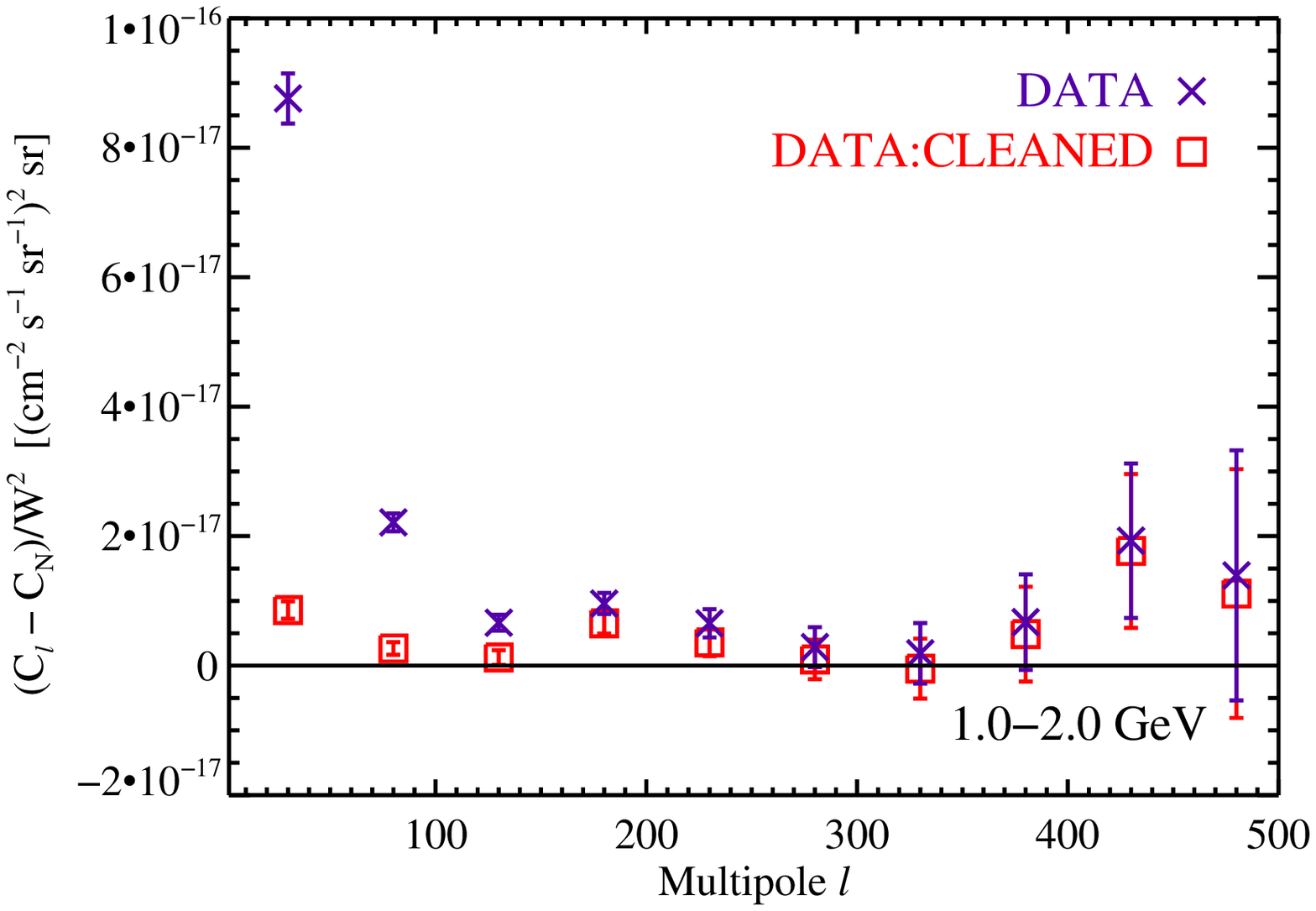}
\includegraphics[width=0.45\textwidth]{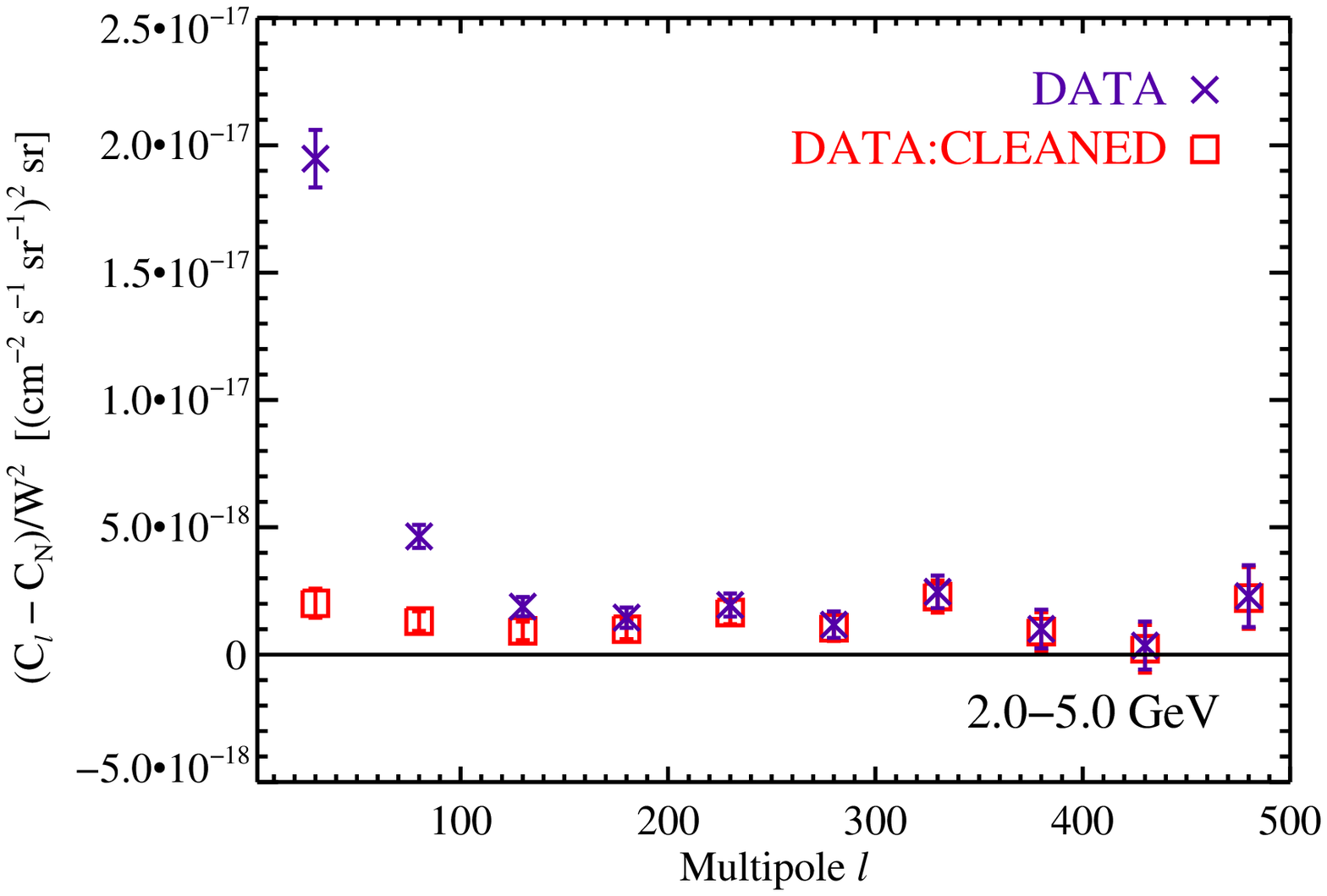}
\includegraphics[width=0.45\textwidth]{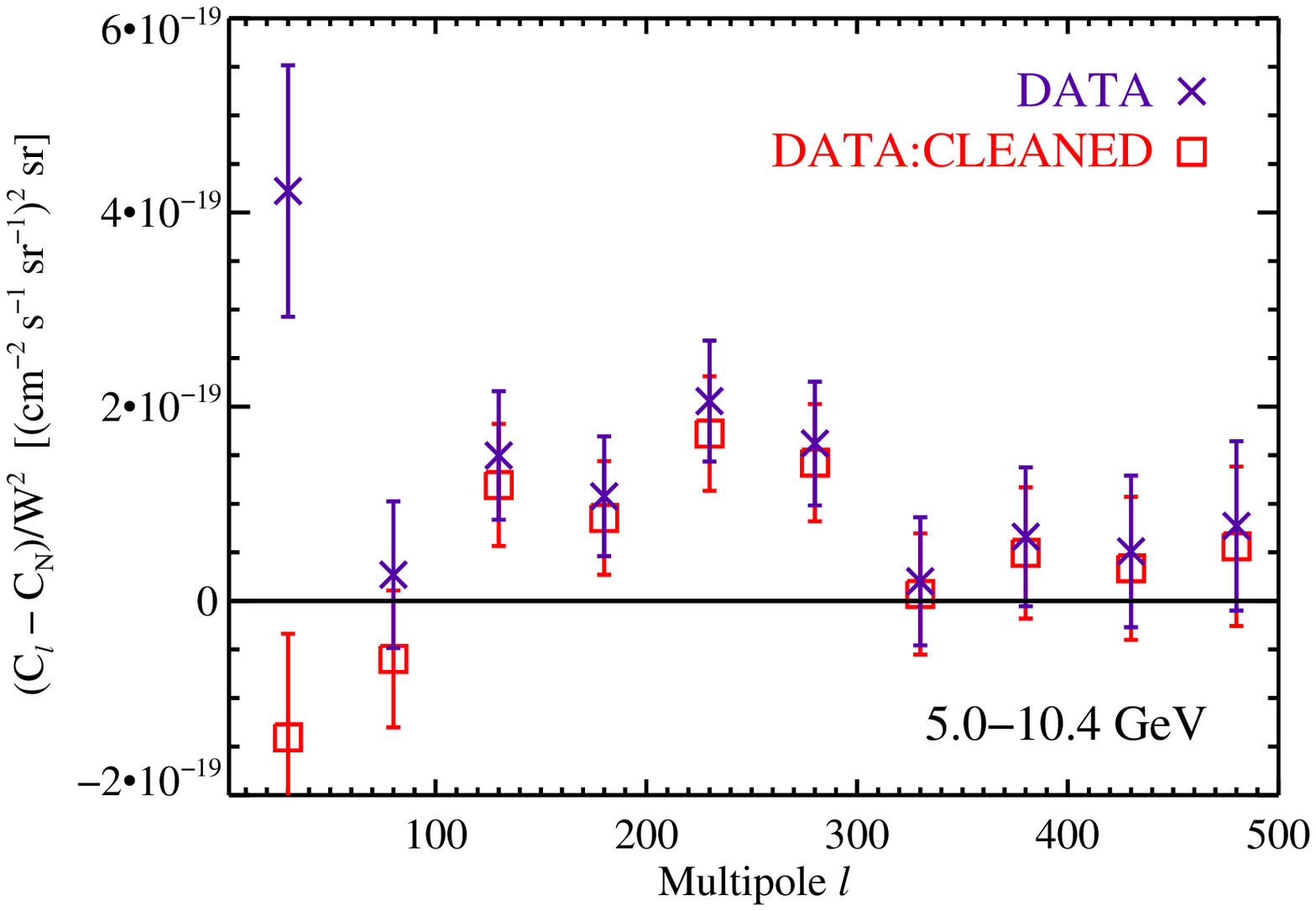}
\includegraphics[width=0.45\textwidth]{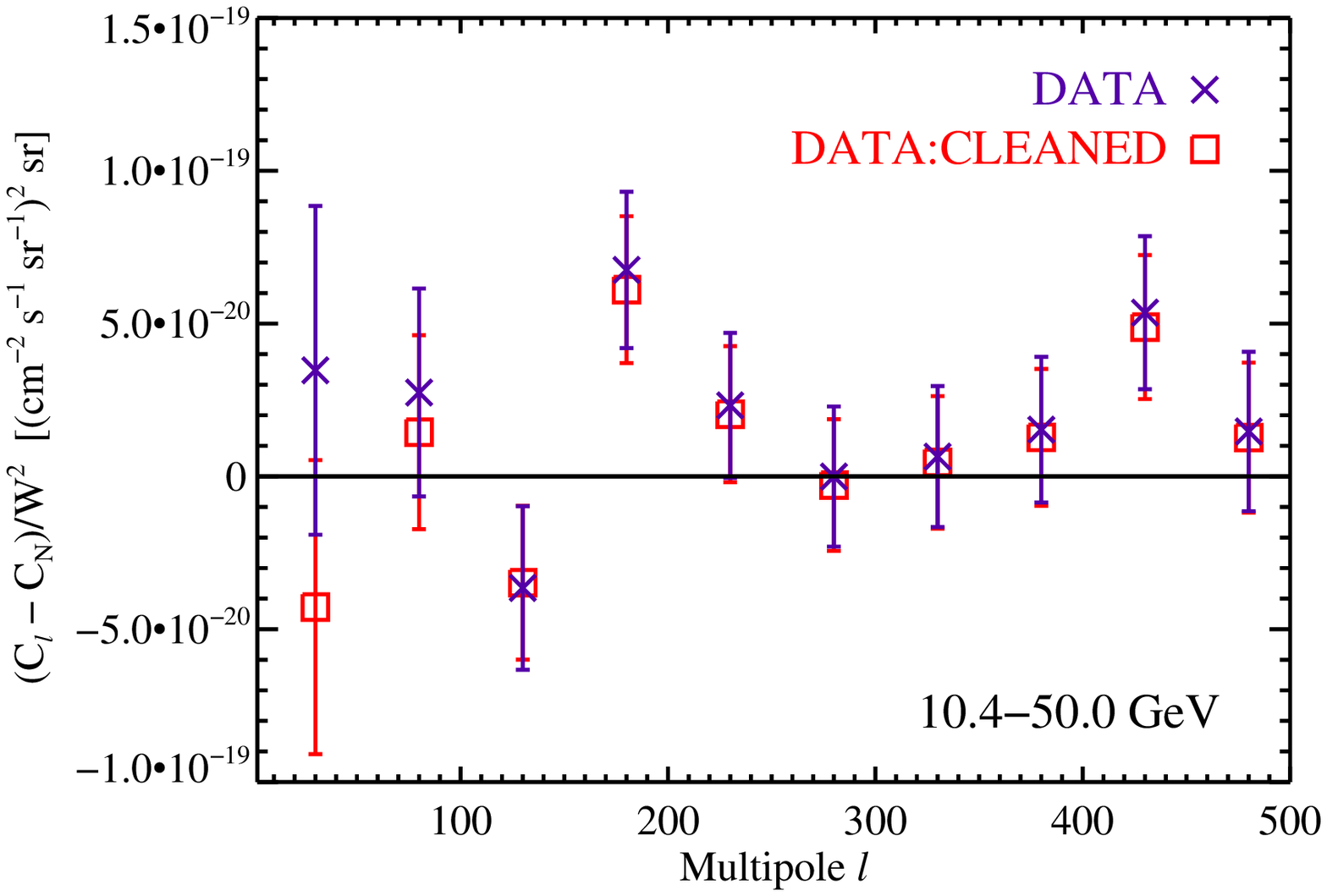}
\caption{Comparison of intensity angular power spectra of the data and Galactic-foreground--cleaned data.  For $\ell \ge 155$ the measured power at all energies is approximately constant in multipole, suggesting that it originates from one or more unclustered source populations.  The large increase in angular power in the default data at $\ell < 155$ in the 1--2 and 2--5~GeV bins is likely attributable largely to contamination from Galactic diffuse emission.   In these two energy bins, foreground cleaning primarily reduces angular power at $\ell < 155$, with the most significant reductions at $\ell < 105$.  At energies greater than 5 GeV the effect of foreground cleaning is small for $\ell \ge 55$.  Expanded versions of the top panels are shown in Fig.~\ref{fig:cleandatazoom}.\label{fig:cleandata}}
\end{figure*}

\begin{figure*}
\includegraphics[width=0.45\textwidth]{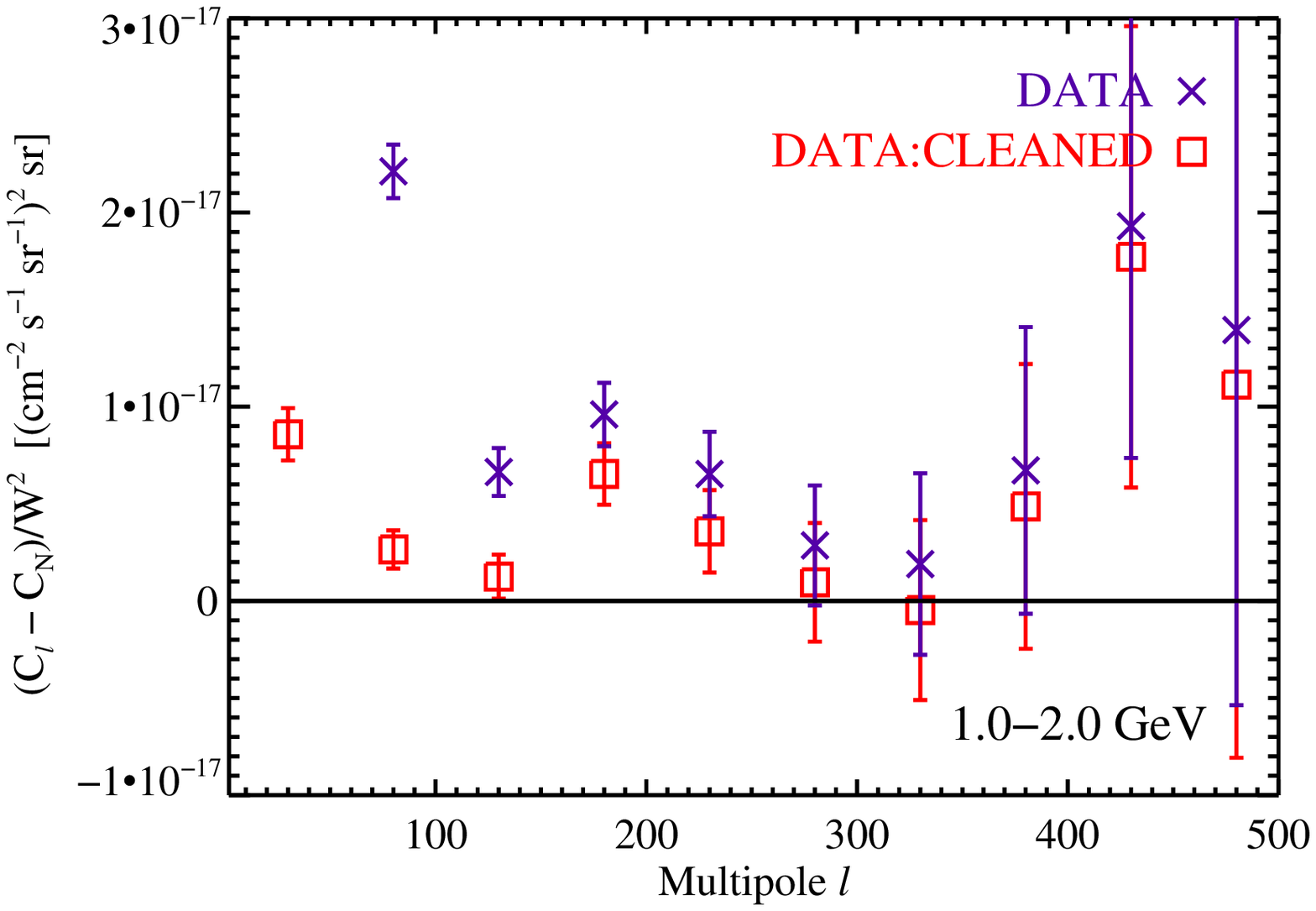}
\includegraphics[width=0.45\textwidth]{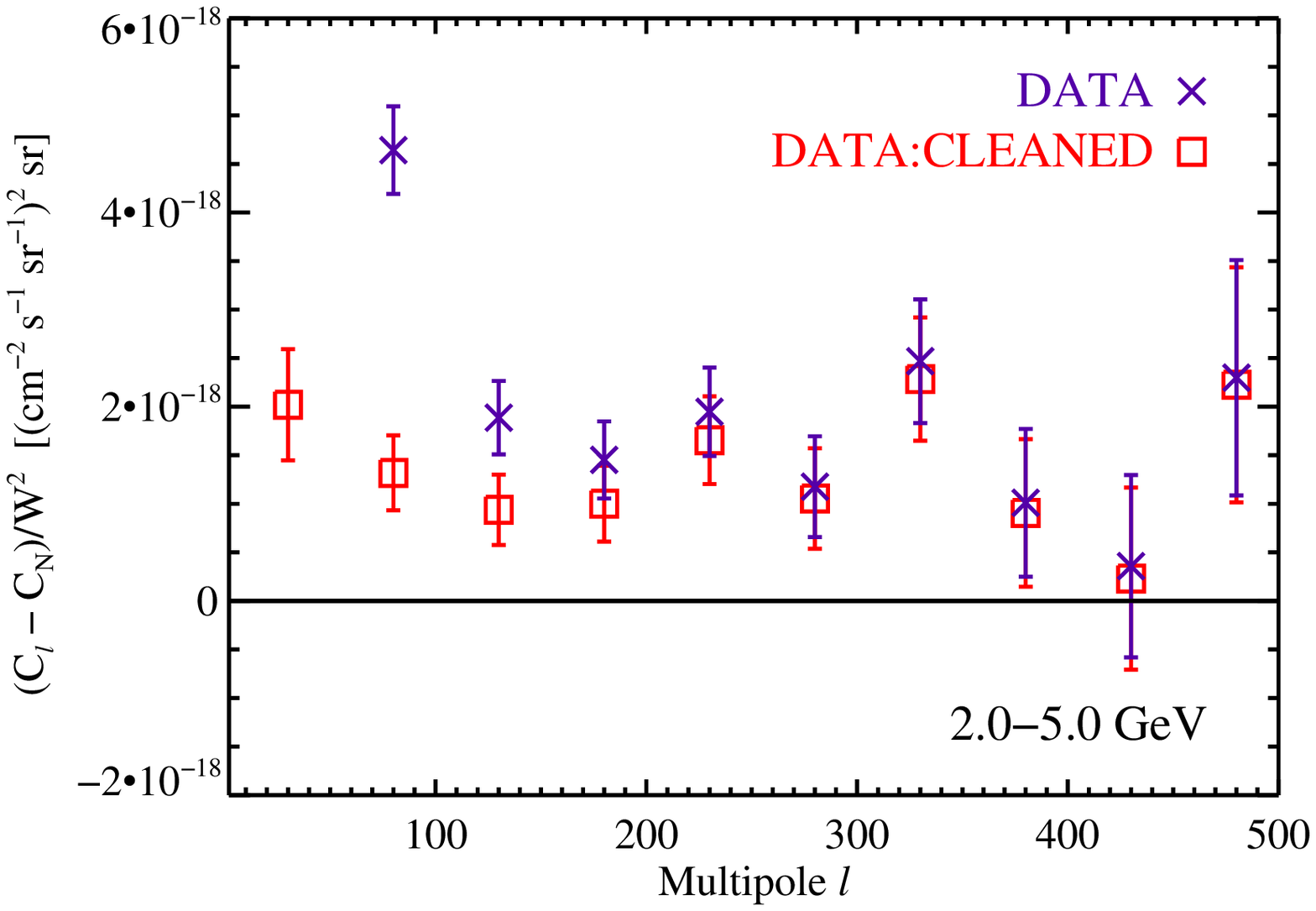}
\caption{Expanded versions of top panels of Fig.~\ref{fig:cleandata}, focusing on the high-multipole angular power. \label{fig:cleandatazoom}}
\end{figure*}

We present the angular power spectra of the data before and after Galactic foreground cleaning in Fig.~\ref{fig:cleandata}; expanded versions of the angular power spectra for the 1--2~GeV and 2--5~GeV bins focusing on the high-multipole data are shown in Fig.~\ref{fig:cleandatazoom}.  In both analyses, angular power at $\ell \ge 155$ is measured in the data in all energy bins considered, and the angular power spectra for the default and cleaned data are in good agreement in this multipole range.  In the default data, the large increase in angular power at $\ell < 155$ in the two energy bins spanning 1--5~GeV is likely due to contamination from the Galactic diffuse emission which features correlations on large angular scales, but may also be attributable in part to the effects of the source mask (see \S\ref{sec:maskeffects}).  

At $\ell \ge 155$ the measured angular power does not exhibit a clear scale dependence in any energy bin.  The results of fitting the unbinned signal angular power spectrum estimator for $155 \le \ell \le 504$ in each energy bin to a power law $C_{\ell}^{\rm signal} \propto (\ell/\ell_{0})^n$ with $\ell_{0}=155$ are given in Table~\ref{tab:scalefits} for the default data analysis.  
In each energy bin, the angular power spectrum for $155 \le \ell \le 504$ is consistent with a Poisson spectrum (constant in multipole, i.e., $n=0$ falls within the 95\% CL range of the best-fit power-law index), as expected for the angular power spectrum of one or more unclustered source populations.  However, we emphasize that the uncertainty in the scale dependence is appreciable, particularly for the 10--50~GeV bin.

\begin{table}[th]
\caption{\label{tab:scalefits}
Multipole dependence of intensity angular power in the data (default analysis) for $155 \le \ell \le 504$ in each energy bin.  The best-fit power-law index $n$ in each energy bin is given with the associated $\chi^{2}$ per degree of freedom (d.o.f.) of the fit.}
\begin{ruledtabular}
\begin{tabular}{cccc}
\multicolumn{1}{c}{$E_{\rm min}$}&
\multicolumn{1}{c}{$E_{\rm max}$}&
\multicolumn{1}{c}{$n$}&
\multicolumn{1}{c}{$\chi^{2}/$\textrm{d.o.f.}}\\
\colrule
1.04 & 	1.99 & 	$-1.33 \pm 0.78$ &	0.38\\
1.99 & 	5.00 & 	$-0.07 \pm 0.45$ &	0.43\\
5.00 & 	10.4 & 	$-0.79 \pm 0.76$ &	0.37\\
10.4 & 	50.0 & 	$-1.54\pm 1.15$ &	0.39\\
\end{tabular}
\end{ruledtabular}
\end{table}

In light of the scale independence of the angular power at $\ell \ge 155$, we associate the signal in this multipole range with a Poisson angular power spectrum and determine the best-fit constant value of the angular power $C_{\rm P}$ and the fluctuation angular power $C_{\rm P}/\langle I \rangle^{2}$ over $155 \le \ell \le 504$ in each energy bin, by weighted averaging of the unbinned measurements.  These results for the default and cleaned data are summarized in Table~\ref{tab:cpfits}, along with the results obtained for the data using an updated source catalog to define the source mask and for a simulated model, which will be discussed in \S\ref{sec:1fgl2fgl} and \S\ref{sec:datamodel}, respectively.  

\begin{table*}[th]
\caption{\label{tab:cpfits}
Best-fit values of the angular power $C_{\rm P}$ and fluctuation angular power $C_{\rm P}/\langle I \rangle^{2}$ in each energy bin over the multipole range $155 \le \ell \le 504$.  Results are shown for the data processed with the default analysis pipeline, the foreground-cleaned data, the data analyzed with the 2FGL source mask, and the default simulated model.  Significance indicates the measured angular power expressed in units of the measurement uncertainty $\sigma$; the measurement uncertainties can be taken to be Gaussian.}
\begin{ruledtabular}
\begin{tabular}{ccccccc}
&
\multicolumn{1}{c}{$E_{\rm min}$}&
\multicolumn{1}{c}{$E_{\rm max}$}&
\multicolumn{1}{c}{$C_{\rm P}$}&
\multicolumn{1}{c}{\textrm{Significance}}&
\multicolumn{1}{c}{$C_{\rm P}/\langle I \rangle^{2}$}&
\multicolumn{1}{c}{\textrm{Significance}}\\
&
\multicolumn{1}{c}{\textrm{[GeV]}}&
\multicolumn{1}{c}{\textrm{[GeV]}}&
\multicolumn{1}{c}{\textrm{[(cm$^{-2}$ s$^{-1}$ sr$^{-1}$)$^{2}$ sr]}}&
&
\multicolumn{1}{c}{\textrm{[$10^{-6}$ sr]}}&
\\
\colrule
DATA & 			1.04 & 	1.99 & 	$7.39 \pm 1.14 \times 10^{-18}$ & 		$6.5\sigma$ &	$10.2 \pm 1.6$ & 	$6.5\sigma$\\
 & 				1.99 & 	5.00 & 	$1.57 \pm 0.22 \times 10^{-18}$ & 		$7.2\sigma$ & 	$8.35 \pm 1.17$ & 	$7.1\sigma$\\
 & 				5.00 & 	10.4 & 	$1.06 \pm 0.26 \times 10^{-19}$ & 		$4.1\sigma$ &	$9.83 \pm 2.42$ & 	$4.1\sigma$\\
 & 				10.4 & 	50.0 & 	$2.44 \pm 0.92 \times 10^{-20}$ & 		$2.7\sigma$ &	$8.00 \pm 3.37$ & 	$2.4\sigma$\\
\colrule
DATA:CLEANED & 	1.04 & 	1.99 & 	$4.62 \pm 1.11 \times 10^{-18}$ & 		$4.2\sigma$ &	$6.38 \pm 1.53$ & 	$4.2\sigma$\\
 & 				1.99 & 	5.00 & 	$1.30 \pm 0.22 \times 10^{-18}$ & 		$6.0\sigma$ &	$6.90 \pm 1.16$ & 	$5.9\sigma$\\
 & 				5.00 & 	10.4 & 	$8.45 \pm 2.46 \times 10^{-20}$ & 	$3.4\sigma$ &	$8.37 \pm  2.41$ & 	$3.5\sigma$\\
 & 				10.4 &	 50.0 & 	$2.11 \pm 0.86 \times 10^{-20}$ & 		$2.4\sigma$ &	$7.27 \pm 3.36$ & 	$2.2\sigma$\\
 \colrule
 DATA:2FGL & 	1.04 & 	1.99 & 	$5.18 \pm 1.17 \times 10^{-18}$ & 		$4.4\sigma$ &	$7.23 \pm 1.61$ & 	$4.5\sigma$\\
 & 				1.99 & 	5.00 & 	$1.21 \pm 0.28 \times 10^{-18}$ & 		$5.3\sigma$ &	$6.49 \pm 1.22$ & 	$5.3\sigma$\\
 & 				5.00 & 	10.4 & 	$8.38 \pm 2.72 \times 10^{-20}$ & 	$3.1\sigma$ &	$7.67 \pm  2.54$ & 	$3.0\sigma$\\
 & 				10.4 &	 50.0 & 	$8.00 \pm 9.57 \times 10^{-21}$ & 		$0.8\sigma$ &	$2.28 \pm 3.52$ & 	$0.6\sigma$\\
 \colrule
MODEL & 		1.04 & 	1.99 & 	$1.89 \pm 1.08 \times 10^{-18}$ & 		$0.7\sigma$ &	$2.53 \pm 1.47$ & 	$1.7\sigma$\\
 & 				1.99 & 	5.00 & 	$1.92 \pm 2.10 \times 10^{-19}$ & 		$0.9\sigma$ &	$0.99 \pm 1.12$ & 	$0.9\sigma$\\
 & 				5.00 & 	10.4 & 	$3.41 \pm 2.60 \times 10^{-20}$ & 		$1.3\sigma$ &	$3.04 \pm  2.34$ & 	$1.3\sigma$\\
 & 				10.4 & 	50.0 & 	$0.62 \pm 9.63 \times 10^{-21}$ & 		$0.1\sigma$ &	$0.24 \pm 3.02$ & 	$0.1\sigma$\\
\end{tabular}
\end{ruledtabular}
\end{table*}

We note that the associated measurement uncertainties can be taken to be Gaussian, in which case the reported significance quantifies the probability of the measured angular power to have resulted by chance from a truly uniform background.  We consider a 3$\sigma$ or greater detection of angular power ($C_{\rm P}$) in a single energy bin to be statistically significant.  For the default data, the best-fit values of $C_{\rm P}$ indicate significant detections of angular power in the 1--2, 2--5, and 5--10~GeV bins (6.5$\sigma$, 7.2$\sigma$, and 4.1$\sigma$, respectively), while in the 10--50~GeV bin the best-fit $C_{\rm P}$ represents a 2.7$\sigma$ measurement of angular power.  We further note that the best-fit value of the fluctuation angular power over all four energy bins (see \S\ref{sec:energydep} and Table~\ref{tab:edep}) yields a detection with greater than 10$\sigma$ significance for the default data.

For the 1--2~GeV and 2--5~GeV energy bands the cleaning procedure results in a significant decrease in the angular power at low multipoles ($\ell < 105$), and a smaller reduction at higher multipoles.  However, the decrease is small for $\ell \ge 155$, and angular power is still measured at all energies, at slightly lower significances (see Table~\ref{tab:cpfits}).  We emphasize that the detections in the three energy bins spanning 1--10~GeV remain statistically significant, and the best-fit fluctuation angular power over all energy bins is detected at greater than 8$\sigma$ significance.  For energies above 5~GeV the foreground cleaning does not strongly affect the measured angular power spectrum for $\ell \ge 55$.  At all energies the decrease in angular power at low multipoles can be attributed to the reduction of Galactic foregrounds which feature strong correlations on large angular scales.  We conclude that contamination of the data by Galactic diffuse emission does not have a substantial impact on our results at the multipoles of interest ($\ell \ge 155$).  This conclusion is in agreement with that of Ref.~\citep{Cuoco:2010jb}, which found that the Galactic foregrounds have a rapidly declining angular power spectrum above $\ell \sim 100$.

To further study the expected angular power spectrum of Galactic foregrounds, we analyzed the angular power spectrum of the E(B-V) emission map of Ref.~\citep{Schlegel:1997yv} (hereafter SFD map), which is proportional to the column density of the interstellar dust, after masking $|b|<30^{\circ}$ as in our default analysis.  The SFD map is a good tracer of the Galactic interstellar medium (ISM) away from the Galactic plane, the spatial structure of which should be reflected in the diffuse gamma-ray emission produced by interactions of cosmic rays with the ISM\@.  It has an angular resolution of 6~arcminutes, much smaller than the intrinsic resolution of the GAL model map ($\sim 0.5^{\circ}$), and smaller than the map resolution used in this study, and so it accurately represents the small-scale structure of the ISM\@ on the angular scales accessible to this analysis.  We found that the SFD map produces an angular power spectrum with a slightly harder slope than the default GAL model, and consequently features more angular power at high multipoles.  However, like the GAL model, the SFD map angular power spectrum falls off quickly with multipole compared to a Poisson spectrum, and the amplitude of the SFD map angular power is below that measured in the data for $\ell \gtrsim 100$.  This further reinforces the conclusion that Galactic foreground contamination cannot explain the observed high-multipole angular power in the data.

\subsection{Validation with a simulated point source population}
\label{sec:ptsrc}

To ensure that our analysis procedure accurately recovers an input angular power spectrum, and in particular that the result is not biased by instrumental effects, we compare the angular power spectrum calculated for a simulated point source population with the theoretical prediction for that population.  It is straightforward to calculate the expected angular power spectrum
of unclustered point sources, once a flux distribution function,
${\rm d}N/{\rm d}S$ (in units of cm$^2$~s~sr$^{-1}$), and a source detection flux threshold,
$S_c$ (in units of cm$^{-2}$~s$^{-1}$), are provided.  The angular power spectrum of an unclustered point source population is the Poisson component of the angular power $C_{\rm P}$, which takes the same value at all multipoles and is given by
\begin{equation}
\label{eq:cps}
 C_{\rm P}= \int_0^{S_{c}}dS~S^2\frac{{\rm d}N}{{\rm d}S}.
\end{equation}

\begin{figure}
\includegraphics[width=0.45\textwidth]{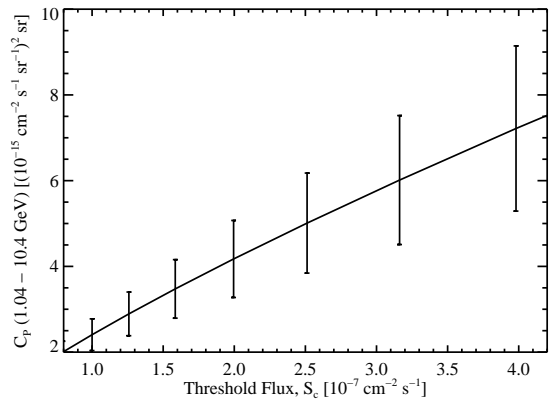}
\caption{Predicted amplitude of the source angular power spectrum, $C_{\rm P}$ (see Eq.~\ref{eq:cps}),
for energies of 1.04--10.4~GeV as a function of a source detection threshold flux, $S_c$.
\label{fig:cps}}
\end{figure}

For our source population model we adopt the best-fit flux distribution for the high-latitude Fermi sources, reported in~\citep{Collaboration:2010gqa:srccounts}, which describes ${\rm d}N/{\rm d}S$ with a broken power-law model:
\begin{eqnarray}
\nonumber
 \frac{dN}{dS}&=& AS^{-\beta_1},\quad S\ge S_b,\\
&=& AS_b^{-\beta_1+\beta_2}S^{-\beta_2}, \quad S< S_b,
\label{eq:dnds}
\end{eqnarray}
which contains four free parameters, $A$, $\beta_1$, $\beta_2$, and
$S_b$. For this form of $dN/dS$, the source power spectrum can be found
analytically (for $S_c>S_b$):
\begin{equation}
\label{eq:cpanalytic}
C_{\rm P}=A\frac{S_c^{3-\beta_1}}{3-\beta_1}
\left[1-\frac{\beta_1-\beta_2}{3-\beta_2}\left(\frac{S_b}{S_c}\right)^{3-\beta_1}\right].
\end{equation}
A fit for the simulated source population for 1.04--10.4~GeV yields 
$A=(1.90\pm 0.48)\times 10^{-13}~(180/\pi)^2$, $\beta_1=2.213\pm 0.073$,
$\beta_2=1.533\pm 0.007$, and $S_b=1.41\times 10^{-9}~{\rm cm^{-2}~s^{-1}}$. 
Note that the errors are correlated. Fig.~\ref{fig:cps} shows the predicted $C_{\rm P}$ for
this source model for threshold fluxes in the range
$S_c=0.8\times 10^{-7} - 4.2\times 10^{-7}$~cm$^{-2}$~s$^{-1}$. 
The error bars are calculated from the full covariance matrix of the above parameters.  Although we have used zero as the lower limit of the integral in Eq.~\ref{eq:cps}, using the actual lower limit of the flux distribution adopted for the simulated population results in a negligible difference in the predicted $C_{\rm P}$.

\begin{figure}
\includegraphics[width=0.45\textwidth]{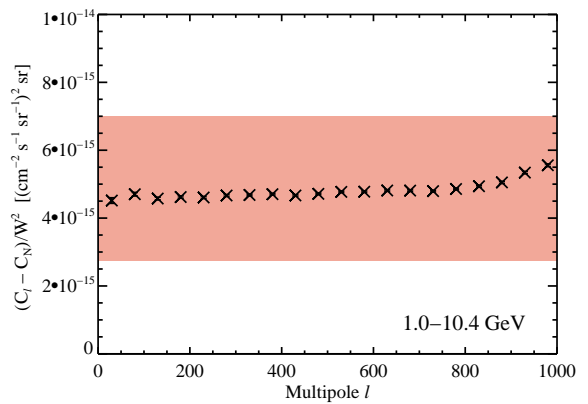}
\caption{Intensity angular power spectrum of a simulated observation of the source population model, compared with the theoretical prediction (\emph{shaded band}).  The angular power spectrum of the simulated population is in excellent agreement with the prediction over a large multipole range.  
\label{fig:simsrc}}
\end{figure}

We simulated this source population model with \emph{gtobssim} using the same procedure as described in \S\ref{sec:sims}.  The simulated population comprises nearly 20k point sources distributed randomly across the entire sky, with each source's flux drawn from the flux distribution specified above.  The photon spectrum of each source is modeled as a power law with a spectral index $\Gamma$ (${\rm d}N/{\rm d}E \propto E^{-\Gamma_{\rm s}}$) drawn from a Gaussian distribution with mean of 2.40 and a standard deviation of 0.28.  The simulated events were processed and the angular power spectrum of this source model calculated using the same procedure as was used for the data and other simulations in this study, except that the energy range of the map was chosen to be 1.04--10.4~GeV, and no mask was applied.

The fluxes of the $\sim$ 20k simulated sources were drawn from a flux distribution in which the maximum possible flux ($E > 100$ MeV) that could be assigned to a source was $10^{-5}$ cm$^{-2}$ s$^{-1}$, however the maximum flux of any source in the simulation, which represents a single realization of this source population, was $\sim 3 \times 10^{-6}$ cm$^{-2}$ s$^{-1}$.  We take these values as the upper and lower bound on the source detection threshold flux ($E > 100$ MeV) corresponding to the simulated model, since we do not impose a source detection threshold by masking or otherwise excluding simulated sources above a specific threshold flux.  A spectral index $\Gamma=2.4$ is assumed to determine the threshold fluxes in the 1.04--10.4~GeV energy band.  From these threshold fluxes we calculate the corresponding upper and lower bound on the predicted $C_{\rm P}$ in the 1.04--10.4~GeV energy band.

The angular power spectrum for the simulated source population calculated via the analysis pipeline used in this study is presented in Fig.~\ref{fig:simsrc}, with the shaded region indicating the predicted range of $C_{\rm P}$ (the mean values of $C_{\rm P}$ at the upper and lower flux threshold); for a given model $C_{\rm P}$ is independent of multipole, thus we expect the recovered angular power spectrum to be independent of multipole with amplitude within the shaded region.  The angular power spectrum recovered from the simulated data is in excellent agreement with the prediction up to multipoles of $\ell \sim 800$.   Above $\ell \sim 800$, the upturn in the measured angular power spectrum is likely due to inaccuracies in the modeling of the beam window function, which can introduce features on very small angular scales.  In the remainder of this study, we present results only for the multipole range $\ell = 5$ to $\ell = 504$.

\subsection{Sensitivity to the exposure map calculation}
\label{sec:datashuff}

\begin{figure*}
\includegraphics[width=0.45\textwidth]{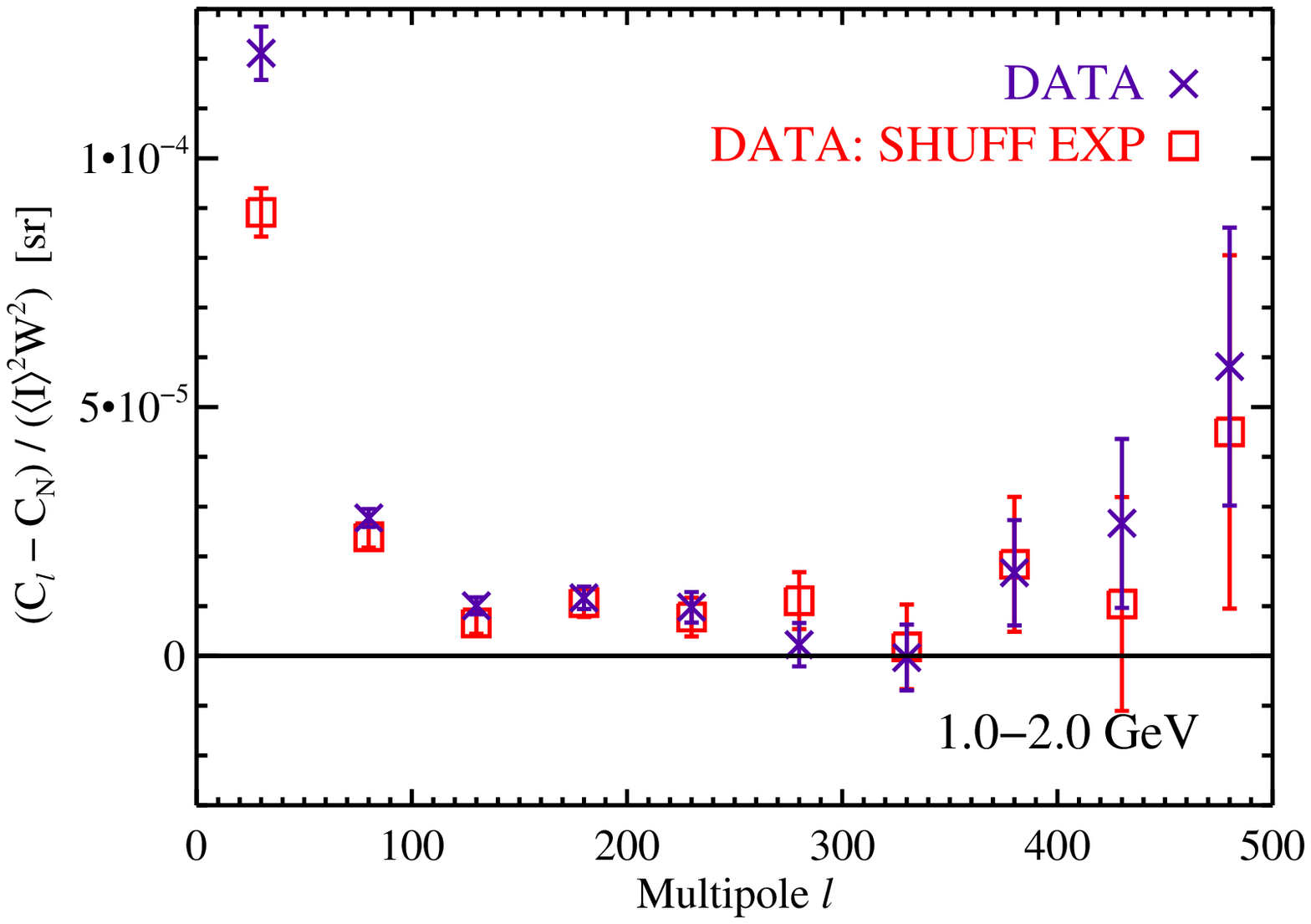}
\includegraphics[width=0.45\textwidth]{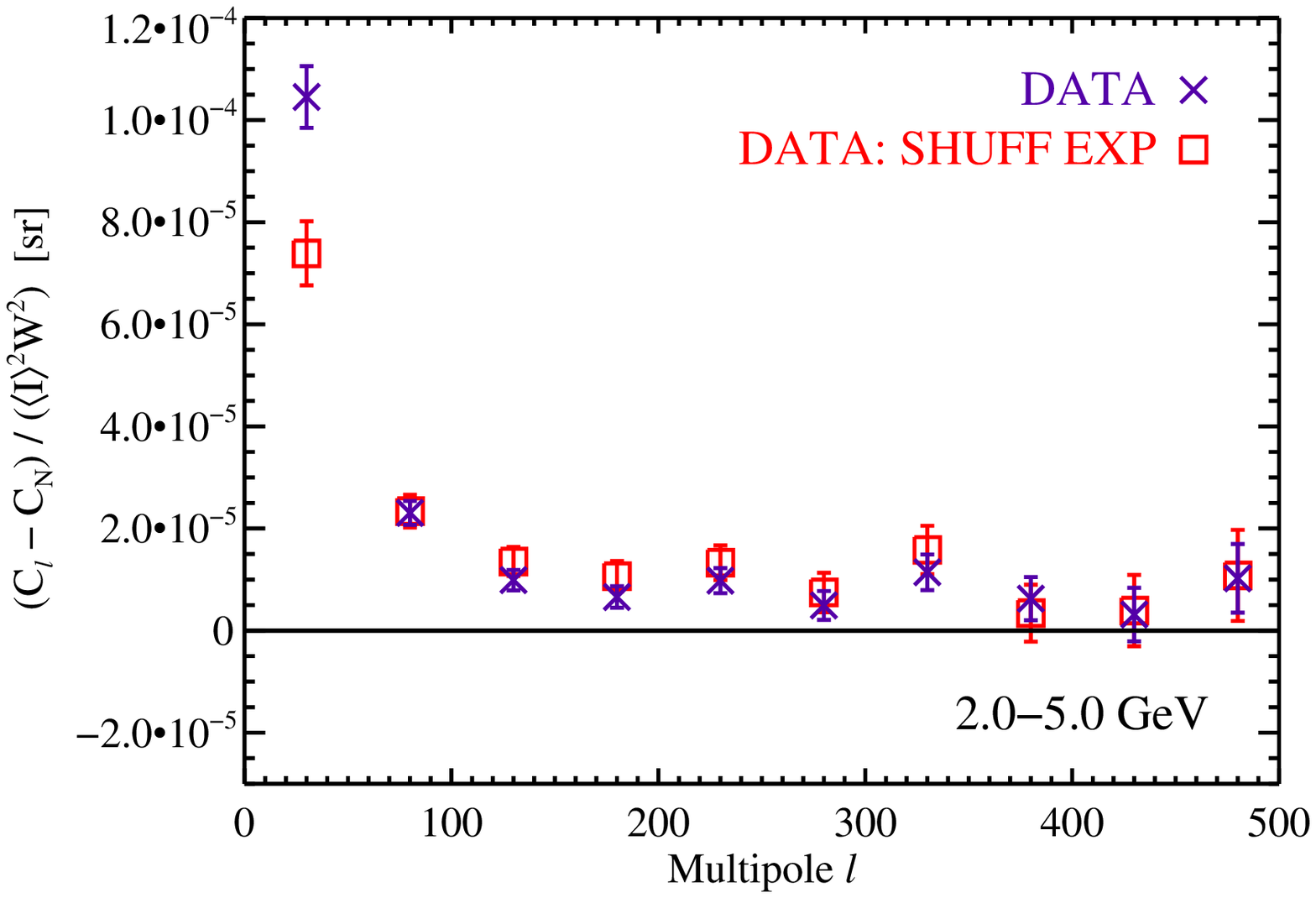}
\includegraphics[width=0.45\textwidth]{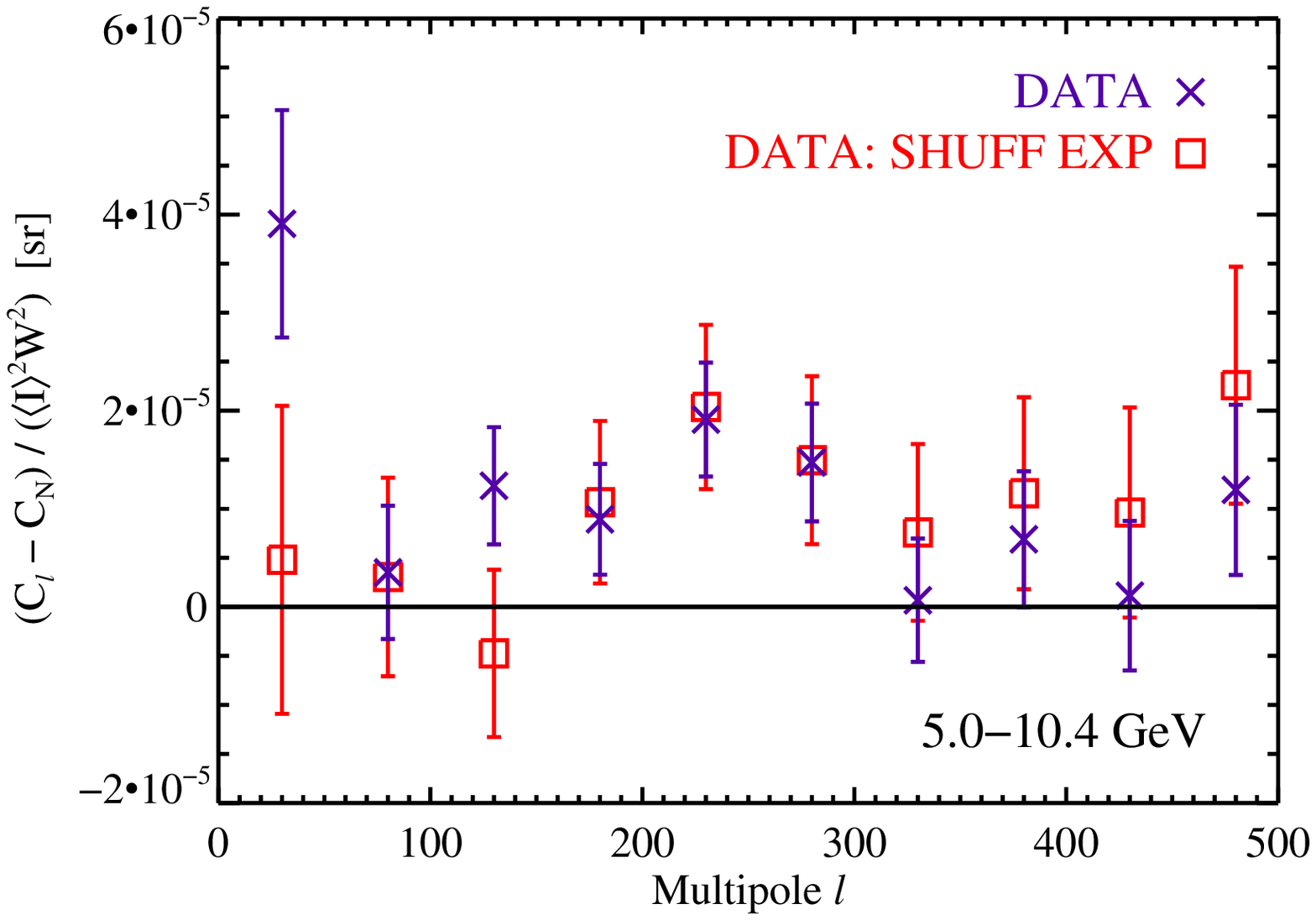}
\includegraphics[width=0.45\textwidth]{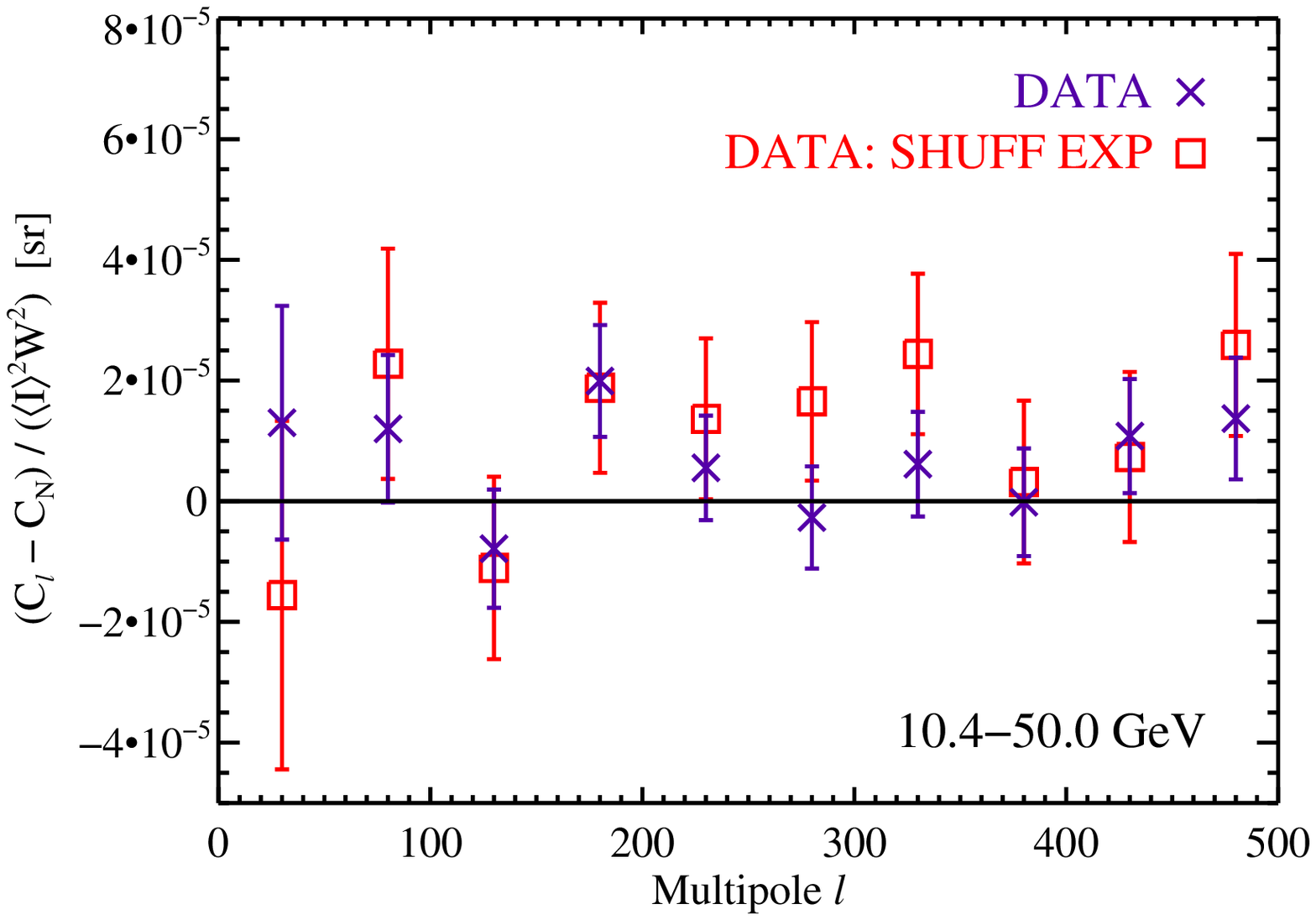}
\caption{Fluctuation angular power spectra $C_{\ell}/\langle I \rangle^{2}$ calculated using the default analysis pipeline compared with those obtained using the exposure map from the event shuffling technique described in \S\ref{sec:shuffle}.  Angular power is measured in all four energy bins by both analysis methods.  The lack of significant differences at the multipoles of interest between the angular power spectra yielded by the two methods demonstrates that any inaccuracies in the exposure map have a negligible impact on the measured angular power spectra.  Expanded versions of the top panels are shown in Fig.~\ref{fig:datashuffzoom}.\label{fig:datashuff}}
\end{figure*}

\begin{figure*}
\includegraphics[width=0.45\textwidth]{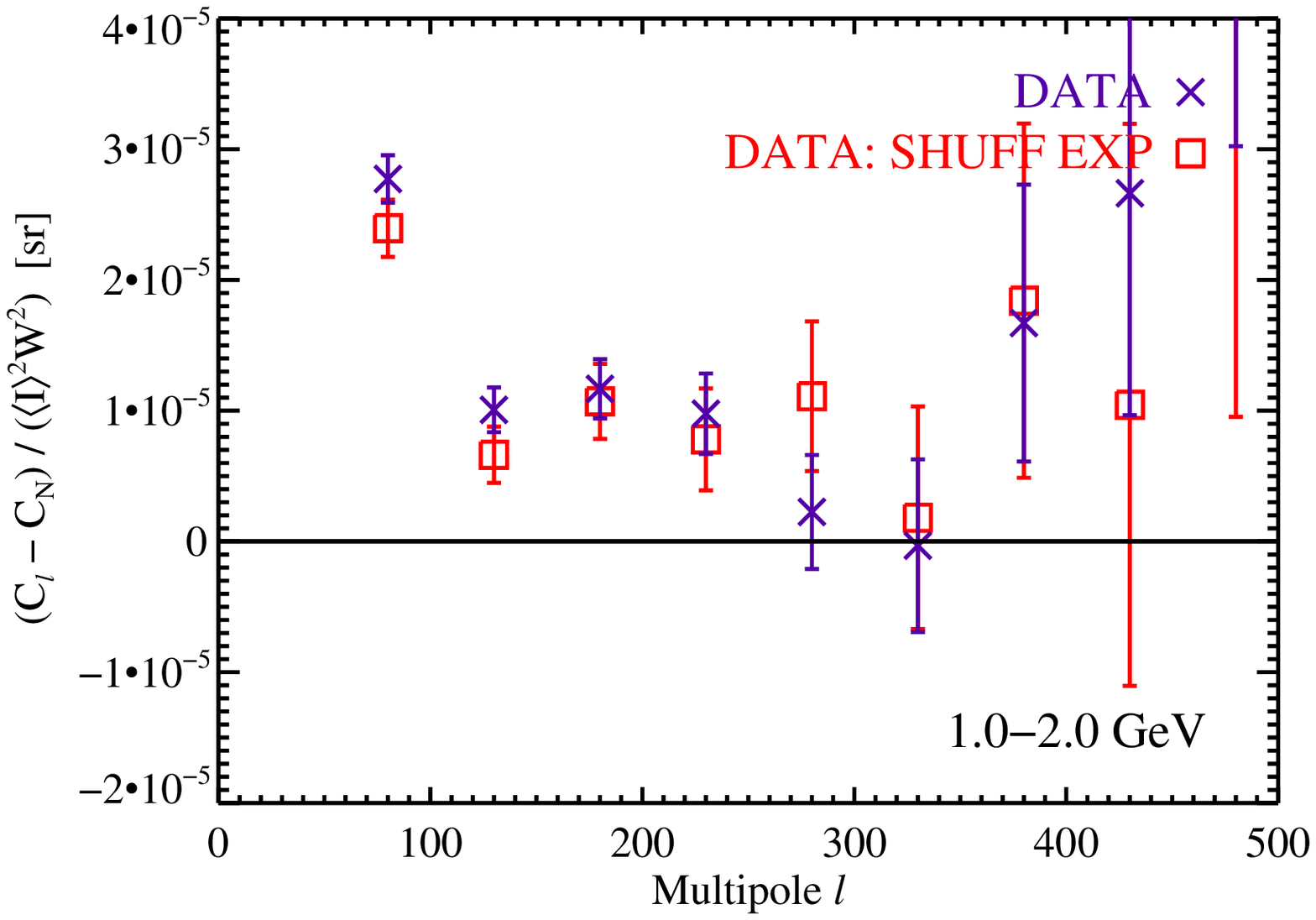}
\includegraphics[width=0.45\textwidth]{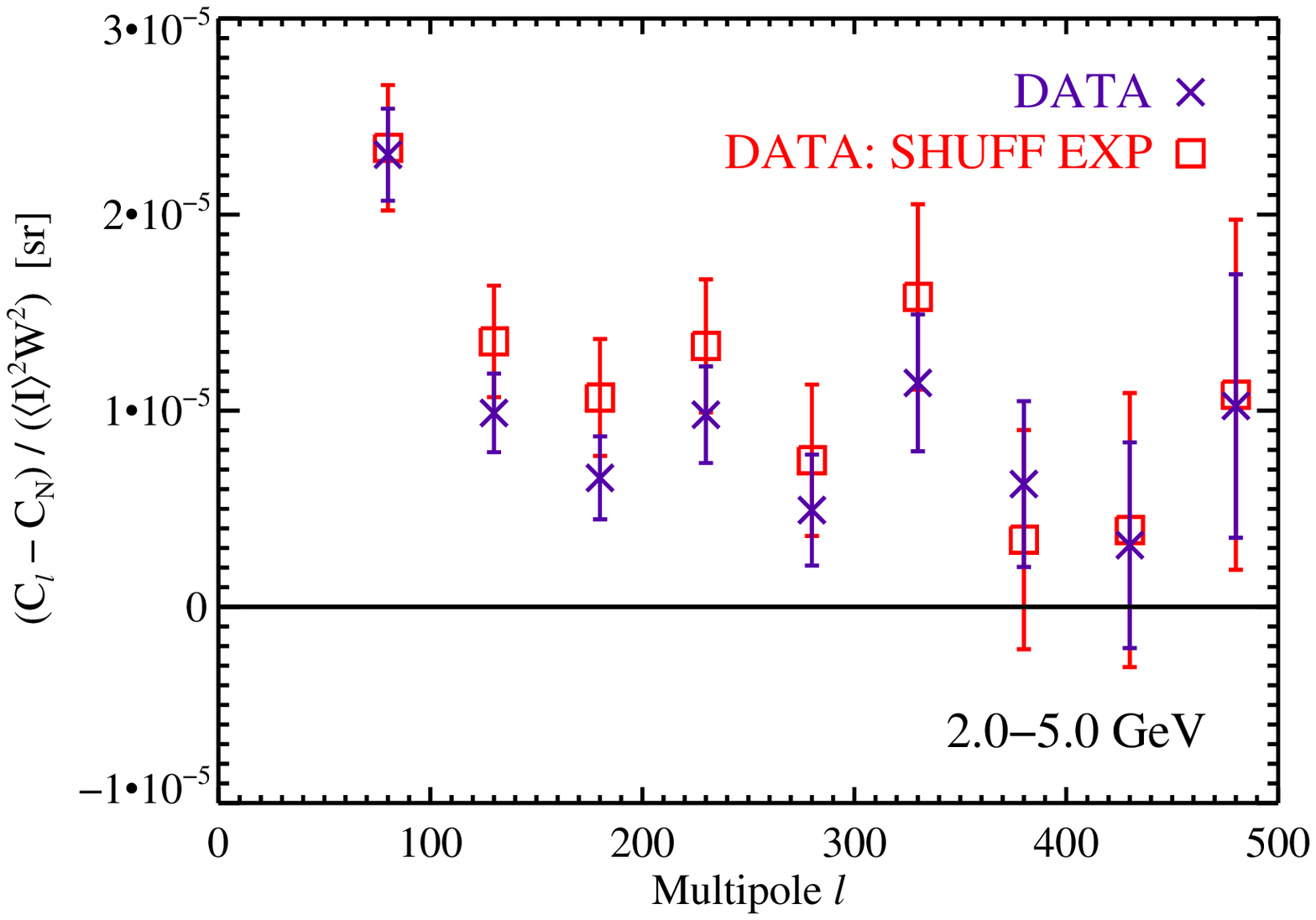}
\caption{Expanded versions of top panels of Fig.~\ref{fig:datashuff}.  
\label{fig:datashuffzoom}}
\end{figure*}

To investigate the possibility that potential inaccuracies in the exposure map calculation for the default analysis might generate spurious anisotropy in the intensity maps, we compare the fluctuation angular power spectra of the data using our default analysis pipeline with the results obtained after replacing the default exposure map with that generated by the event shuffling technique described in~\S\ref{sec:shuffle}.  This is shown in Figs.~\ref{fig:datashuff} and~\ref{fig:datashuffzoom}.  In these two figures only, the results from the default data analysis were obtained from maps of HEALPix resolution $N_{\rm side}=256$ to match the resolution of the maps using the exposure determined from the shuffling technique.  All other results presented in this study were obtained from $N_{\rm side}=512$ maps.  Due to the reduced map resolution, the pixel window function has a small effect on the angular power spectra shown in Figs.~\ref{fig:datashuff} and~\ref{fig:datashuffzoom}, however it affects the results of the default analysis and the analysis using the shuffled exposure map in the same way, and so these results can still be directly compared for the purpose of checking the effect of the exposure map calculation.

The results of the two analysis methods are in good agreement at all energies and multipoles considered, except for slight deviations at $\ell < 55$ for 1--5 GeV.  We caution that at these low multipoles the measured angular power spectra may be strongly affected by the mask, which has features on large angular scales.  The slight differences in the data selection cuts for the analysis using the exposure map from the shuffling technique compared to those for the default data analysis could lead to the observed differences in the low-multipole angular power spectra.  The differences could also result from systematics in the Monte-Carlo--based exposure calculation implemented in the Science Tools, leading to inaccuracies in the exposure map which vary on large angular scales.  As we do not focus on the low-multipole angular power in this study, we defer a full investigation of this issue to future work.  The agreement at $\ell \ge 55$ demonstrates that any potential spatially-dependent inaccuracies in the Science Tools exposure calculation have a negligible impact on the angular power spectra in the multipole range of interest.  In particular, from the consistency of the two methods we conclude that using the Monte-Carlo--based exposure calculation does not induce spurious signal anisotropy in our results.  

\subsection{Dependence on the PSF model}
\label{sec:irfcompare}

\begin{figure*}
\includegraphics[width=0.45\textwidth]{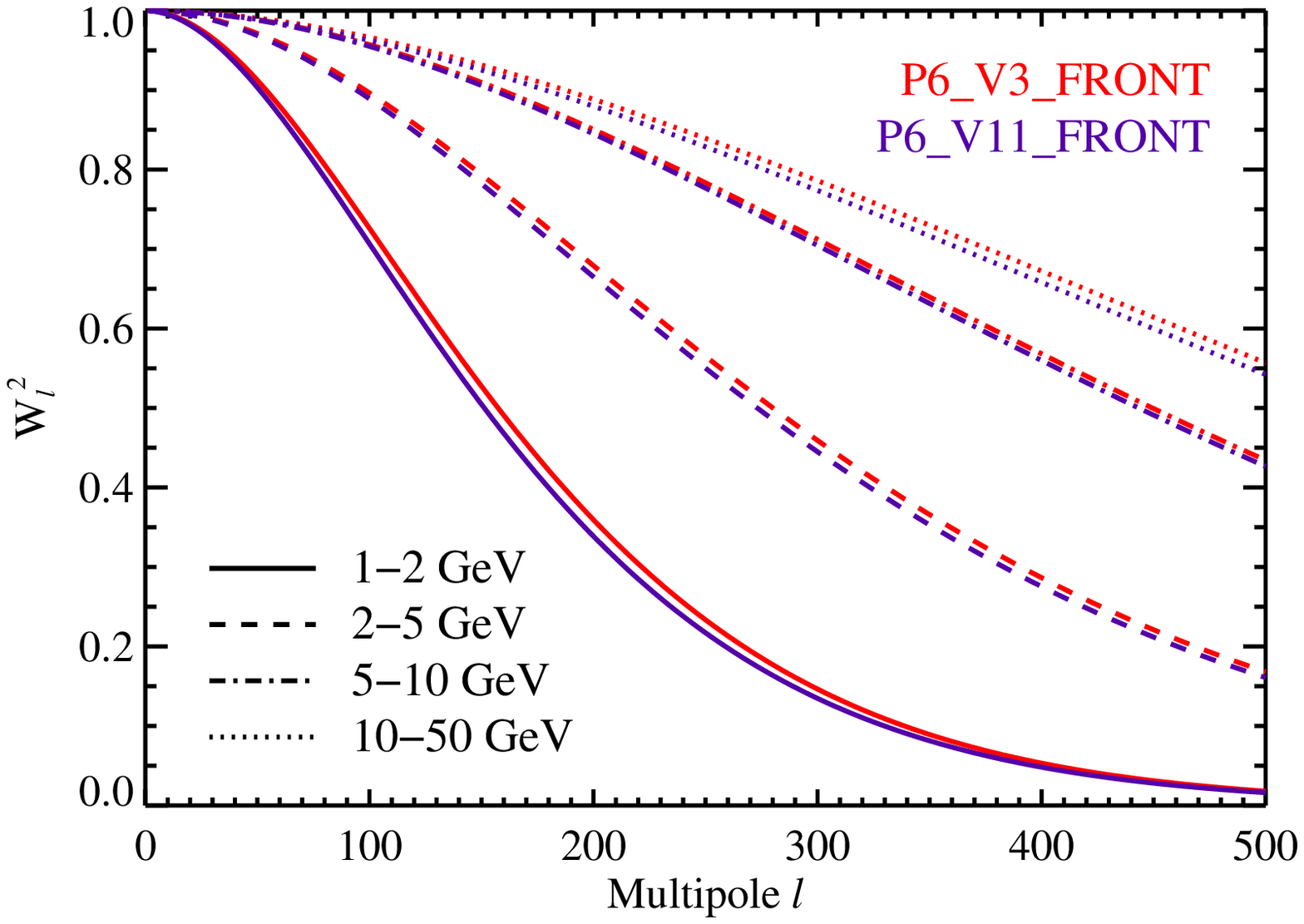}
\includegraphics[width=0.45\textwidth]{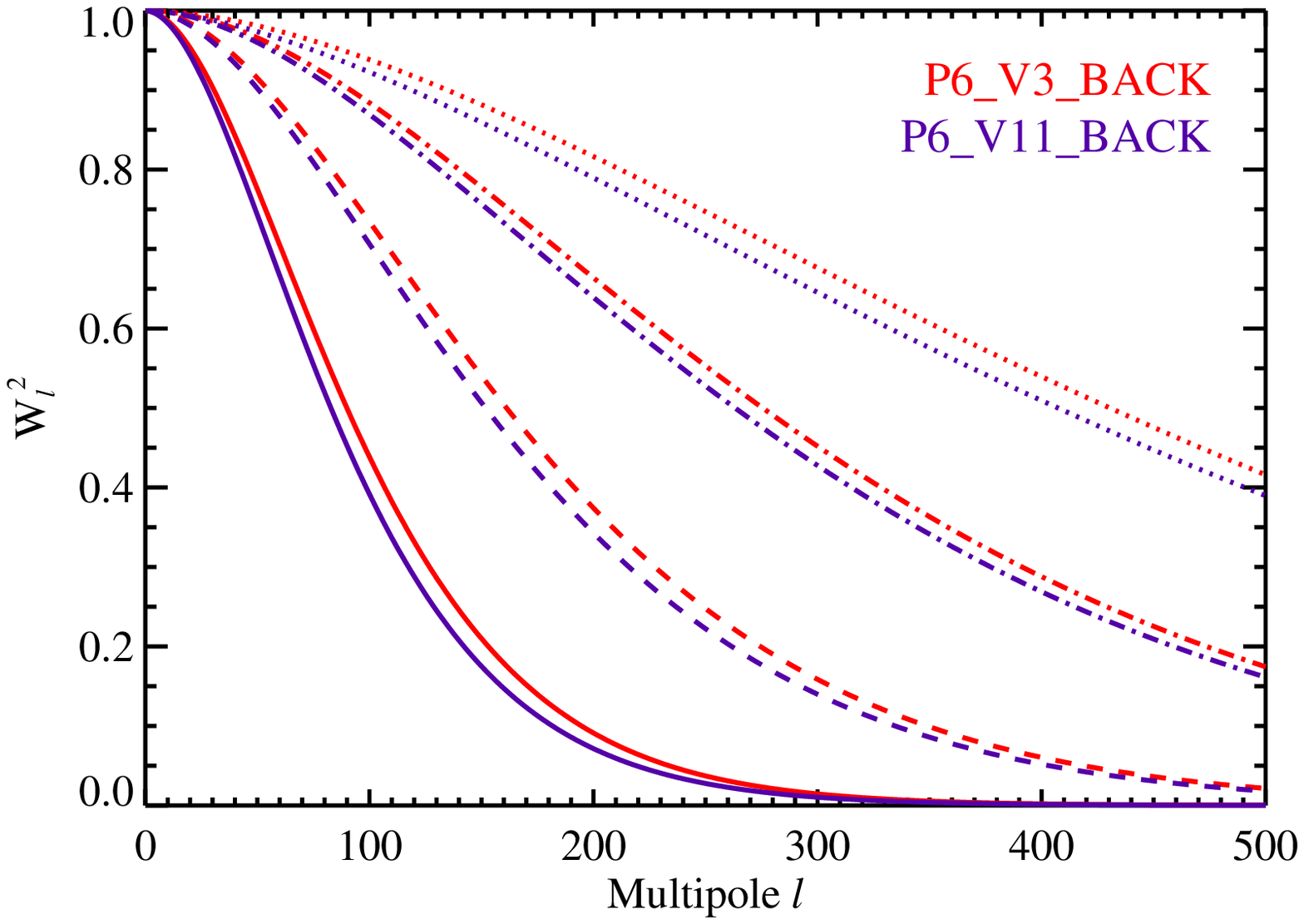}
\caption{Comparison of the beam window functions for the P6\_V3 and P6\_V11 IRFs; the P6\_V3 IRFs are the default used in this analysis.  The quantity $W_{\ell}^{2}$, which is the factor by which the angular power is suppressed due to the finite angular resolution of the instrument, is shown  for the front-converting (\emph{left panel}) and back-converting (\emph{right panel}) events, evaluated at the log-center of each energy bin used in this analysis.  The differences between the $W_{\ell}^{2}$ of these two IRFs are small ($\lesssim$~few percent) at all energies considered, indicating that our results are insensitive to the differences between the PSF models implemented in these IRFs.  
\label{fig:datairf}}
\end{figure*}

We examine the impact of variations in the assumed PSF on the results of the analysis by comparing the beam window functions (Eq.~\ref{eq:beamwindow}) for the PSF implemented in the P6\_V3 IRFs used in this analysis to those for the PSF in the more recently updated P6\_V11 IRFs.
The P6\_V11 IRFs use a modified functional form for the PSF, and for energies above 1~GeV the PSF implemented in P6\_V11 was calibrated using in-flight data, while in P6\_V3 the PSF was based on Monte Carlo simulations.  Fig.~\ref{fig:datairf} shows the beam window functions for the PSF associated with the front- and back-converting events for each set of IRFs, at the log center of each energy bin used in this analysis.  The small variation between the window functions of the two IRFs confirms that differences between the PSF models in these two IRFs are not large enough to affect the anisotropy measurement on the angular scales to which this analysis is sensitive.

\subsection{Dependence on the latitude mask}
\label{sec:latcompare}

\begin{figure*}
\includegraphics[width=0.45\textwidth]{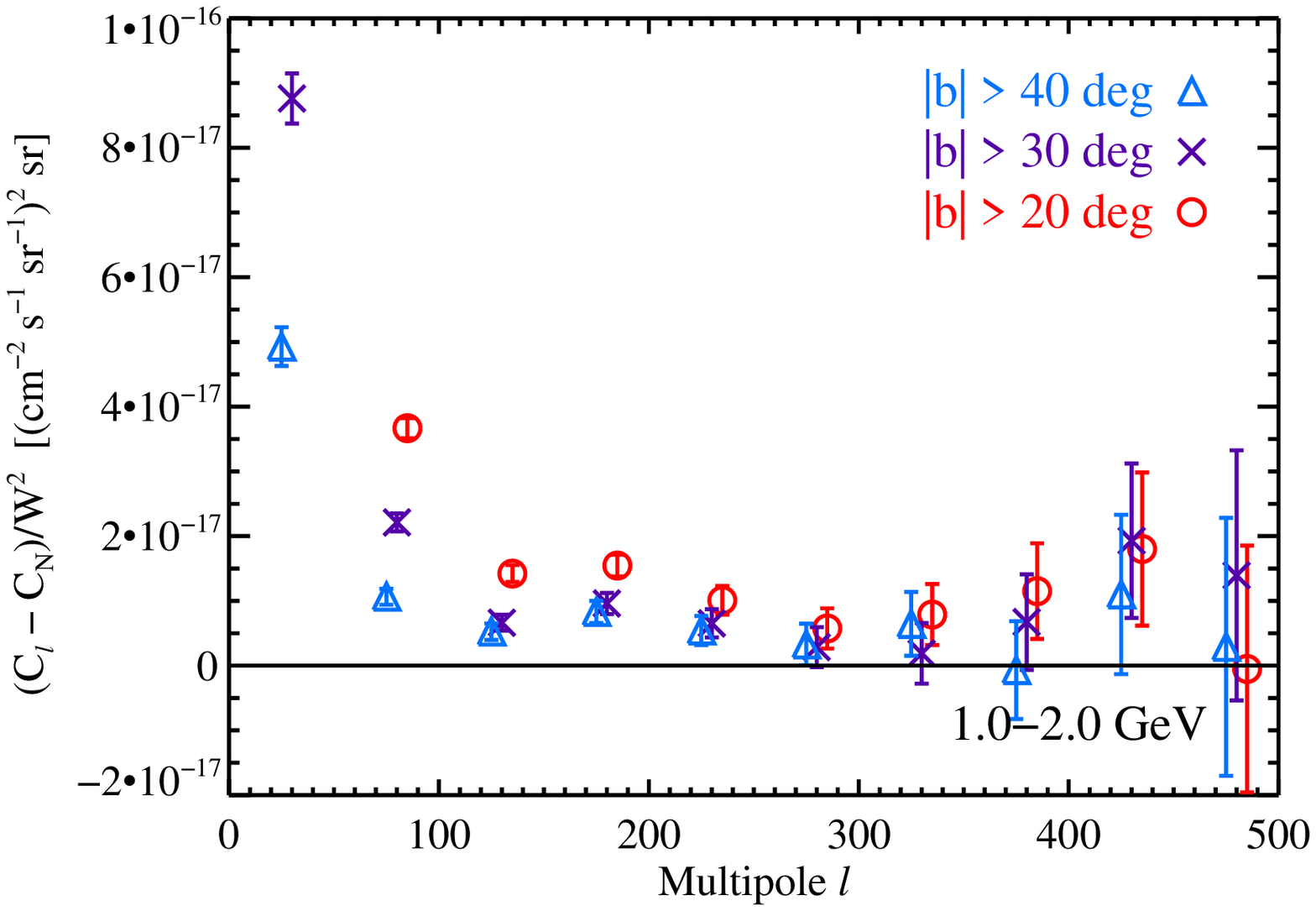}
\includegraphics[width=0.45\textwidth]{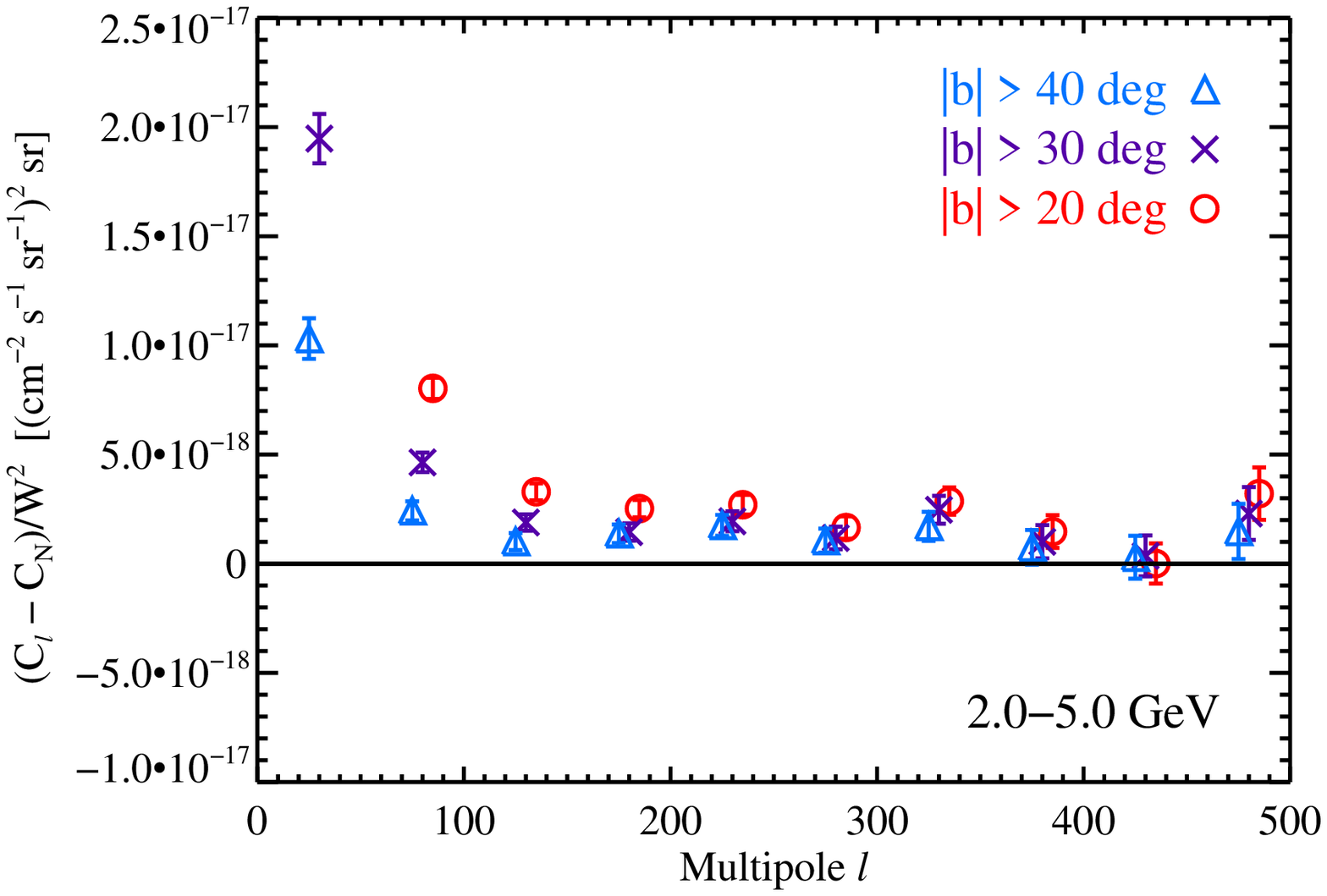}
\includegraphics[width=0.45\textwidth]{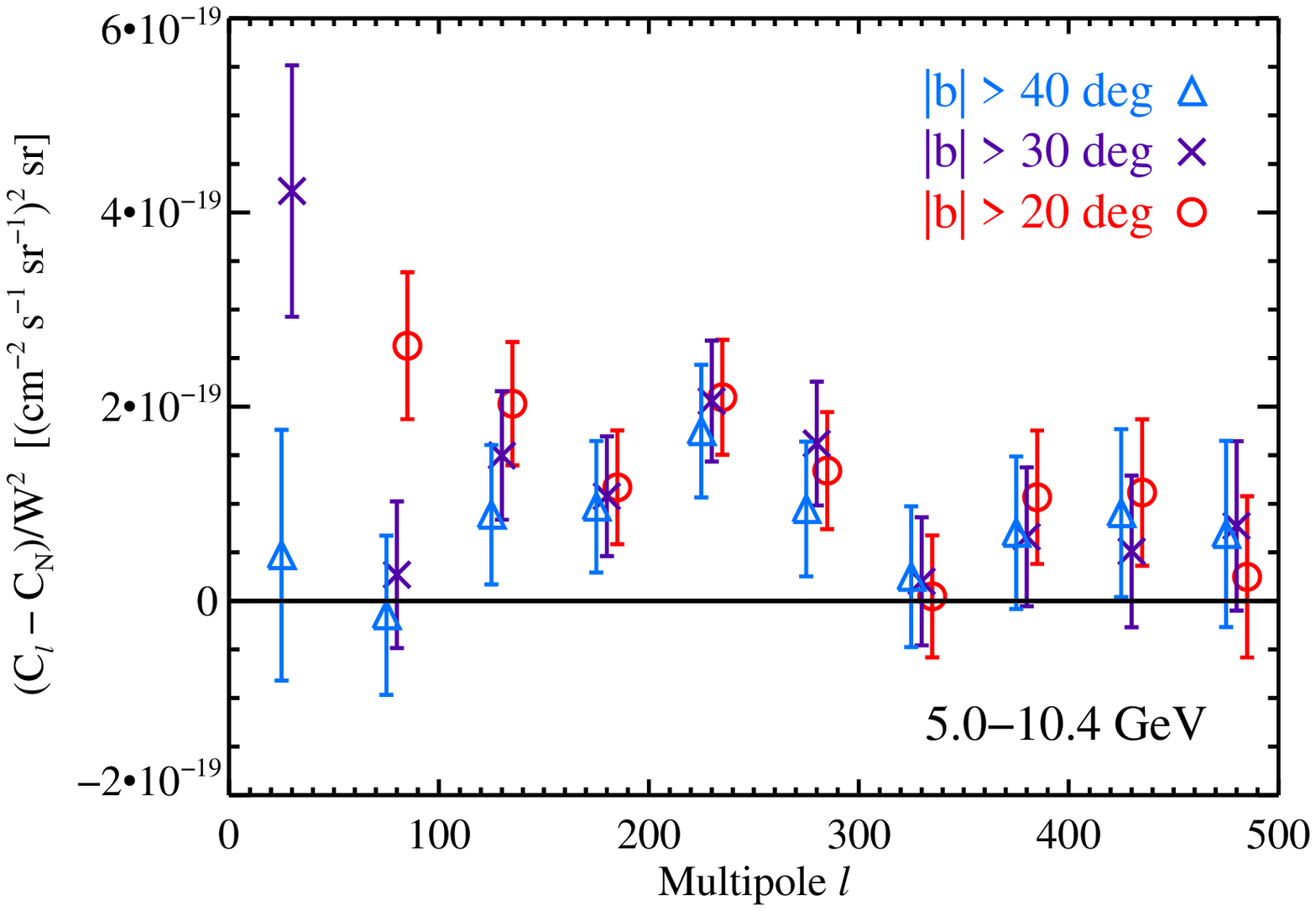}
\includegraphics[width=0.45\textwidth]{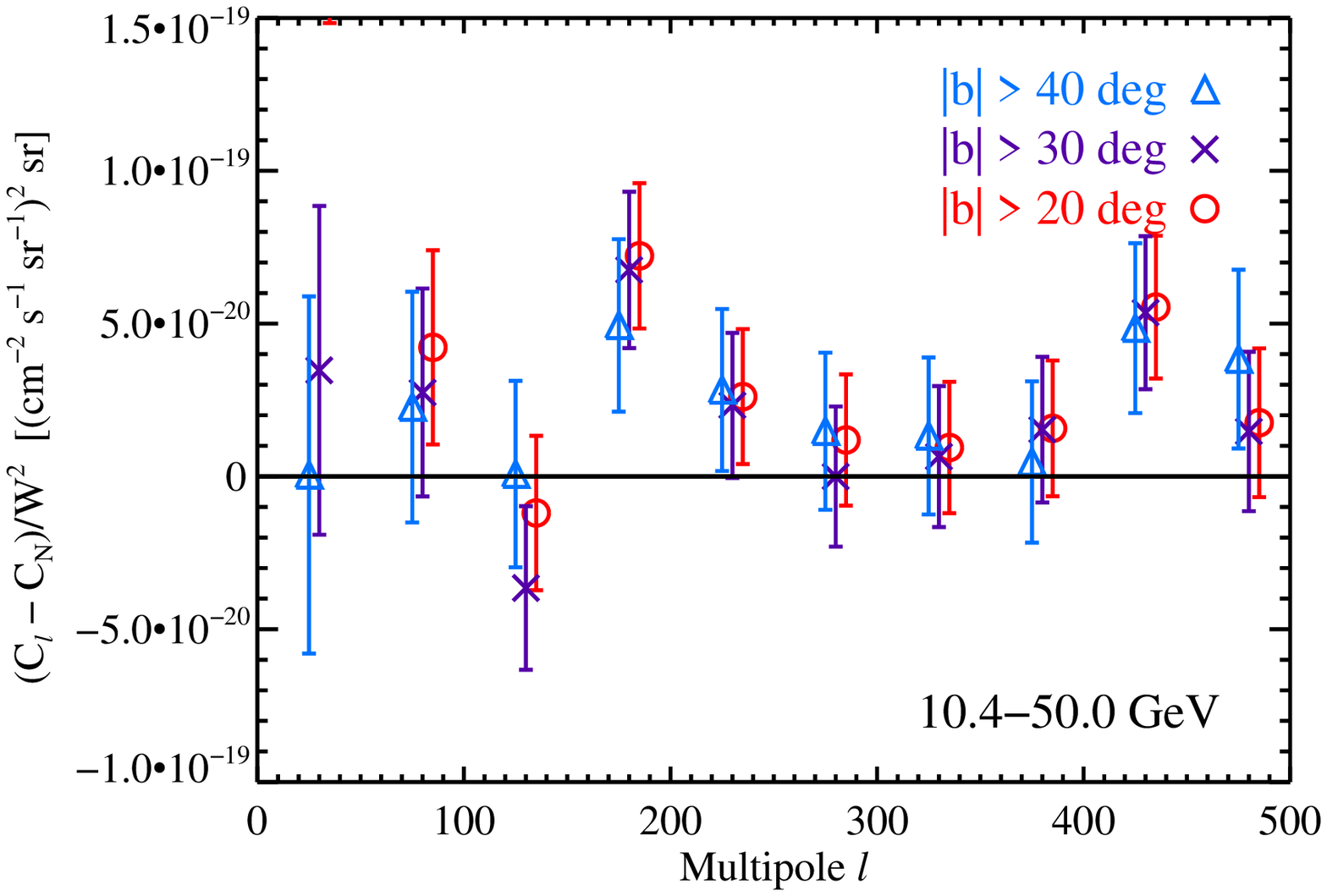}
\caption{Intensity angular power spectra of the data calculated with different latitude cuts.  The point source mask was applied in addition to the latitude mask in all cases.  The differences between the results masking $|b|<30^{\circ}$ (the default latitude cut) and $|b|<40^{\circ}$ are small for $\ell \ge 155$ for all four energy bins, demonstrating that the power observed in the data at these multipoles is not strongly correlated with a component that has a strong latitude dependence in the range $30^{\circ} < |b| < 40^{\circ}$, such as the Galactic diffuse emission.  At energies above 5~GeV convergence is seen for multipoles $\ell \ge 155$ even when masking only $|b| < 20^{\circ}$.  Points from different data sets are offset slightly in multipole for clarity.  The lowest multipole data point for the $|b| < 20^{\circ}$ mask in each panel is above the range shown in the figure.  Expanded versions of the top panels are shown in Fig.~\ref{fig:latcomparezoom}.\label{fig:latcompare}}
\end{figure*}

\begin{figure*}
\includegraphics[width=0.45\textwidth]{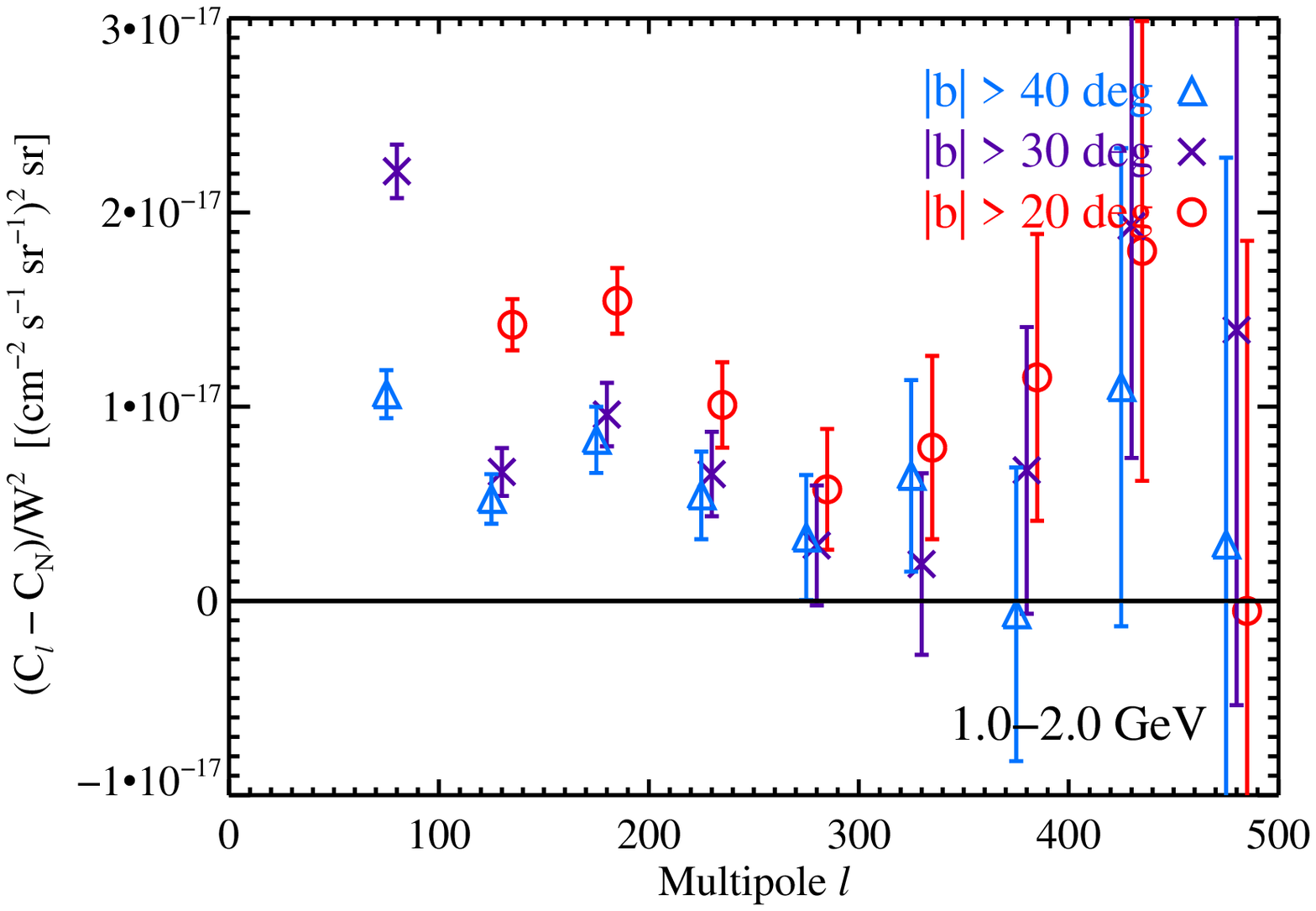}
\includegraphics[width=0.45\textwidth]{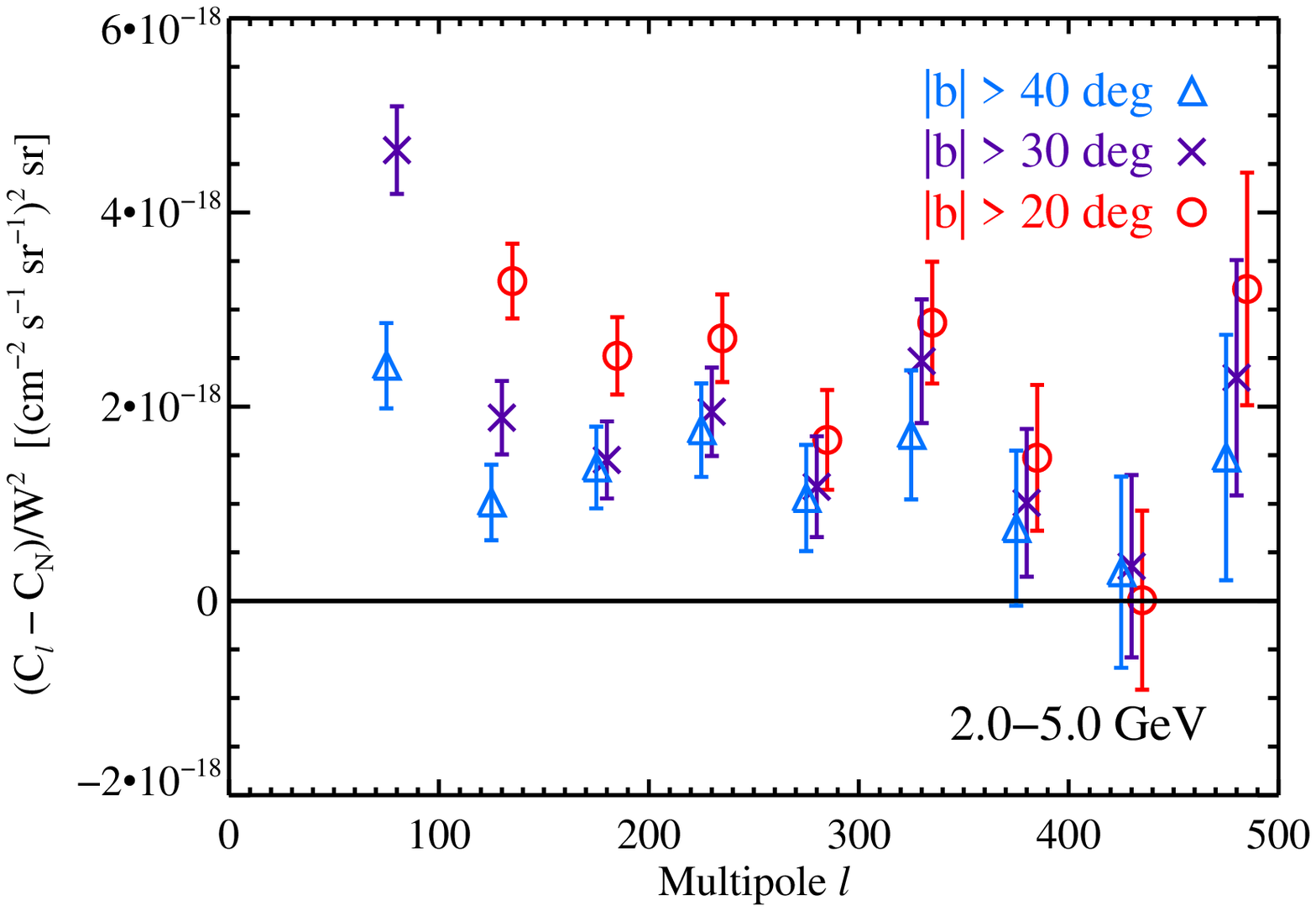}
\caption{Expanded versions of top panels of Fig.~\ref{fig:latcompare}.
\label{fig:latcomparezoom}}
\end{figure*}

In this analysis we apply a generous latitude mask to reduce contamination of the data by Galactic diffuse emission.  The mask is intended to remove enough contamination so that the measured angular power can be attributed to sources whose distribution is  statistically isotropic in the sky region we consider, i.e., a distribution which does not show any preferred direction on the sky.  In particular, we wish to exclude sources whose angular distribution exhibits a strong gradient with Galactic latitude.  The effectiveness of the mask at reducing the contribution to the angular power from a strongly latitude-dependent component can be evaluated by considering the angular power spectrum of the data as a function of latitude cut.  The results are shown in Figs.~\ref{fig:latcompare} and~\ref{fig:latcomparezoom}.

At low multipoles ($\ell \lesssim 100$), increasing the latitude cut significantly reduces the angular power, indicating that in this multipole range the contamination by a strongly latitude-dependent component, such as Galactic diffuse emission, is considerable.  For $155 \le \ell \le 254$ at 1--2~GeV and 2--5~GeV, the angular power measured using the $30^{\circ}$ latitude mask is noticeably smaller than when using the $20^{\circ}$ latitude mask.  However, at all energies there are no significant differences in the angular power measured for $\ell \ge 155$ using the $30^{\circ}$ and $40^{\circ}$ latitude masks, and for energies greater than 5~GeV the $20^{\circ}$ latitude mask also yields consistent results.  We conclude that applying the $30^{\circ}$ latitude mask is sufficient to ensure that no significant amount of the measured angular power at $\ell \ge 155$ originates from the Galactic diffuse emission or from any source class that varies greatly in the region $30^{\circ} < |b| < 40^{\circ}$. 

\subsection{Effects of masking on the power spectrum}
\label{sec:maskeffects}

\begin{figure*}
\includegraphics[width=0.45\textwidth]{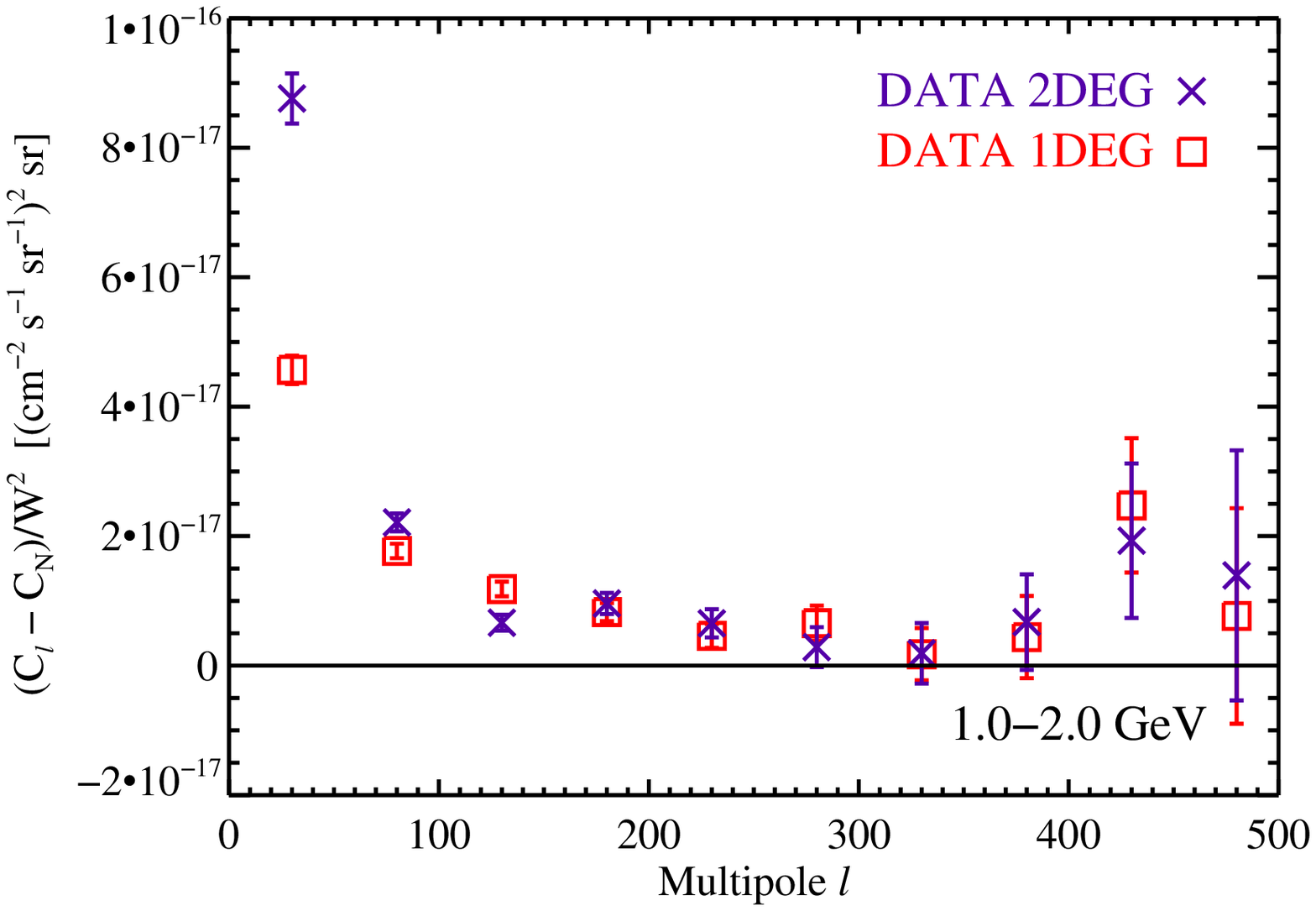}
\includegraphics[width=0.45\textwidth]{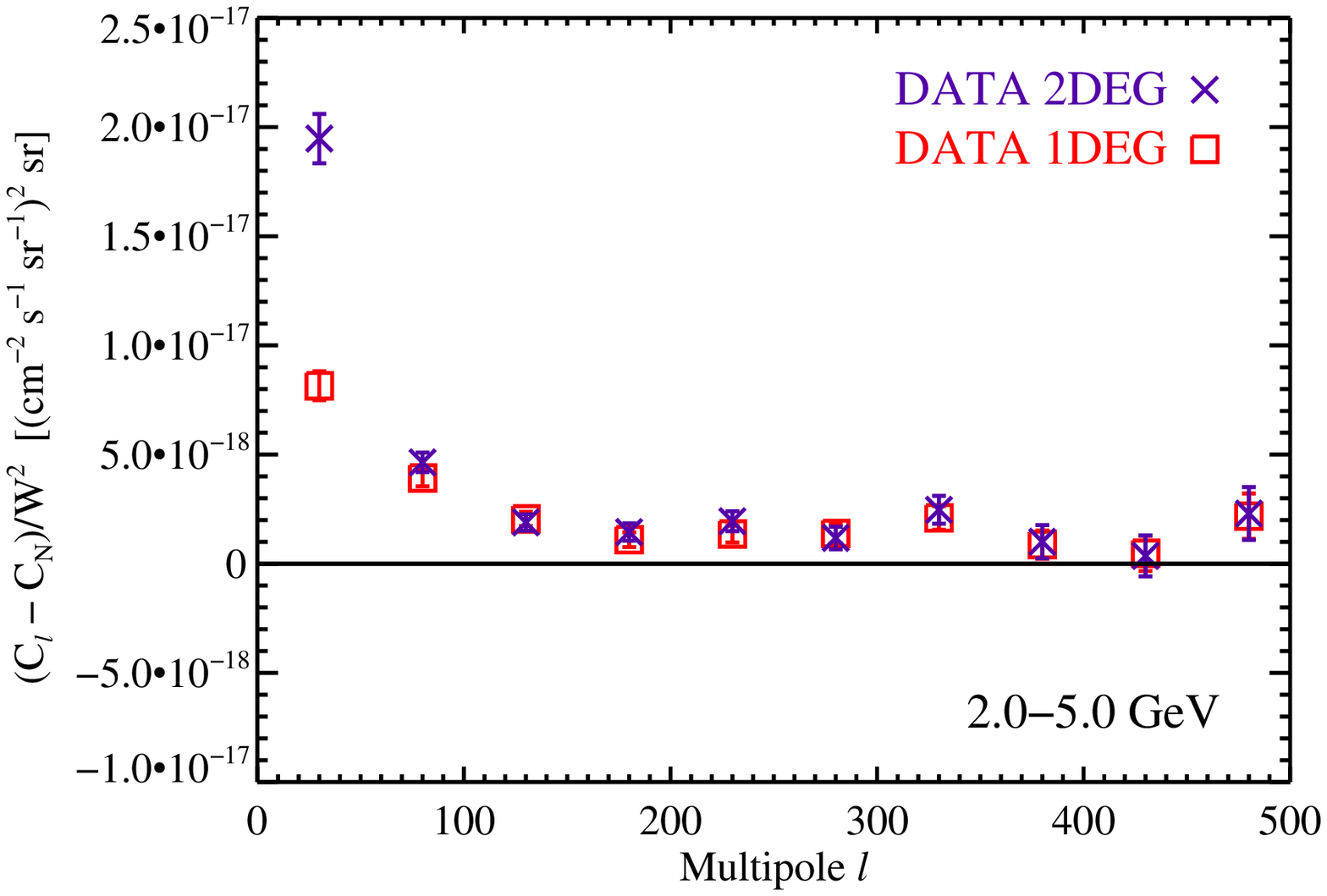}
\includegraphics[width=0.45\textwidth]{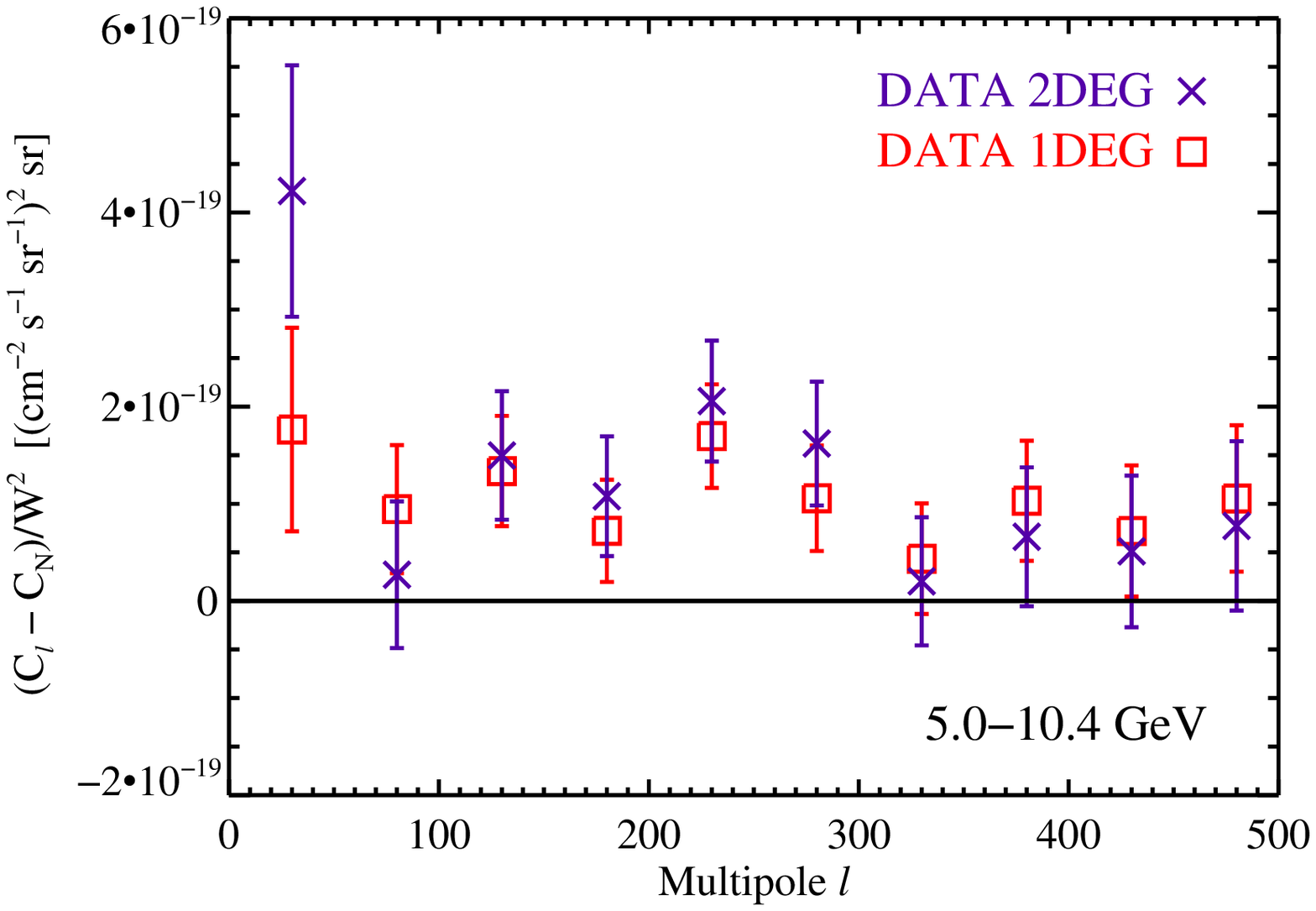}
\includegraphics[width=0.45\textwidth]{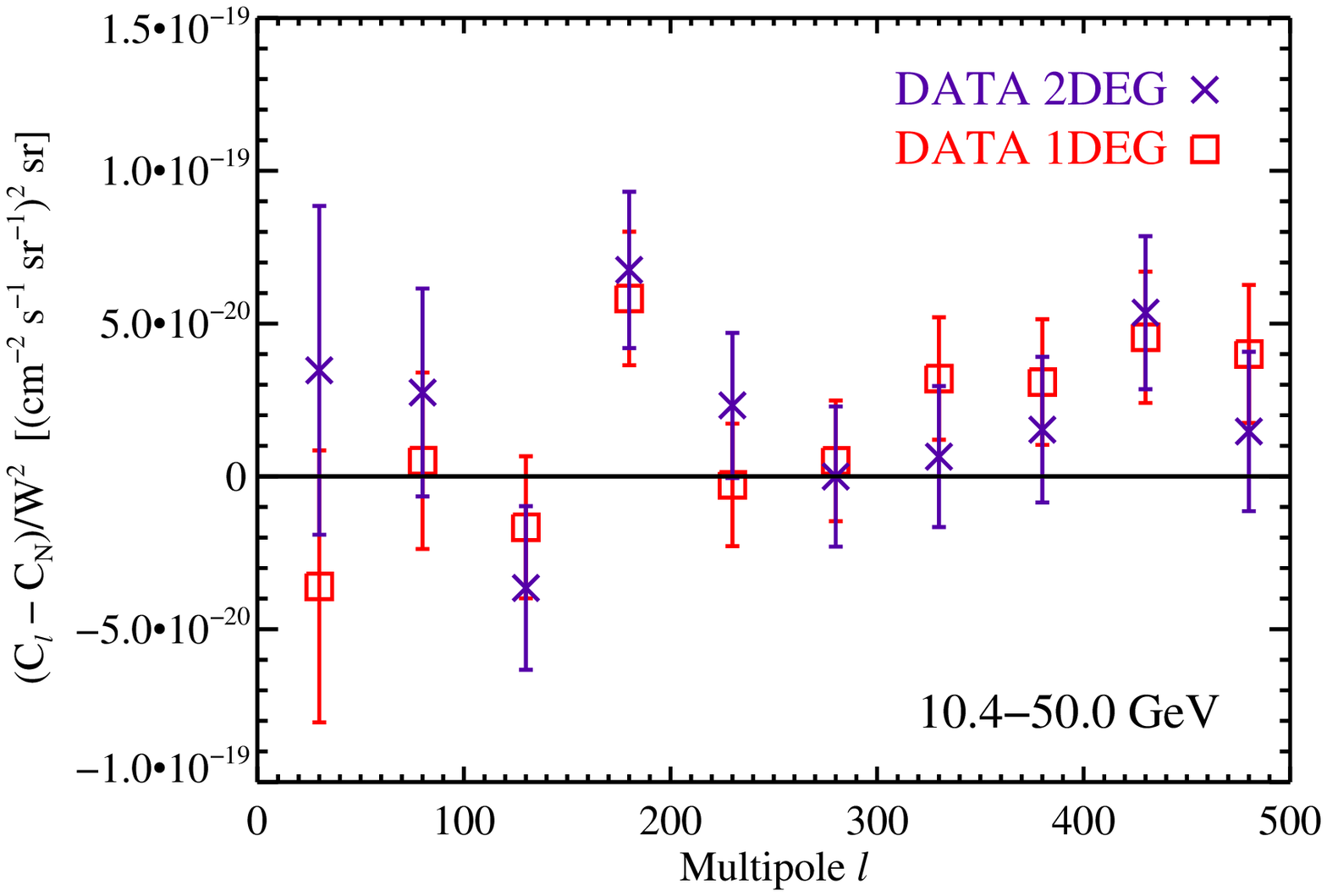}
\caption{Intensity angular power spectra of the data calculated with a mask excluding a 1$^{\circ}$ or 2$^{\circ}$ angular radius around each source; excluding a 2$^{\circ}$ angular radius is the default in this analysis.  The default latitude mask excluding $|b|<30^{\circ}$ was applied in addition to the source mask in all cases.  At all energies the angular power spectra obtained using the different source mask radii are consistent at $\ell \ge 155$ (the multipole range of interest), and above 2~GeV the results are consistent at $\ell \ge 55$.  Expanded versions of the top panels are shown in Fig.~\ref{fig:angradcomparezoom}.\label{fig:angradcompare}}
\end{figure*}

\begin{figure*}
\includegraphics[width=0.45\textwidth]{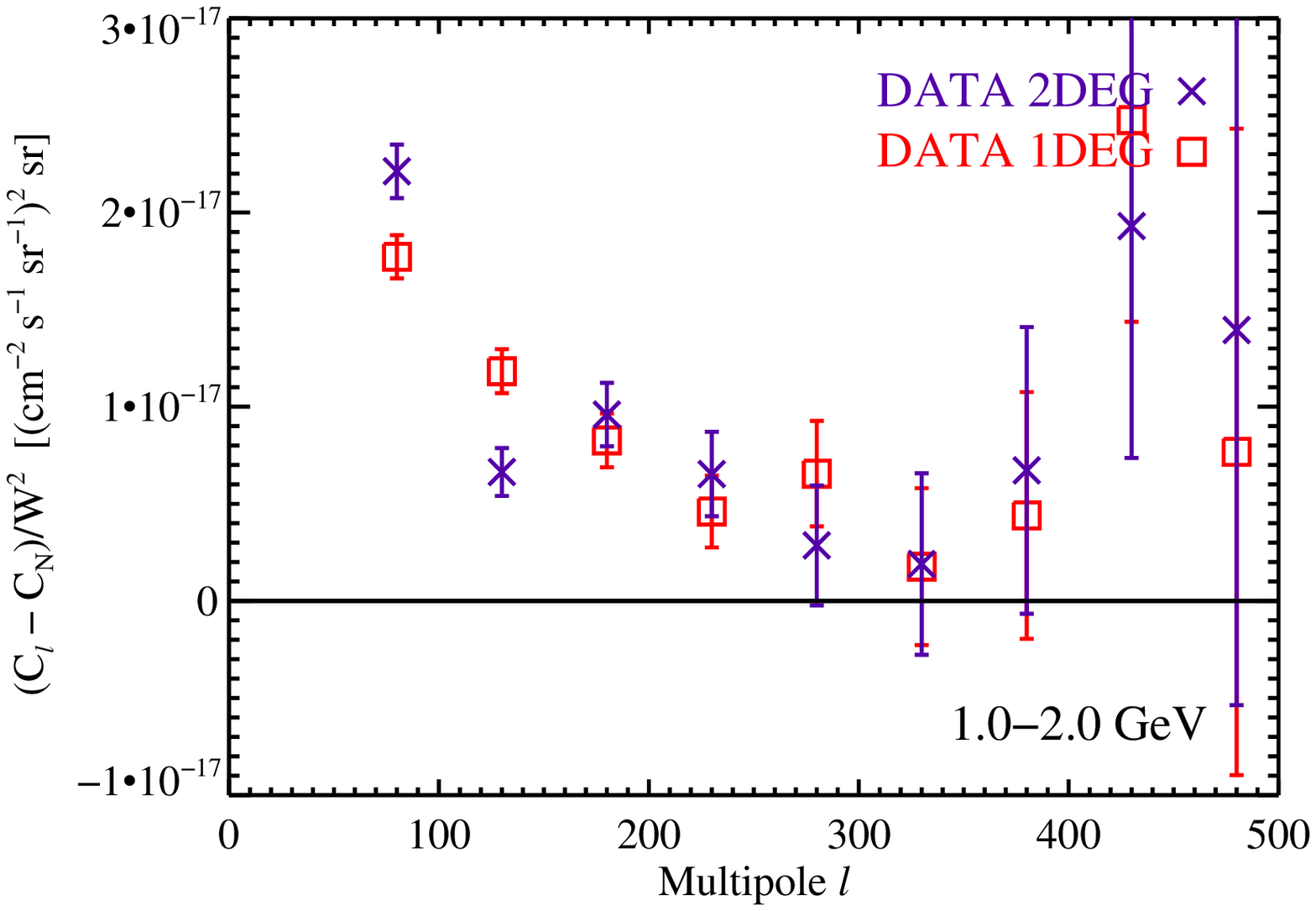}
\includegraphics[width=0.45\textwidth]{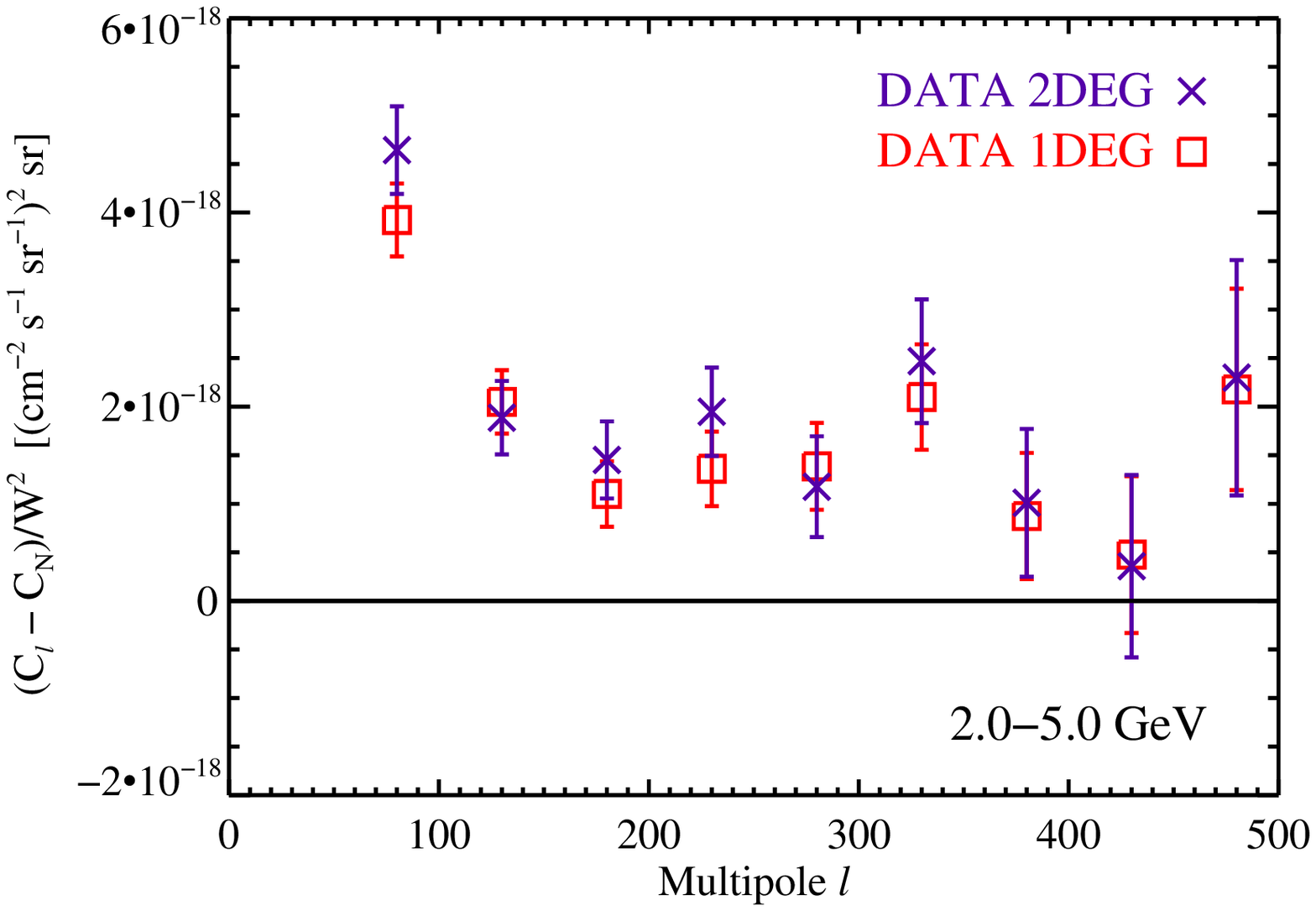}
\caption{Expanded versions of top panels of Fig.~\ref{fig:angradcompare}.
\label{fig:angradcomparezoom}}
\end{figure*}

To verify that the results do not depend sensitively on the angular radius of the source mask, in Figs.~\ref{fig:angradcompare} and~\ref{fig:angradcomparezoom} we compare the results when masking a 1$^{\circ}$ angular radius around each source with those when masking the 2$^{\circ}$ radius used as the default in this work.

In the 1--2~GeV energy bin the results show significant differences at $\ell < 155$, however for $\ell \ge 155$ (the multipole range of interest) the angular power spectra for the 1$^{\circ}$ and 2$^{\circ}$ source mask cases agree within the error bars.  In the higher energy bins the angular power spectra in all except the first multipole bin ($5 \le \ell < 55$, well below the range of interest) agree within the error bars.  Since varying the angular size of the region masked around each source does not significantly change the measured angular power at $\ell \ge 155$, we conclude that any features that may be induced in the angular power spectra by the morphology of the source mask are confined to low multipoles and therefore do not affect the measurements of $C_{\rm P}$ reported in this work.

In addition, we have confirmed that the angular power spectra of the front- and back-converting events are in good agreement within each energy bin in the multipole range of interest ($\ell \ge 155$), and are generally consistent at $\ell < 155$ even in the 1--2~GeV energy bin where the 95\% containment radius of the PSF of the back-converting events is comparable to the angular radius used for the source mask.   Consequently, although the PSF associated with the back-converting events is larger than that of the front-converting events, the consistency of their angular power spectra implies that the source masking is sufficiently effective even at low energies.

The sharp latitude cut used in this analysis also has the potential to induce features in the angular power spectrum, although these would be expected to appear on the large angular scales characteristic of the morphology of the mask.  We therefore note that the stability of the angular power spectra at $\ell \ge 155$ for latitude cuts masking at least $|b| \le 30^{\circ}$, discussed in \S\ref{sec:latcompare} and demonstrated in Figs.~\ref{fig:latcompare} and~\ref{fig:latcomparezoom}, indicates that the latitude mask does not induce features in the power spectrum at the angular scales of interest.

The analysis of the simulated isotropic component, presented in \S\ref{sec:datamodel} and in Figs.~\ref{fig:modelcomps} and~\ref{fig:modelcompszoom}, provides another means of assessing the impact of the mask on the angular power spectra.  Since the isotropic component should only contribute to the monopole ($\ell=0$) term of the power spectrum, statistically significant deviations from zero power at $\ell > 0$ can be attributed to the use of the mask.  We emphasize that the consistency of the angular power of the isotropic component with zero at $\ell \ge 155$ indicates that, despite the complex morphology of the total mask, the mask does not induce features in the angular power spectrum at the multipoles of interest ($\ell \ge 155$). 

\subsection{Dependence on the set of masked sources}
\label{sec:1fgl2fgl}

The recently-released second Fermi LAT source catalog (2FGL)~\citep{Collaboration:2011bm} is an update to the 1FGL catalog used to define the default source mask adopted in this work.  The 2FGL catalog reports the detection of 1873 sources, compared to the 1451 included in the 1FGL catalog.

We briefly comment that one motivation for using the 1FGL catalog, rather than the 2FGL catalog, to define the source mask in our default analysis is that the 1FGL catalog was also used in the Fermi LAT source count distribution analysis~\citep{Collaboration:2010gqa:srccounts}.  The results of that study are closely related to the interpretation of the results of the current analysis, and so our choice to mask that same source list in our default analysis allows the results of the two analyses to be used together straightforwardly.  However, it is natural to ask to what extent the measured angular power reported in the data may be attributable to the additional sources resolved in the 2FGL catalog.

We address this question by analyzing the data using a source mask defined by the 2FGL sources and comparing the results to those obtained using the 1FGL source mask.  We repeat the analysis of the data using the default pipeline, changing only the source mask; the total mask is defined by the source mask combined with the default latitude cut masking $|b|<30^{\circ}$.  When combined with the default latitude cut, the 2FGL source mask results in an unmasked sky fraction $f_{\rm sky}=0.295$, a small decrease compared to $f_{\rm sky}=0.325$ when using the 1FGL source mask.

\begin{figure*}
\includegraphics[width=0.45\textwidth]{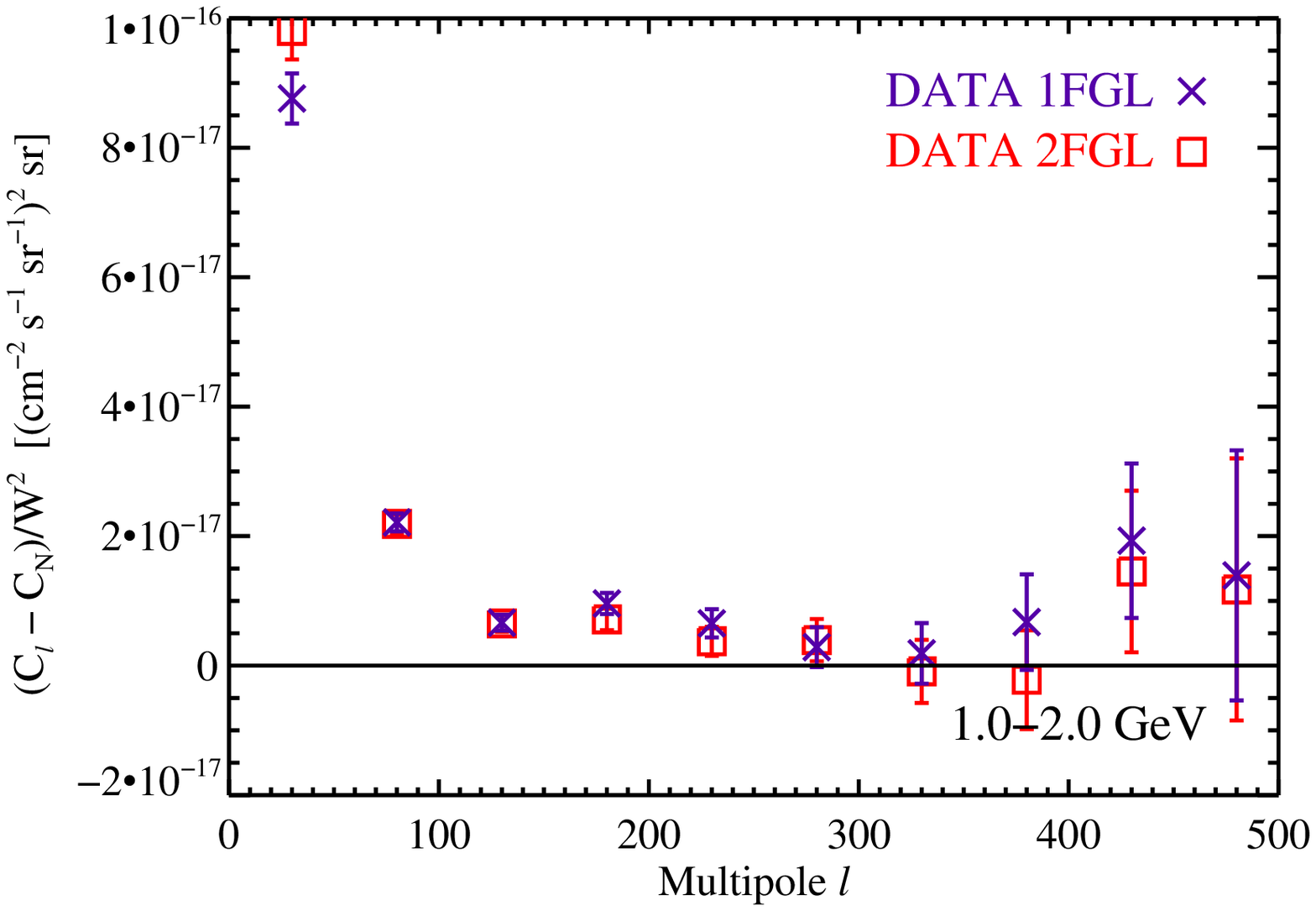}
\includegraphics[width=0.45\textwidth]{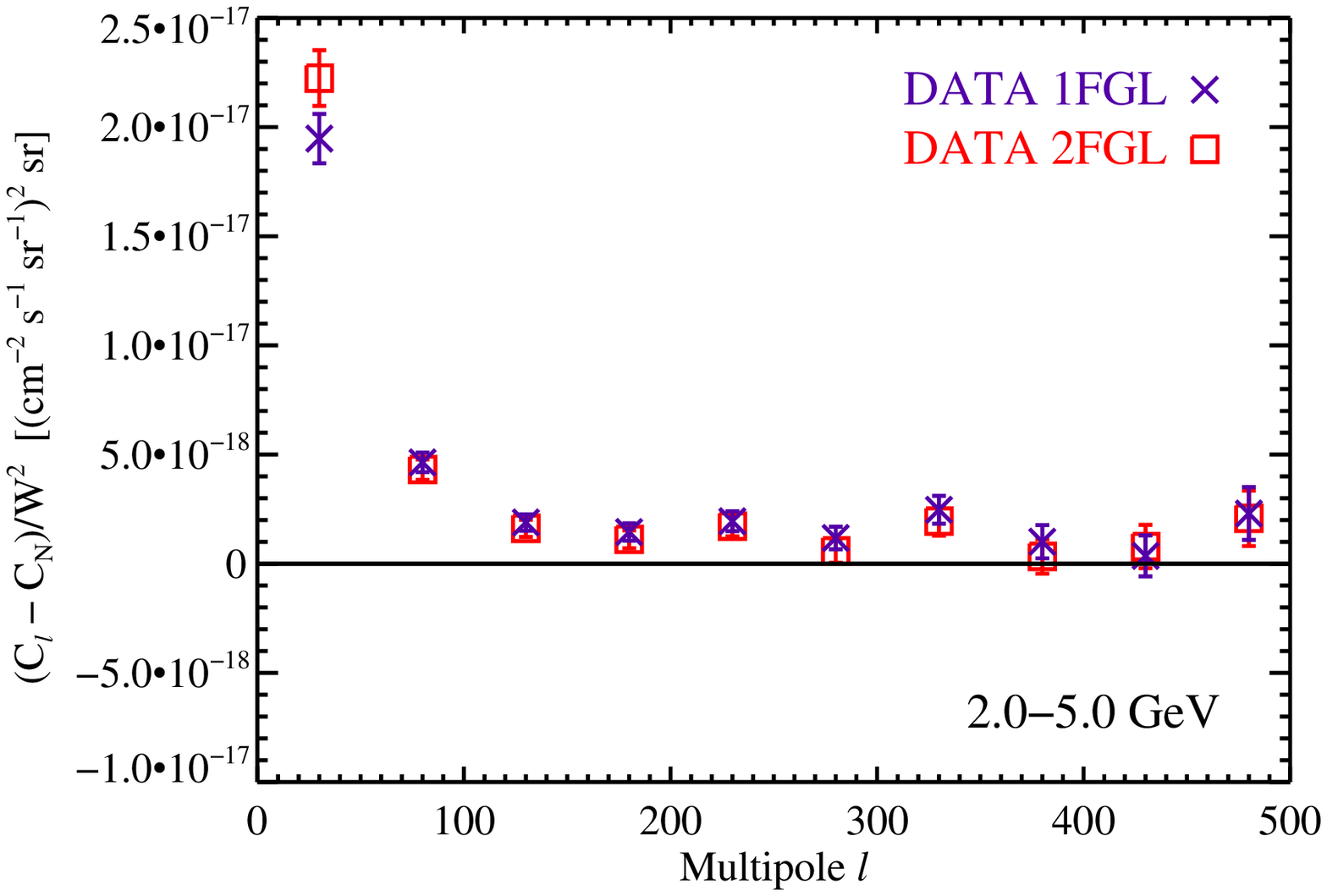}
\includegraphics[width=0.45\textwidth]{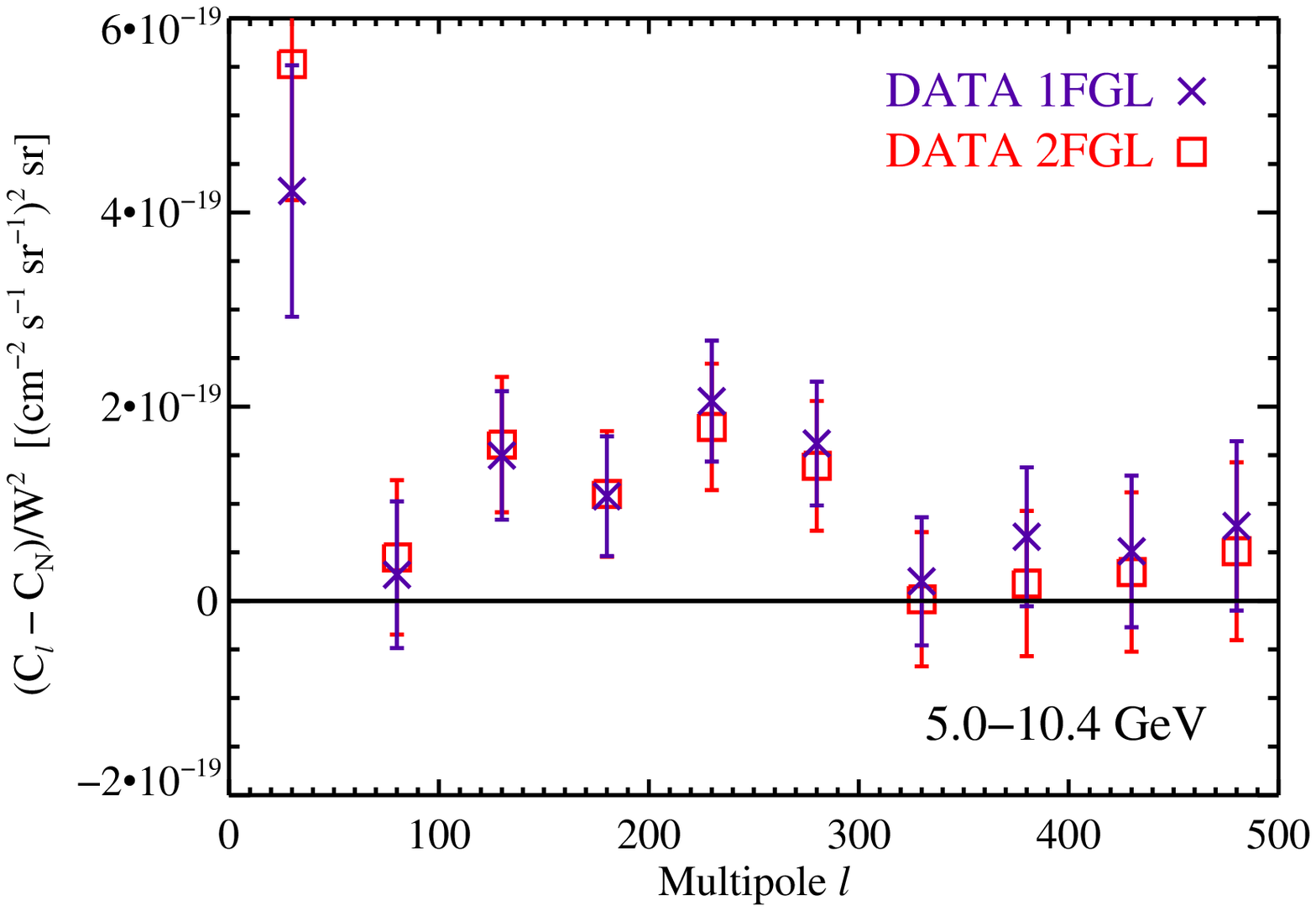}
\includegraphics[width=0.45\textwidth]{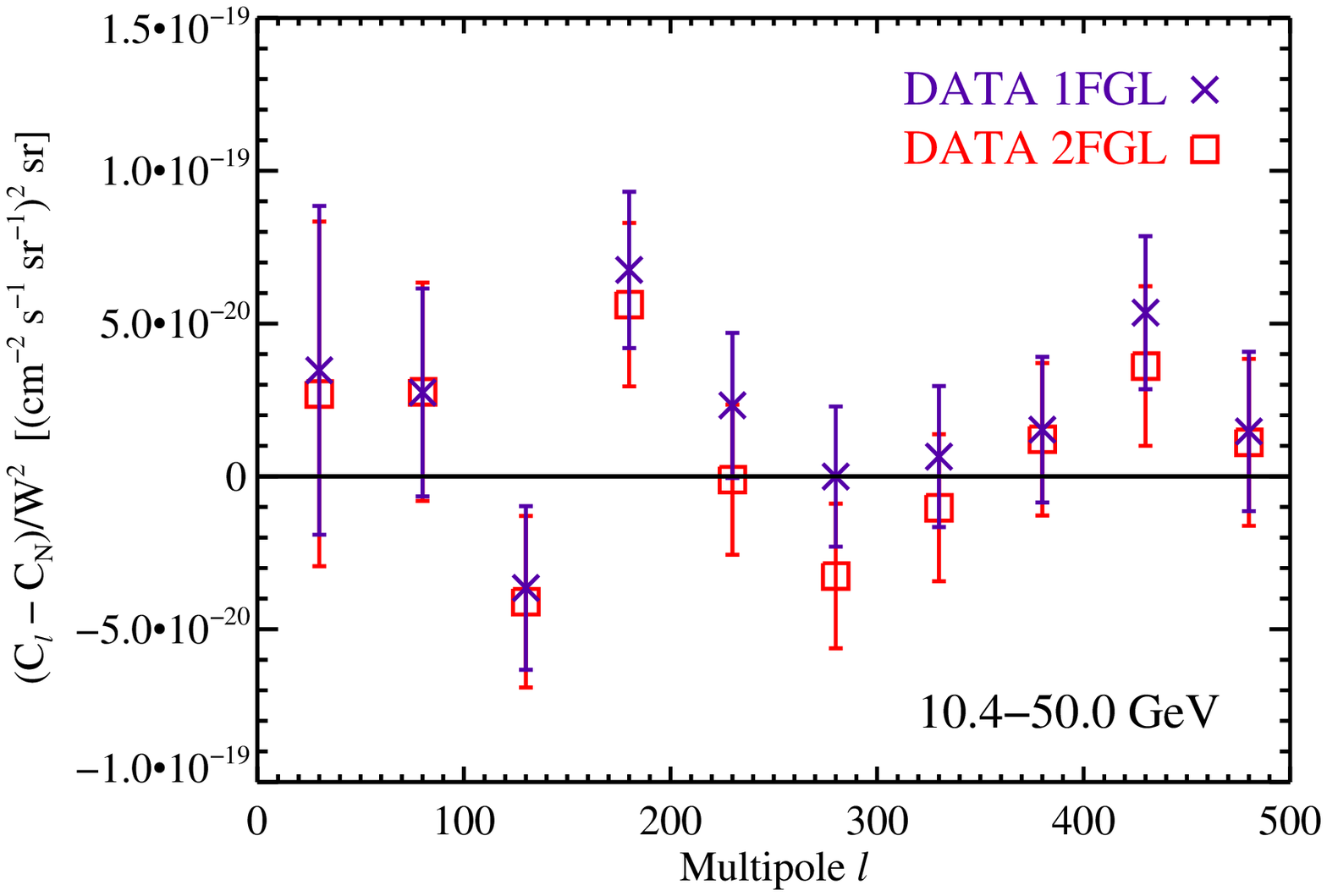}
\caption{Intensity angular power spectra of the data calculated using the source mask defined by the 2FGL catalog compared with the results using the 1FGL catalog; the source mask defined by the 1FGL catalog is the default used in this analysis.  The angular power at $\ell \ge 155$ is smaller in the 2FGL case by $\sim 20$--30\% in the bins spanning 1--10~GeV and by $\sim 70$\% at 10--50~GeV.  Expanded versions of the top panels are shown in Fig.~\ref{fig:1fgl2fglcomparezoom}.\label{fig:1fgl2fglcompare}}
\end{figure*}

\begin{figure*}
\includegraphics[width=0.45\textwidth]{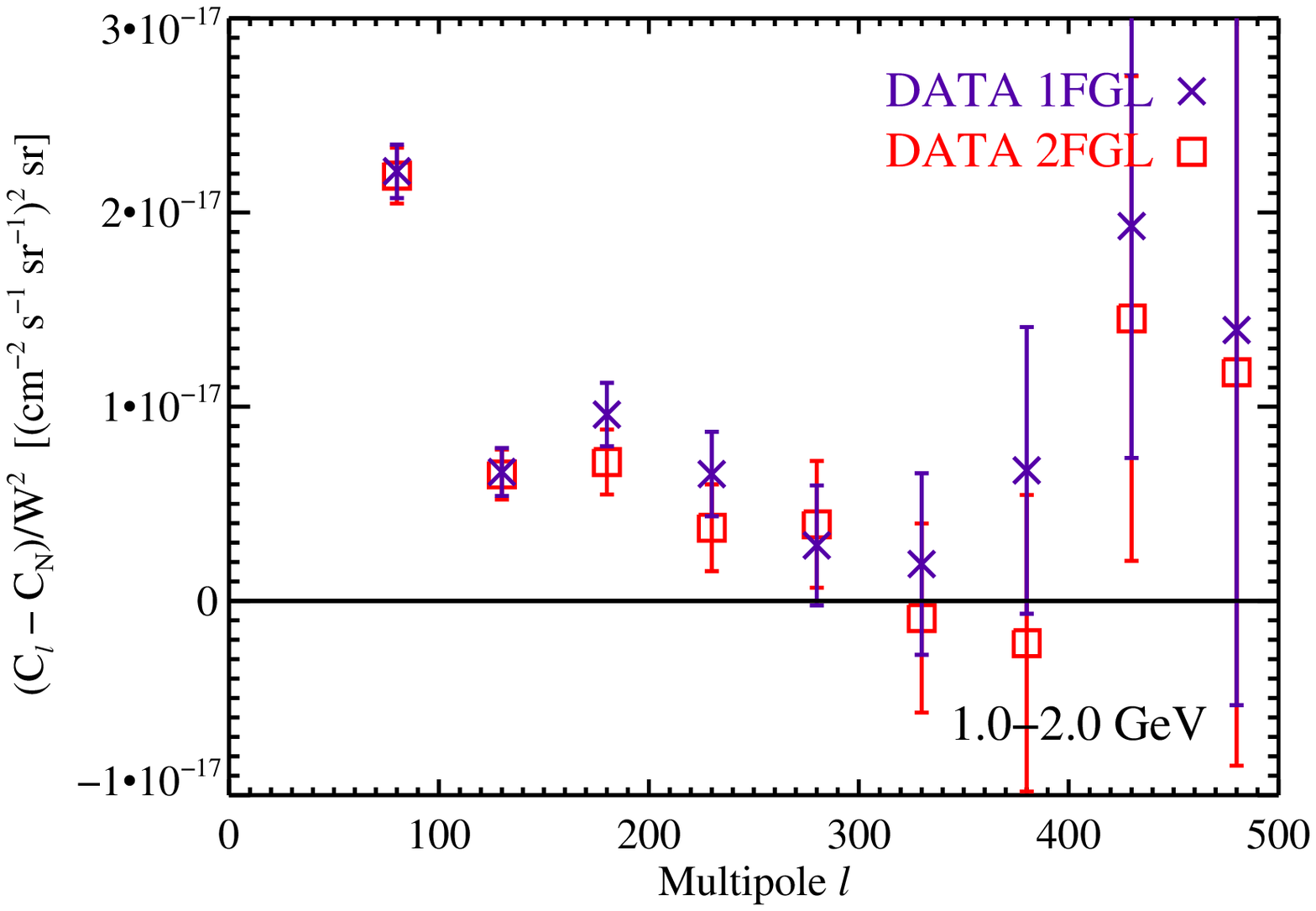}
\includegraphics[width=0.45\textwidth]{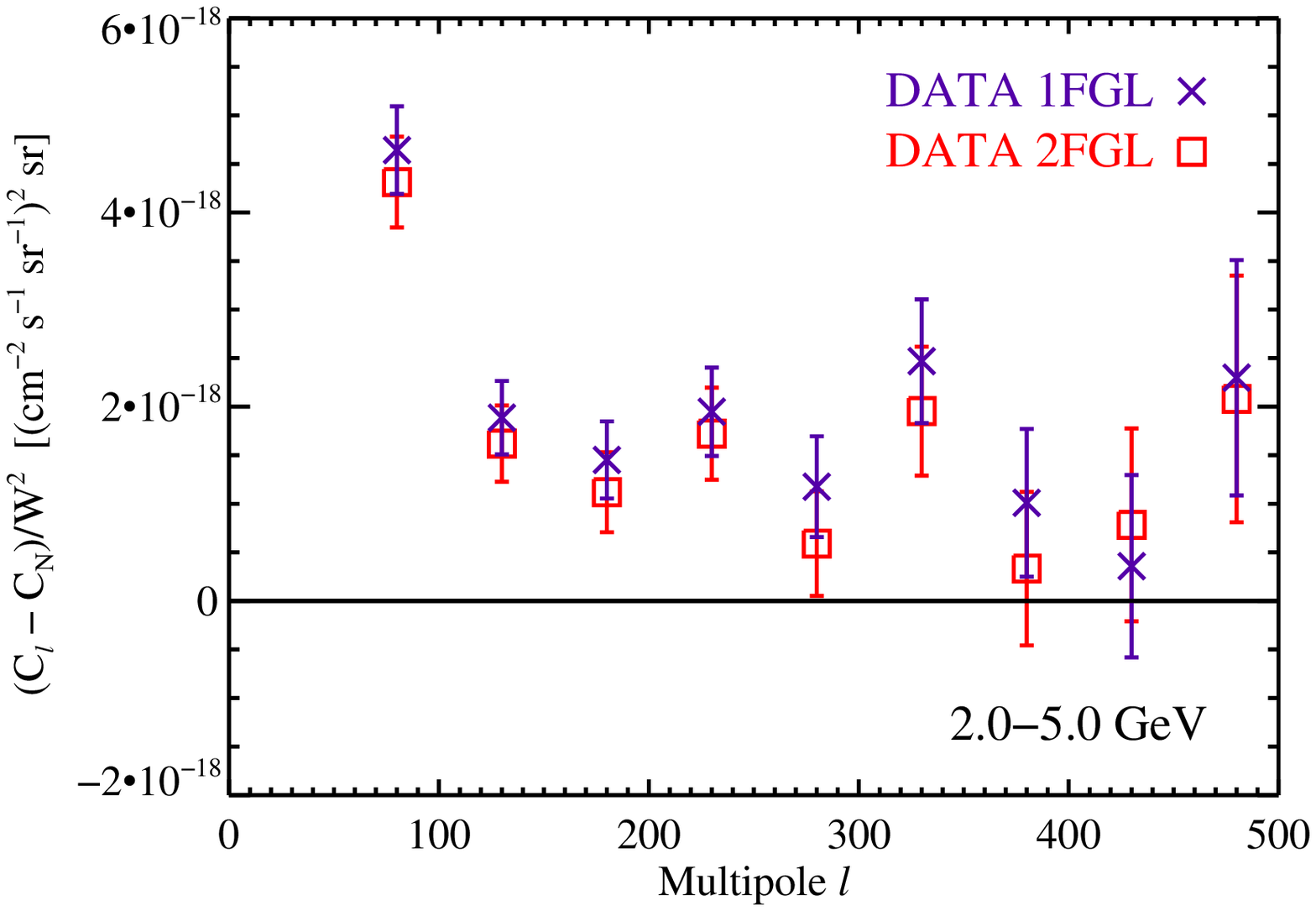}
\caption{Expanded versions of top panels of Fig.~\ref{fig:1fgl2fglcompare}.
\label{fig:1fgl2fglcomparezoom}}
\end{figure*}

The angular power spectra of the data analyzed using the 2FGL catalog to define the source mask are shown in Figs.~\ref{fig:1fgl2fglcompare} and~\ref{fig:1fgl2fglcomparezoom}, compared with the results of the default data analysis which uses the 1FGL catalog.  The angular power $C_{\rm P}$ measured in the data using the 2FGL source mask is reduced relative to the 1FGL case (see Table~\ref{tab:cpfits}), while the measurement uncertainties remain roughly the same as in the 1FGL case.  The decrease in $C_{\rm P}$ is $\sim20$--$30$\% in the 1--2, 2--5, and 5--10~GeV energy bins, however significant detections ($>3\sigma$) are still found in these three bins.  A $\sim 70$\% decrease in $C_{\rm P}$ is seen in the 10--50~GeV bin, and due to the large measurement uncertainty the significance of the measurement in this bin falls from 2.7$\sigma$ to 0.8$\sigma$.  The significance of the detected fluctuation angular power over all four energy bins remains greater than 7$\sigma$.

We can estimate the expected decrease in angular power when masking the 2FGL sources by calculating the difference in angular power produced when the source detection threshold is reduced from the 1FGL to the 2FGL catalog level, following the approach used to calculate the angular power of the simulated sources in \S\ref{sec:ptsrc}.  We assume the sources follow the flux distribution function ${\rm d}N/{\rm d}S$ given by Eq.~\ref{eq:dnds} with the same parameters given in that section as the best-fit for the high-latitude Fermi sources in the 1.04--10.4~GeV band.  Calculating $C_{\rm P}$ via Eq.~\ref{eq:cpanalytic} for an assumed flux threshold of $\sim 5 \times 10^{-10} {\rm cm^{-2}~s^{-1}}$, appropriate for the 1FGL catalog~\citep{Collaboration:2010ru:1fgl}, yields $C_{\rm P} \sim  9.4 \times 10^{-18}$ (cm$^{-2}$ s$^{-1}$ sr$^{-1}$ GeV$^{-1}$)$^2$ sr.  Using a lower flux threshold of $\sim 4 \times 10^{-10} {\rm cm^{-2}~s^{-1}}$, appropriate for the 2FGL catalog~\citep{Collaboration:2011bm}, gives $C_{\rm P} \sim 6.8 \times 10^{-18}$ (cm$^{-2}$ s$^{-1}$ sr$^{-1}$ GeV$^{-1}$)$^2$ sr, which is indeed a roughly 30\% decrease in $C_{\rm P}$, as observed in the data.

\subsection{Comparison of data and simulated models}
\label{sec:datamodel}

\begin{figure*}
\includegraphics[width=0.45\textwidth]{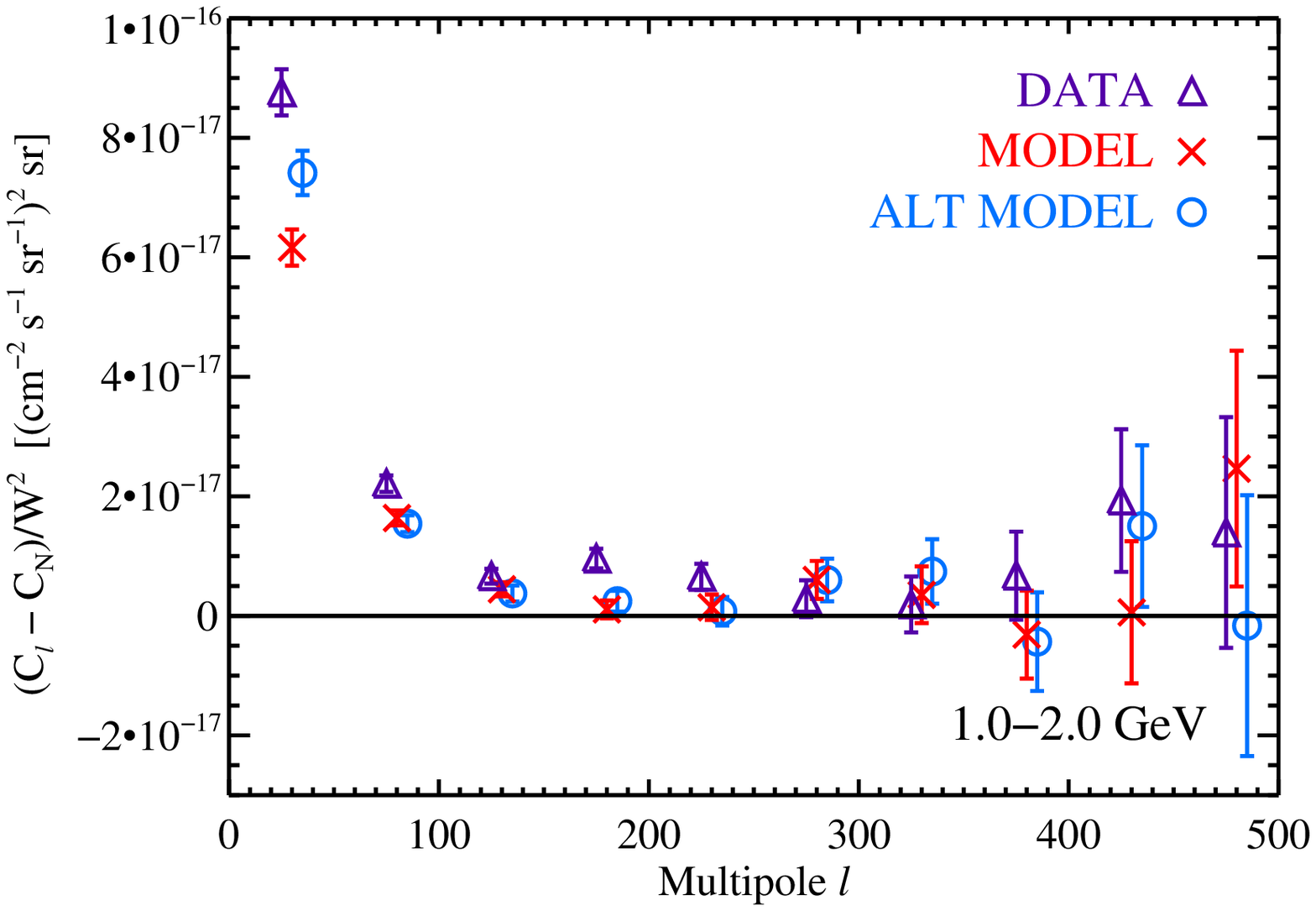}
\includegraphics[width=0.45\textwidth]{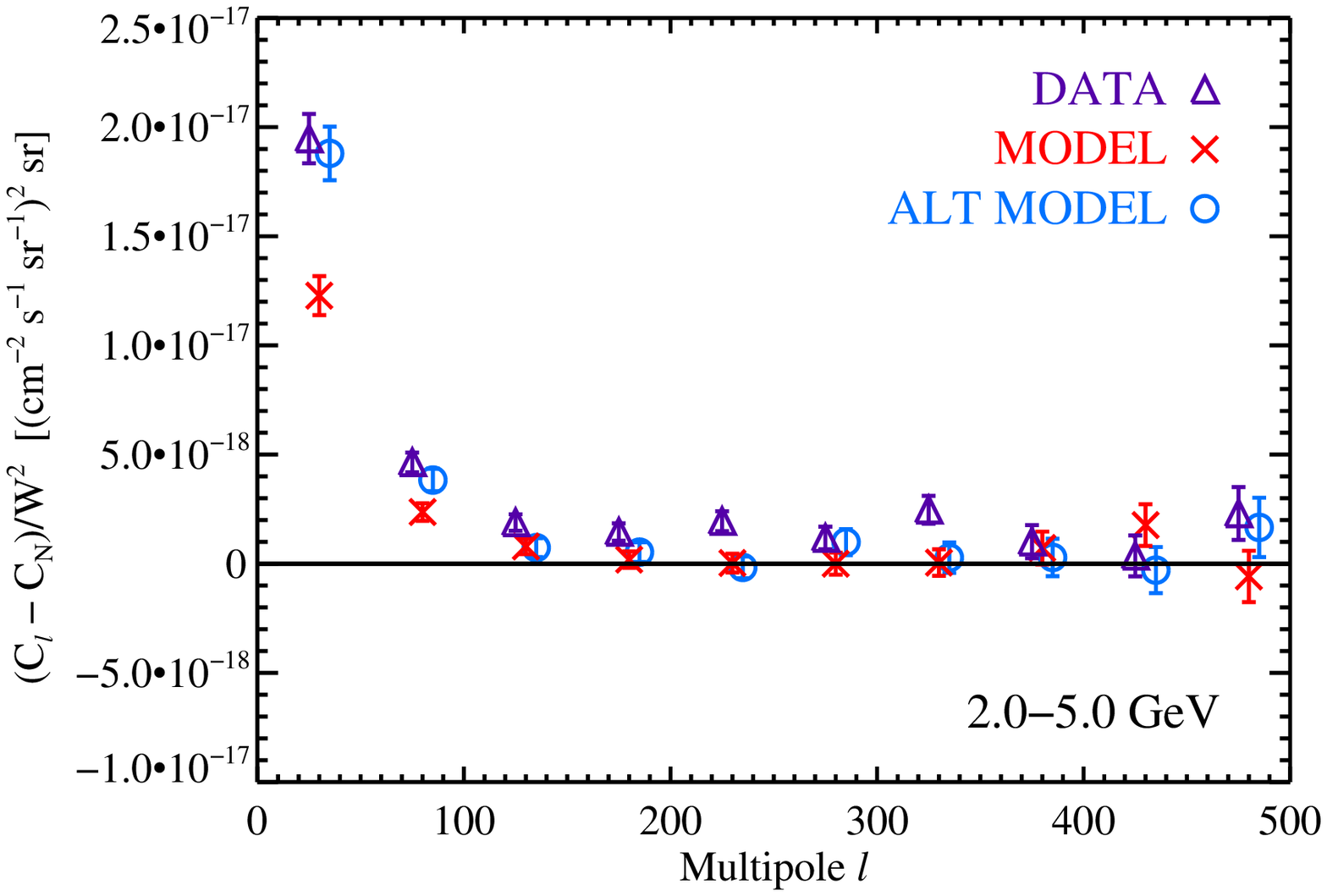}
\includegraphics[width=0.45\textwidth]{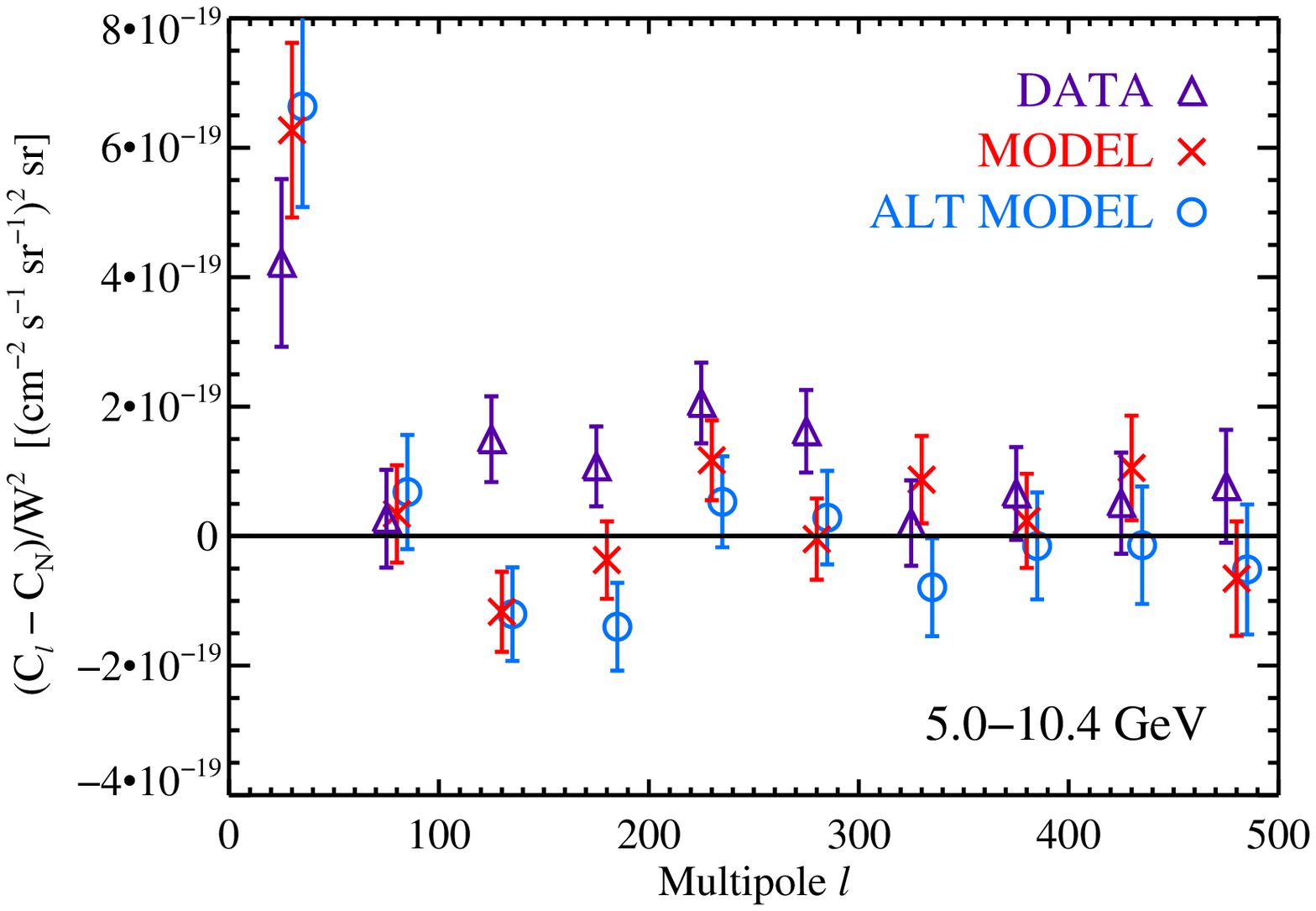}
\includegraphics[width=0.45\textwidth]{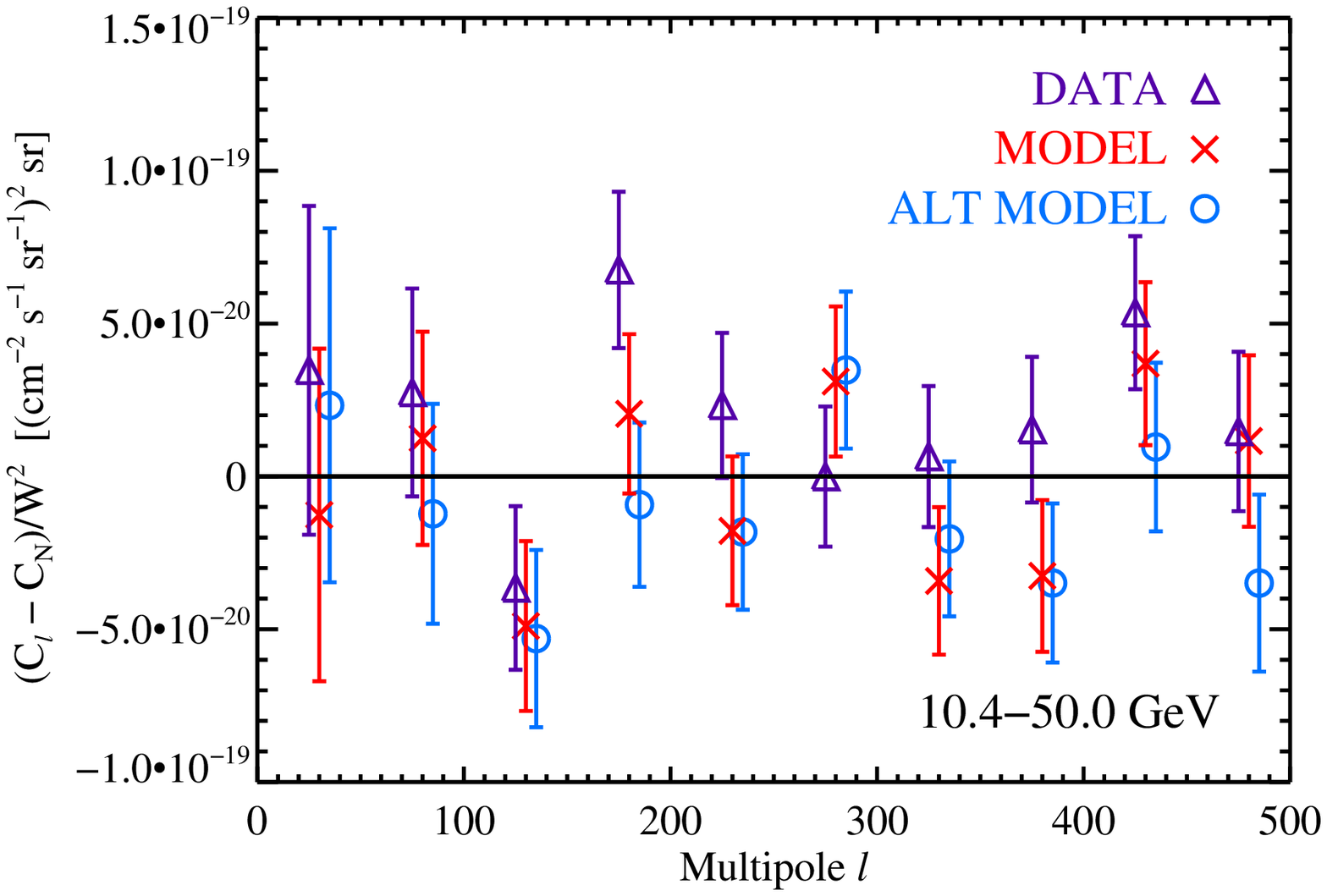}
\caption{Angular power spectra of the data, the default simulated model (MODEL), and the alternate simulated model (ALT MODEL).  The angular power spectra of the two models are in good agreement in all energy bins.  The smaller amplitude angular power at $\ell \ge 155$ measured at lower significance in both models is inconsistent with the angular power observed in the data at all energies.  Points from different data sets are offset slightly in multipole for clarity.  Expanded versions of the top panels are shown in Fig.~\ref{fig:datamodelzoom}.\label{fig:datamodel}}
\end{figure*}

\begin{figure*}
\includegraphics[width=0.45\textwidth]{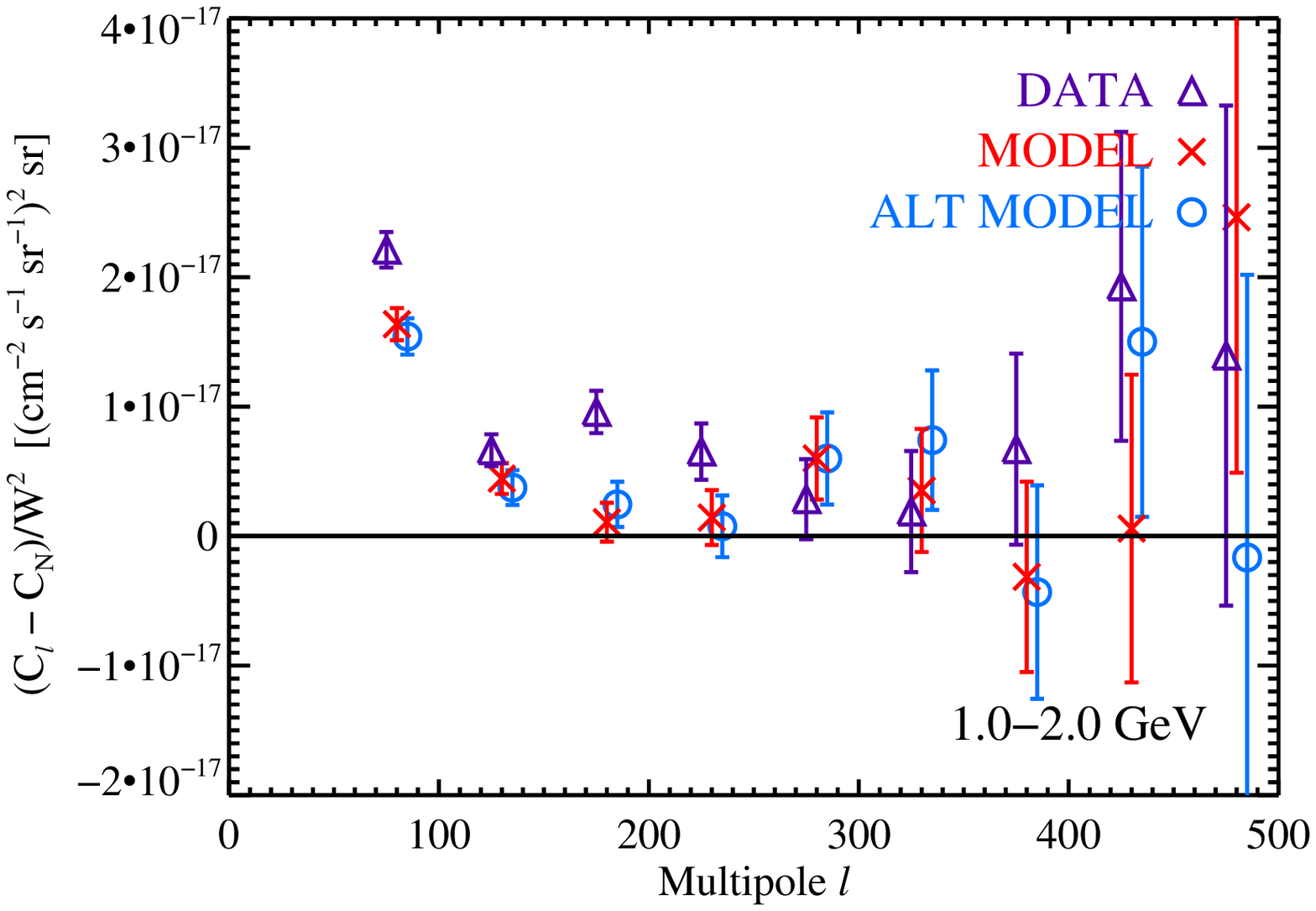}
\includegraphics[width=0.45\textwidth]{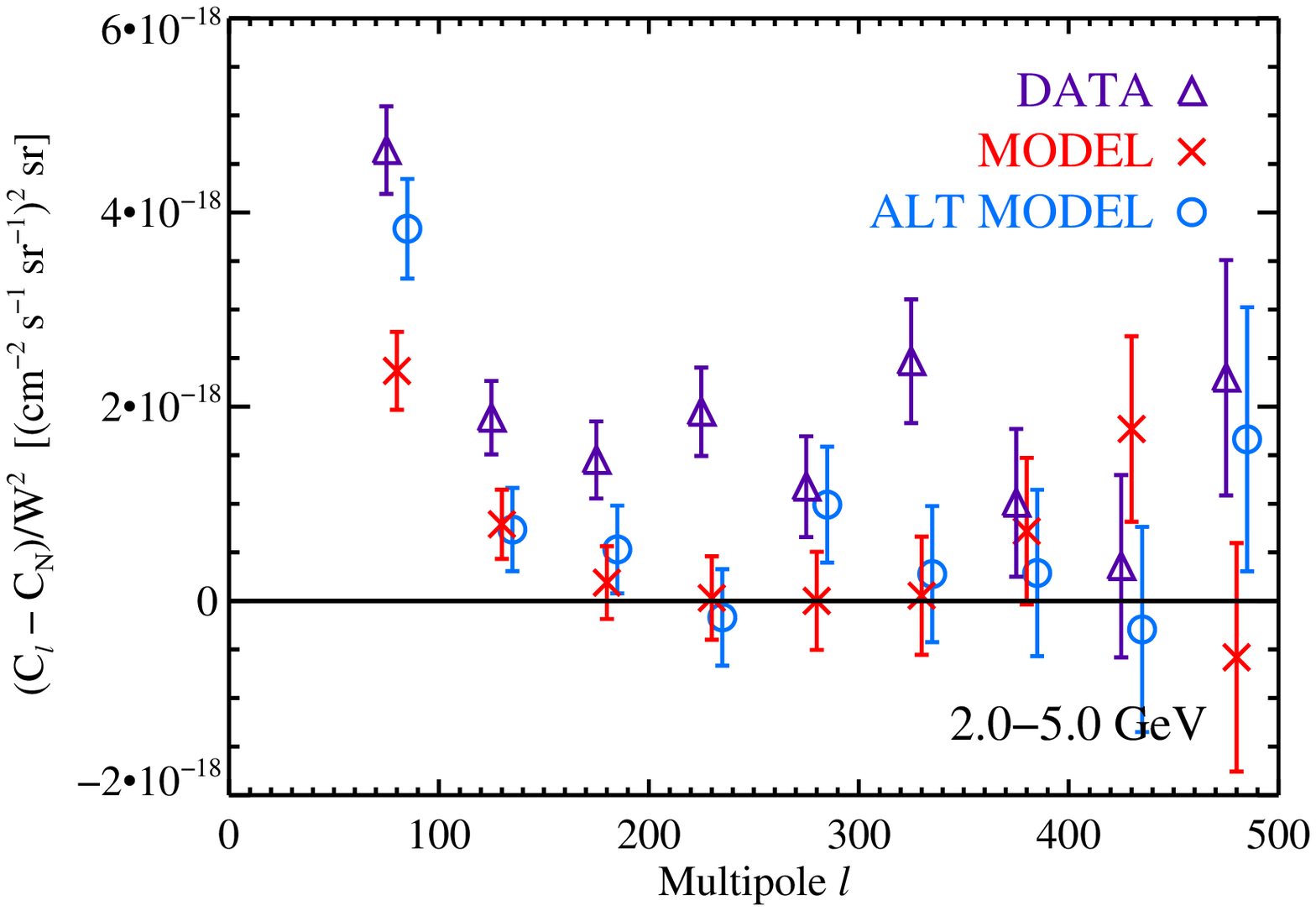}
\caption{Expanded versions of top panels of Fig.~\ref{fig:datamodel}.\label{fig:datamodelzoom}}
\end{figure*}

To understand the origin of the angular power measured in the data, we compare the angular power spectra of the default data to those of the default simulated model and the alternate simulated model, described in \S\ref{sec:sims}.  The simulated models were processed and their angular power spectra calculated using the same analysis pipeline as the data, and thus we expect the angular power spectra of the data and models to be consistent if the models accurately reflected the statistical properties of the emission on the relevant angular scales.

Figs.~\ref{fig:datamodel} and~\ref{fig:datamodelzoom} present the angular power spectra of the data and models.  The angular power spectra of the two models agree very well at all energies at multipoles above $\ell = 105$.  At all energies and scales, both models exhibit less angular power than the data.  Moreover, the amplitude of the detected angular power in both models is inconsistent with that of the data at $> 95$\%~CL in the three energy bins spanning 1--10~GeV, and at $>90$\%~CL in the 10--50~GeV bin (see Table~\ref{tab:datamodeldiffsig}).  The lack of significant power at high multipoles in either simulated model indicates that the Galactic diffuse emission, as implemented in these models, provides a negligible contribution to the anisotropy $\ell \ge 155$.  At lower multipoles, the discrepancy between the data and models and between the two models may be due to the presence of large-scale features in the data which are not included in the models, however we defer a full investigation of the origin of the low-multipole angular power to future work.

\begin{table}[th]
\caption{\label{tab:datamodeldiffsig}
Significance of the difference $\Delta C_{\rm P}$ between intensity angular power $C_{\rm P}$ for $155 \le \ell \le 504$ in the default data and the default simulated model in each energy bin.  The associated measurement uncertainties can be taken to be Gaussian.}
\begin{ruledtabular}
\begin{tabular}{ccc}
\multicolumn{1}{c}{$E_{\rm min}$}&
\multicolumn{1}{c}{$E_{\rm max}$}&
\multicolumn{1}{c}{\textrm{Significance} of $\Delta C_{\rm P}$}\\
\colrule
1.04 & 	1.99 & 	$3.5\sigma$\\
1.99 & 	5.00 & 	$4.5\sigma$\\
5.00 & 	10.4 & 	$2.0\sigma $\\
10.4 & 	50.0 & 	$1.7\sigma$\\
\end{tabular}
\end{ruledtabular}
\end{table}

\begin{figure*}
\includegraphics[width=0.45\textwidth]{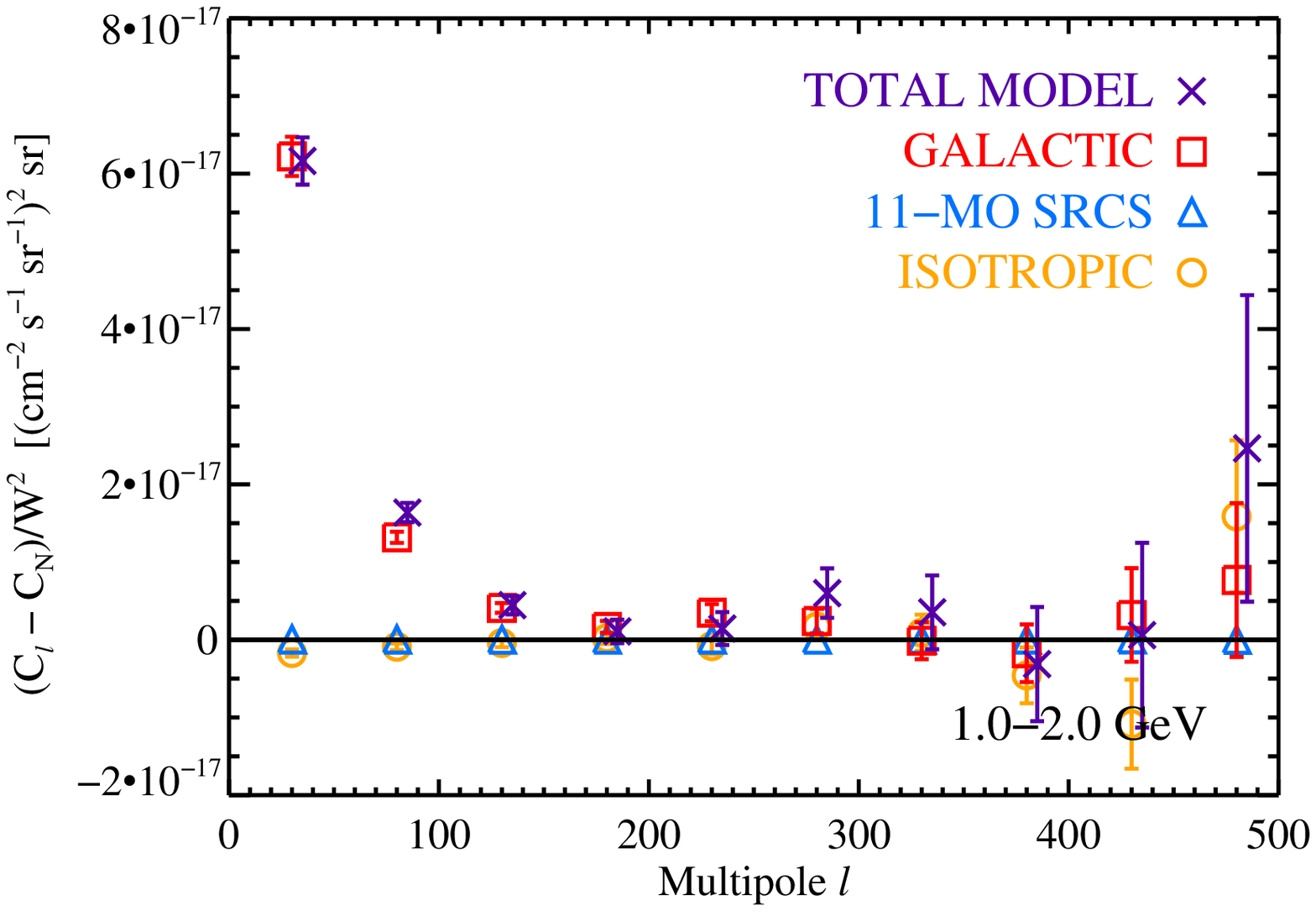}
\includegraphics[width=0.45\textwidth]{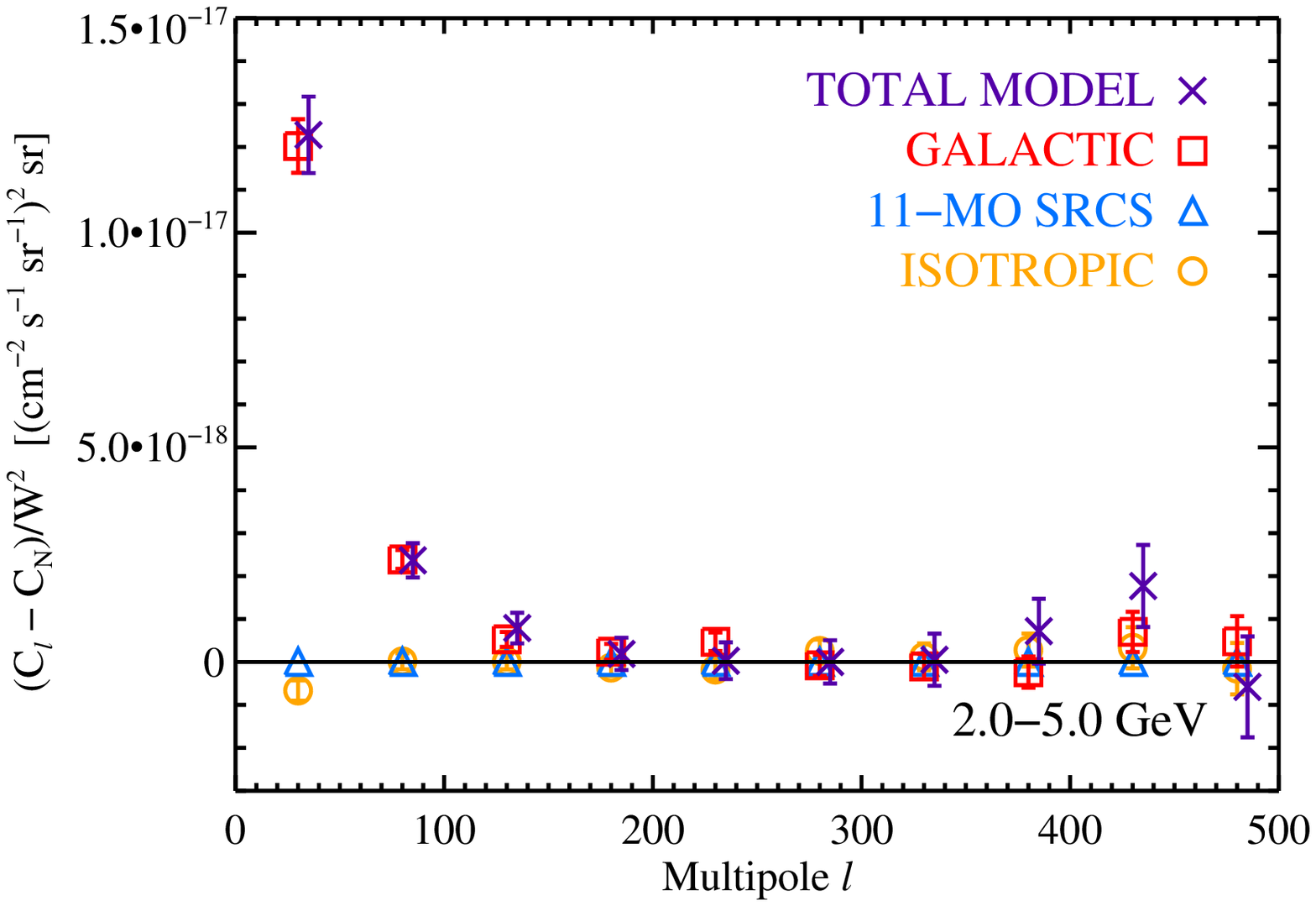}
\includegraphics[width=0.45\textwidth]{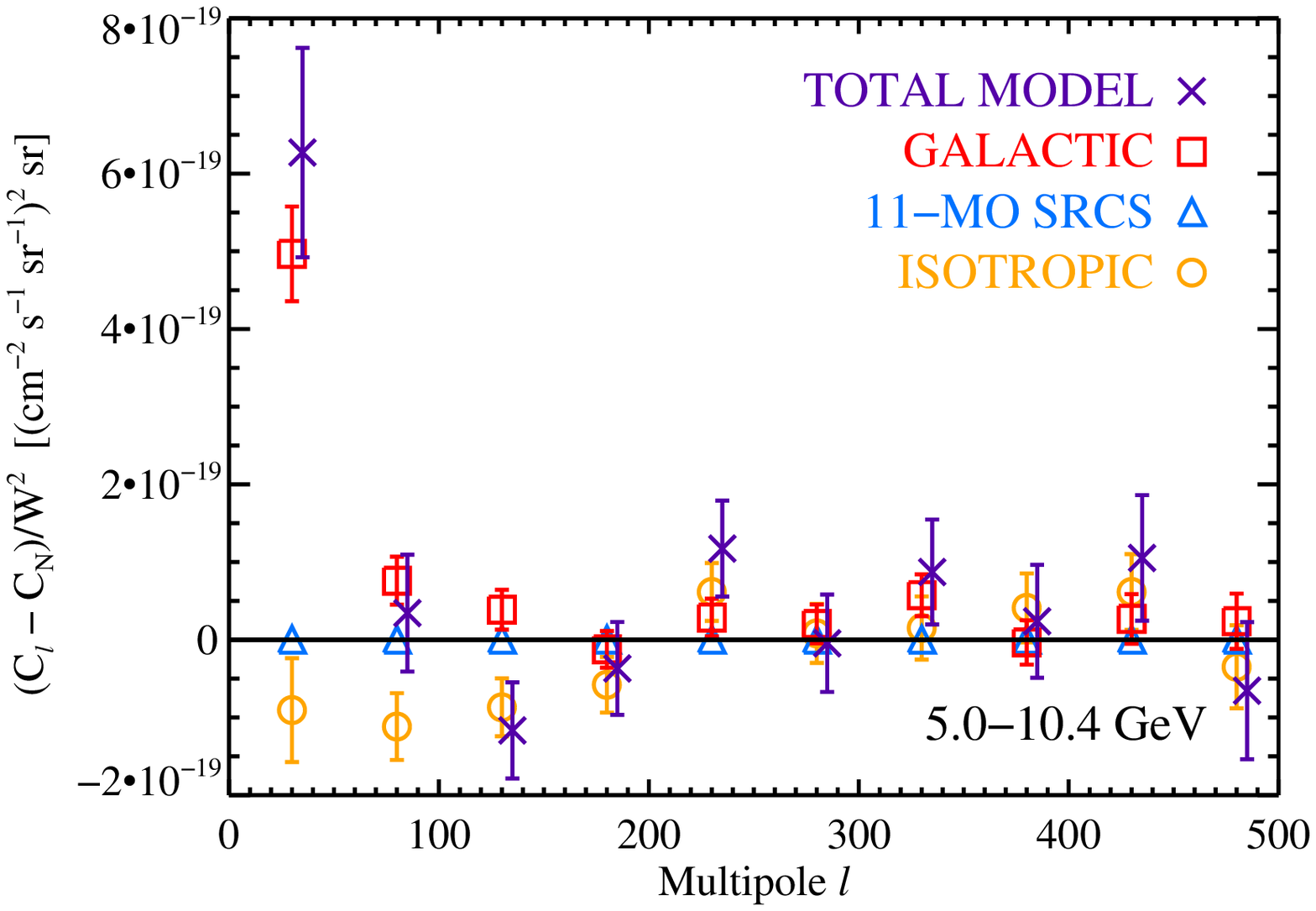}
\includegraphics[width=0.45\textwidth]{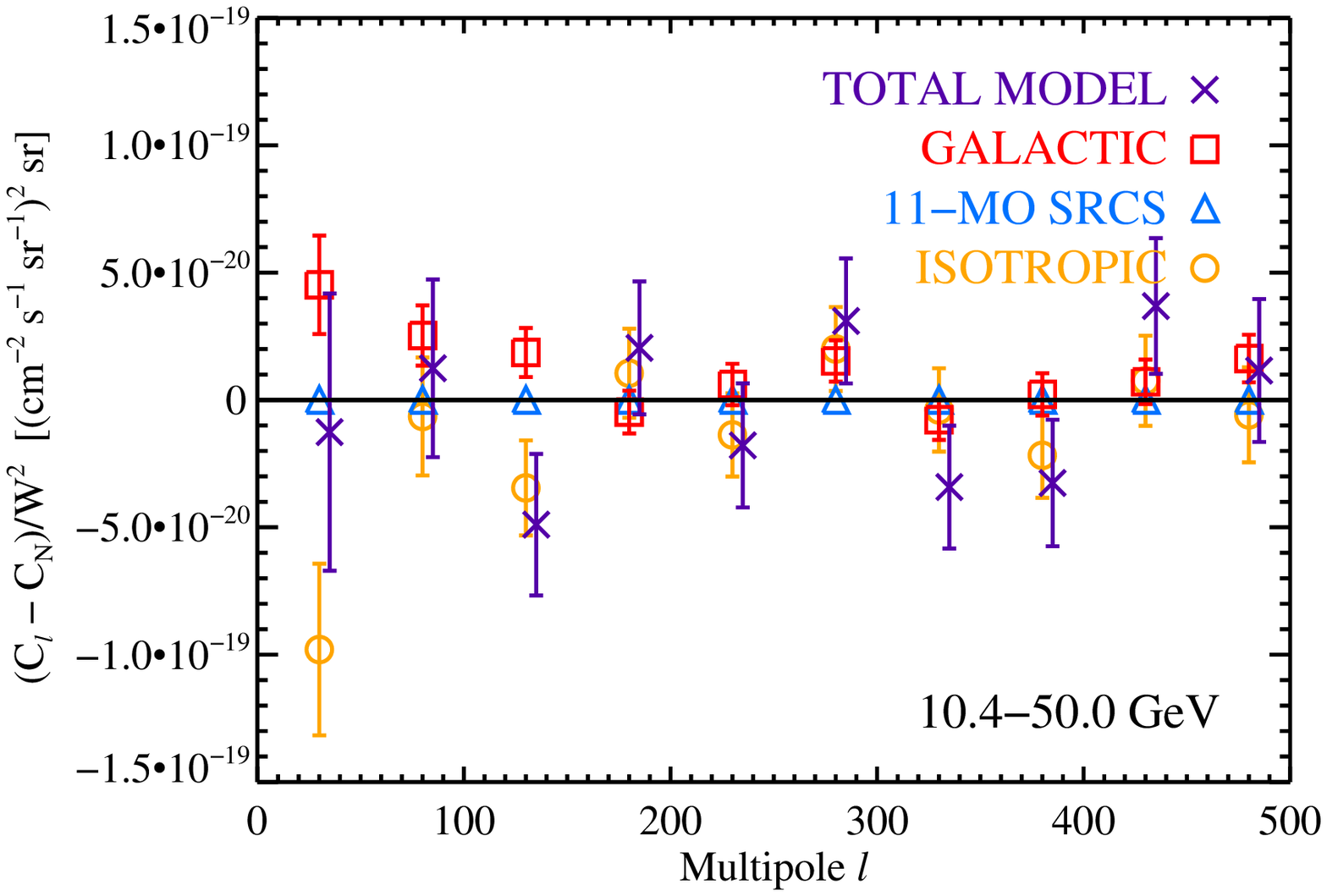}
\caption{Angular power spectra of the components of the default simulated model (MODEL).  As expected, most of the total angular power at all multipoles (TOTAL MODEL) is due to the GAL component.  By construction, the isotropic component (ISO) component contributes no significant angular power; likewise, the source component (CAT) provides no contribution because all sources were masked.  Points corresponding to the TOTAL MODEL are offset slightly in multipole for clarity.  Expanded versions of the top panels are shown in Fig.~\ref{fig:modelcompszoom}.\label{fig:modelcomps}}
\end{figure*}   

\begin{figure*}
\includegraphics[width=0.45\textwidth]{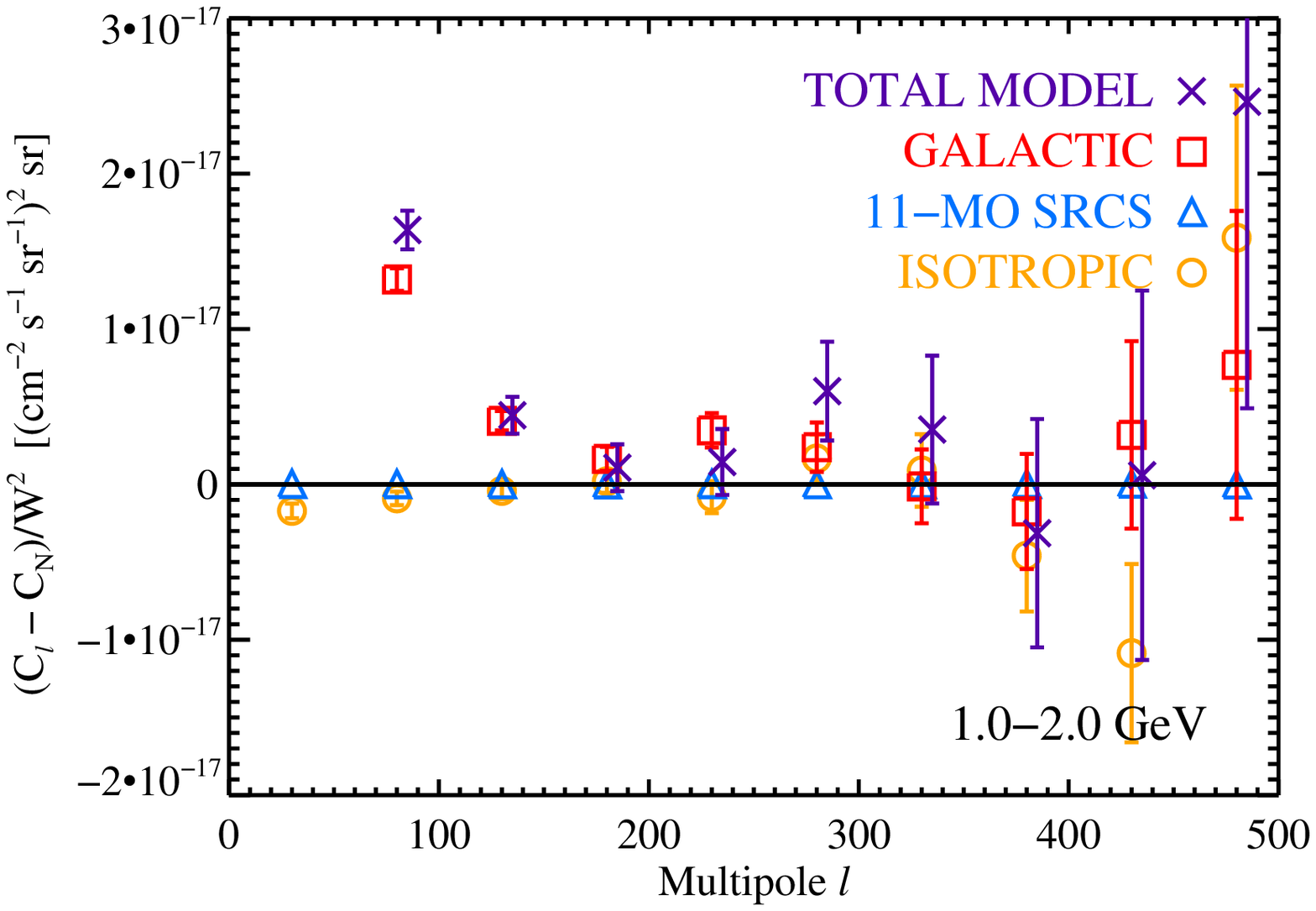}
\includegraphics[width=0.45\textwidth]{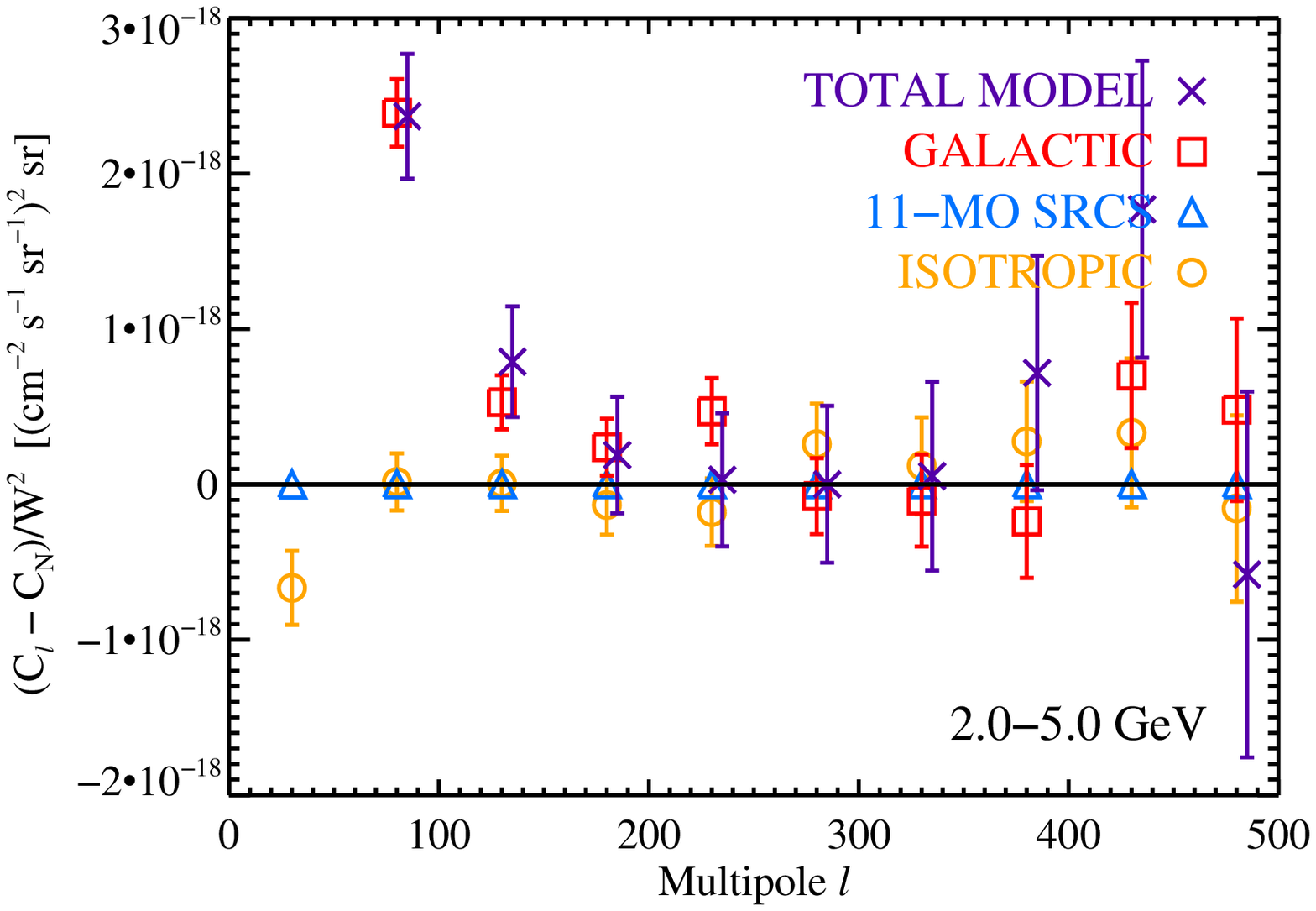}
\caption{Expanded versions of top panels of Fig.~\ref{fig:modelcomps}.
\label{fig:modelcompszoom}}
\end{figure*}   

The contributions to the angular power spectrum of the individual components of the default model are shown in Figs.~\ref{fig:modelcomps} and~\ref{fig:modelcompszoom}.  At all energies the only component contributing significantly to the total power is the Galactic diffuse emission.  The contribution from the isotropic component is negligible, since this component is isotropic by construction and thus, after the photon noise is subtracted, it should only contribute to the monopole ($\ell = 0$) term.  The deviations from zero of the isotropic component in the lowest multipole bin ($5 \le \ell \le 54$) may be due to imperfect correction of the effects of the mask in this multipole regime.  The source catalog component contributes zero power at all energies and multipoles since the emission maps of this simulated component contain only events from sources which are masked in the analysis.  The consistency of the source catalog angular power with zero indicates that the source masking is effective.  We remark that in general the angular power spectra of distinct components are not linearly additive due to contributions from cross-correlations between the components.  The total power of the model is, however, very consistent with the total power in the Galactic component, with slight discrepancies likely arising from masking effects, since the Galactic and isotropic components should have no cross-correlation power and the simulated sources were fully masked.

\section{Energy dependence of anisotropy in the data}
\label{sec:energydep}

\begin{figure}
\includegraphics[width=0.45\textwidth]{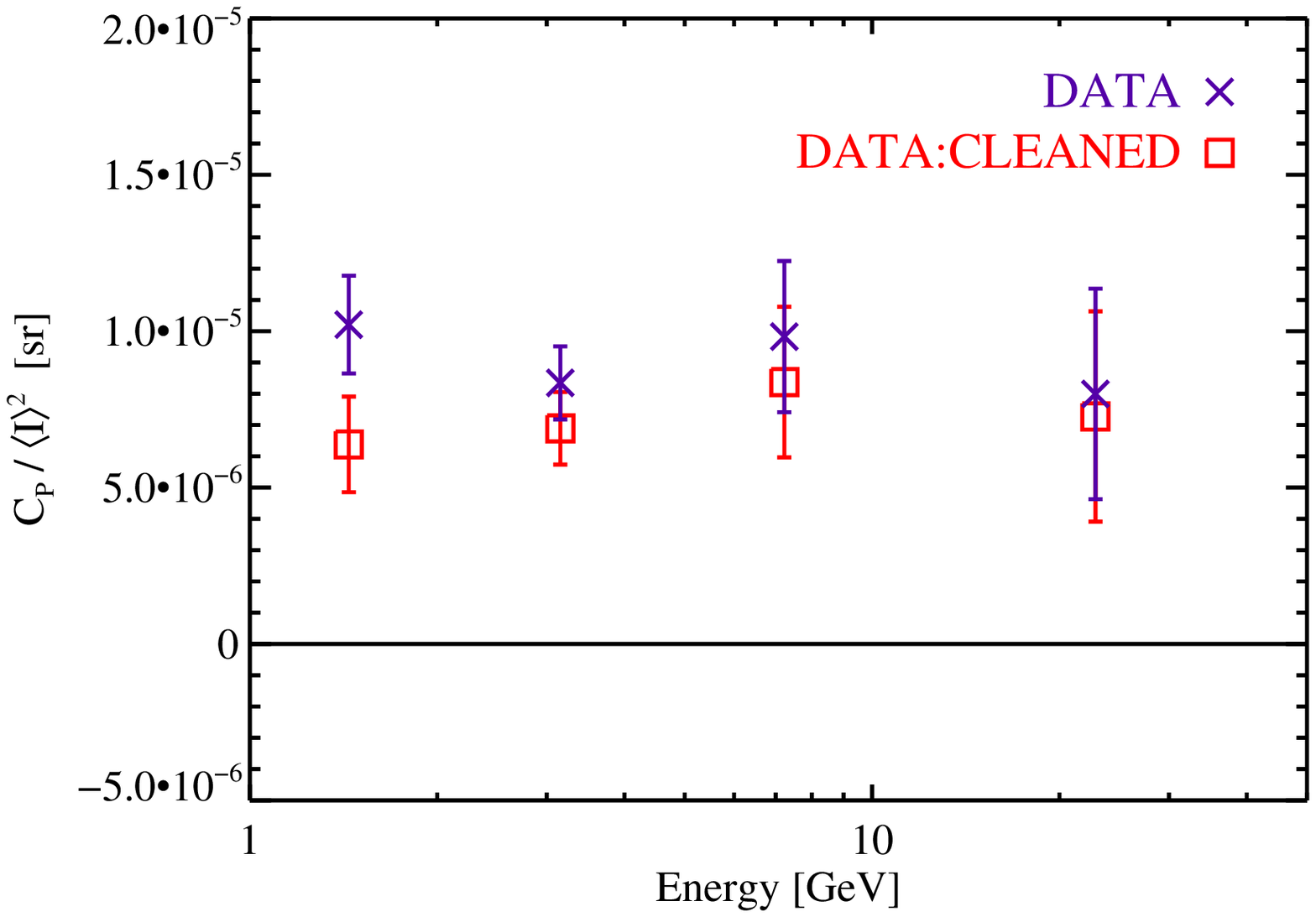}
\includegraphics[width=0.45\textwidth]{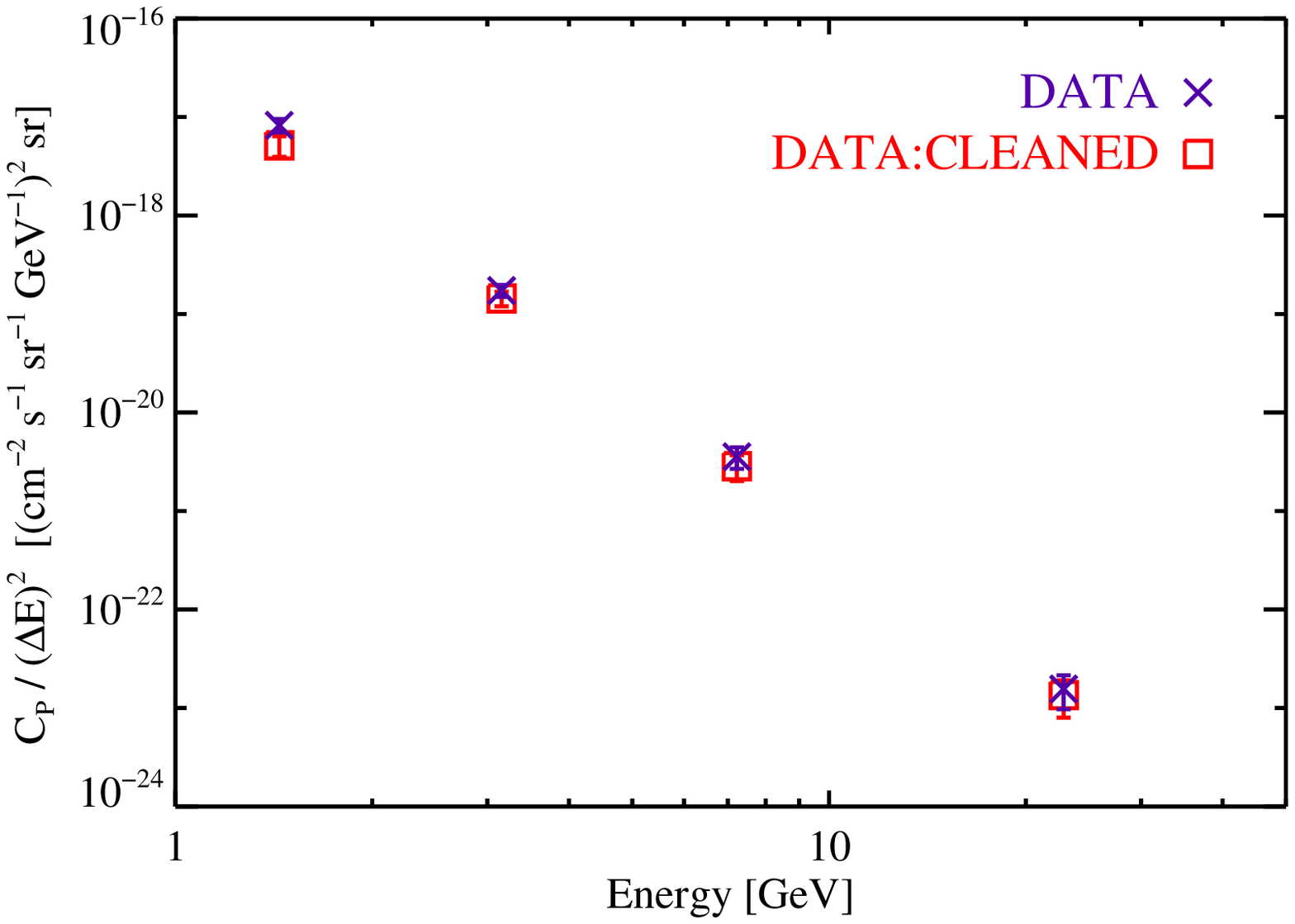}
\caption{Anisotropy energy spectra of the data.  \emph{Top:} Fluctuation anisotropy energy spectrum.  The data are consistent with no energy dependence over the energy range considered, although a mild energy dependence is not excluded.  \emph{Bottom:} Differential intensity anisotropy energy spectrum.  The energy dependence is consistent with that arising from a single source population with a power-law intensity energy spectrum with spectral index $\Gamma_{\rm s} = 2.40 \pm 0.07$ for the default data ($2.33 \pm 0.08$ for the cleaned data).    
\label{fig:anisoenergyspect}}
\end{figure}

\begin{table*}[th]
\caption{\label{tab:edep}
Energy dependence of angular power for $155 \le \ell \le 504$ in each energy bin for the data processed with the default analysis pipeline and the Galactic-foreground--cleaned data.  The best-fit constant value of the fluctuation angular power $\langle C_{\rm P}/\langle I \rangle^{2} \rangle$ over 1--50~GeV is obtained by weighted averaging of $C_{\rm P}/\langle I \rangle^{2}$ of the four energy bins.  The best-fit parameters and associated $\chi^{2}$ per degree of freedom (d.o.f.) are given for fits of the fluctuation angular power to $C_{\rm P}/\langle I \rangle^{2} = A_{\rm F} (E/E_{0})^{-\Gamma_{\rm F}}$ and the differential intensity angular power to $C_{\rm P}/(\Delta E)^{2}=A_{\rm I} (E/E_{0})^{-\Gamma_{\rm I}}$, with $E_{0}=1$~GeV.  The value of $A_{\rm I}$ is given in terms of $A_{\rm I}/A_{{\rm I},0}$ where $A_{{\rm I},0}= 10^{-18}$ (cm$^{-2}$ s$^{-1}$ sr$^{-1}$ GeV$^{-1}$)$^2$ sr.}
\begin{ruledtabular}
\begin{tabular}{cccccccc}
&
\multicolumn{1}{c}{$\langle C_{\rm P}/\langle I \rangle^{2}  \rangle$}&
\multicolumn{1}{c}{$A_{\rm F}$}&
\multicolumn{1}{c}{$\Gamma_{\rm F}$}&
\multicolumn{1}{c}{$\chi^{2}/$\textrm{d.o.f.}}&
\multicolumn{1}{c}{$A_{\rm I}/A_{{\rm I},0}$}&
\multicolumn{1}{c}{$\Gamma_{\rm I}$}&
\multicolumn{1}{c}{$\chi^{2}/$\textrm{d.o.f.}}\\
&
\multicolumn{1}{c}{[$10^{-6}$ sr]}&
\multicolumn{1}{c}{[$10^{-6}$ sr]}&
&
&
&
&
\\
\colrule
DATA & 			 	$9.05 \pm 0.84$ &		$9.85 \pm 1.73$ &	$0.076 \pm 0.139$ &	$0.41$ &	$45.1 \pm 7.8$ &	$4.79 \pm 0.13$ &	$0.19$\\
DATA:CLEANED & 		$6.94 \pm 0.84$ &		$6.31 \pm 1.44$ &	$-0.082 \pm 0.158$ &		$0.12$ &	$29.4 \pm 6.6$ & 	$4.66 \pm 0.15$ & 	$0.035$\\
\end{tabular}
\end{ruledtabular}
\end{table*}

The energy dependence of the fluctuation angular power can be used to identify the presence of multiple distinct contributors to the emission \citep{SiegalGaskins:2009ux}.  Because the fluctuation angular power characterizes only the angular distribution of the emission, independent of the intensity normalization, it is exactly energy-independent for a single source class as long as the members of the class have the same observed energy spectrum.  In general, the fluctuation angular power of a single source class may show energy dependence due to large variation of the energy spectra of individual sources within a population, and, for cosmological source classes, the effects of redshifting and attenuation of high-energy gamma rays by the extragalactic background light (EBL).  Redshifting and EBL attenuation is expected to be important only for populations for which a significant fraction of the observed intensity originates from high-redshift members, with EBL attenuation relevant only at observed energies of several tens of GeV.  All of these effects are most prominent when the source spectra have hard features such as lines or cut-offs; smoothly-varying source spectra, such as power-law energy spectra, typically generate more mild energy dependence in the fluctuation angular power.

The fluctuation anisotropy energy spectrum of the data is shown in the top panel of Fig.~\ref{fig:anisoenergyspect}.  The fluctuation angular power $C_{\rm P}/\langle I \rangle^{2}$ in each energy bin was obtained by weighted averaging of the unbinned fluctuation angular power spectrum over $155 \le \ell \le 504$, weighting the measured angular power at each multipole by its measurement uncertainty; these values are reported in Table~\ref{tab:cpfits}.  Each point is located at the logarithmic center of the energy bin.  

A power-law fit of the fluctuation angular power as a function of energy $C_{\rm P}/\langle I \rangle^{2} \propto E^{-\Gamma_{\rm F}}$ yields $\Gamma_{\rm F} = 0.076 \pm 0.139$ ($-0.082 \pm 0.158$ for the cleaned data), consistent with no energy-dependence over the energy range considered.  The best-fit constant value of $C_{\rm P}/\langle I \rangle^{2}$ across all four energy bins is $9.05 \pm 0.84 \times 10^{-6}$ sr ($6.94 \pm 0.84 \times 10^{-6}$ sr for the cleaned data).  The results of these fits for the data with and without foreground cleaning are summarized in Table~\ref{tab:edep}, along with the results for the energy dependence of the intensity angular power, discussed below.  The lack of a clear energy dependence in the fluctuation angular power is consistent with a single source class providing the dominant contribution to the anisotropy and a constant fractional contribution to the intensity over the energy range considered, although due to the large measurement uncertainties contributions from additional source classes cannot be excluded.  This is especially true for sources whose contribution to the intensity peaks at $E \gtrsim 10$~GeV.  Furthermore, due to the coarseness of the energy binning, this analysis is not sensitive to features in the anisotropy energy spectrum localized to narrow energy bands.

If a single source class dominates the anisotropy at all energies considered, the differential intensity angular power spectrum $C_{\ell}/(\Delta E)^{2}$ scales with energy as the intensity energy spectrum squared $({\rm d}N/{\rm d}E)^{2}$ of that source class.  For example, for a source class with a power-law photon spectrum ${\rm d}N/{\rm d}E \propto E^{-\Gamma_{\rm s}}$, $C_{\ell}/(\Delta E)^{2} \propto E^{-2\Gamma_{\rm s}}$.  We can therefore use this energy scaling to constrain the energy spectrum of the dominant contributor to the anisotropy, under the assumption that the measured angular power (but not necessarily the total measured intensity) originates from a single source class.  

Here we obtain the differential intensity angular power $C_{\rm P}/(\Delta E)^{2}$ by dividing the intensity angular power $C_{\rm P}$ in each energy bin by the bin size squared.  The differential intensity anisotropy energy spectrum of the data is shown in the bottom panel of Fig.~\ref{fig:anisoenergyspect}.  The $C_{\rm P}$ are the best-fit values for $155 \le \ell \le 504$, i.e., the weighted average of $C_{\ell}$ in that multipole range, reported in Table~\ref{tab:cpfits}, and each data point is located at the logarithmic center of the energy bin.  The results of fitting $C_{\rm P}/(\Delta E)^{2} \propto E^{-\Gamma_{\rm I}}$ are given in Table~\ref{tab:edep}.  Identifying $\Gamma_{\rm I} = 2\Gamma_{\rm s}$, the best fit of the energy dependence suggests that the anisotropy is contributed by a source class with a power-law photon spectrum characterized by $\Gamma_{\rm s} = 2.40 \pm 0.07$ ($2.33\pm 0.08$ for the cleaned data), assuming only one source class contributes appreciably to the anisotropy.  As the single power-law energy dependence provides a very good fit to the data, attributing the anisotropy to a single source class is a plausible interpretation.  

We note that the spectral index implied for the dominant source class contributing to the anisotropy is in excellent agreement with the  mean intrinsic spectral index of blazars as inferred from the Fermi-detected members~\citep{Collaboration:2010gqa:srccounts}, strongly supporting the interpretation of the measured anisotropy as originating from unresolved blazars.  We caution, however, that due to the variation between individual blazars' spectral indices, as well as possible effects of EBL attenuation and redshifting,  the fluctuation angular power from blazars could exhibit some energy dependence in the range considered here.  Therefore, assuming that blazars are the dominant source class contributing the anisotropy could lead to tension with the flatness of the measured fluctuation anisotropy energy spectrum.  Additional support for a blazar interpretation could be provided by a detailed study of the energy-dependent anisotropy arising from specific blazar population models, calibrated to match the properties of Fermi-detected blazars, and the consistency of the predicted anisotropy of these models with the measured amplitude of the angular power.  We defer a careful treatment of this subject to future work.

\section{Discussion}
\label{sec:discussion} 

Prior work has generated predictions for the angular power spectra of several source populations which may contribute to the IGRB\@.  In most cases, the predictions for the anisotropy of the emission from a single source class have been cast in terms of fluctuation angular power $C_{\ell}/\langle I \rangle^{2}$, where $C_{\ell}$ is the intensity angular power spectrum of the source class and $\langle I \rangle$ its mean collective intensity in a specified energy range.  Since the intensity contributions of most gamma-ray source classes to the IGRB are subject to large uncertainties, it is convenient to consider the fluctuation angular power, since this quantity is independent of the overall normalization of the intensity.  This convention is particularly useful when the spatial distribution, number density, and relative flux distribution of the sources is known or modeled, and the uncertainty in the collective intensity can be translated into a multiplicative factor that uniformly scales the observed intensity in all sky directions.  For this reason, the fluctuation angular power is very well suited for characterizing an indirect dark matter signal since the intensity normalization scales linearly with the assumed annihilation cross-section or decay rate.

By comparing the measured fluctuation angular power with predictions for various source classes, we can place constraints on the fractional contribution from each source class to the total intensity by requiring that the fluctuation angular power of the total emission is not exceeded.  Assuming that each contributing source class is uncorrelated with the others and the Poisson component dominates the angular power spectrum of each source class at the multipoles considered, the intensity angular power of the total emission is given by
\begin{equation}
C_{\rm P, tot}=C_{{\rm P},1}+C_{{\rm P},2}+...
\end{equation}
and so the fluctuation angular power of the total intensity is
\begin{equation}
\frac{C_{\rm P, tot}}{\langle I_{\rm tot} \rangle^{2}} = 
\frac{C_{{\rm P},1}}{\langle I_{\rm tot} \rangle^{2}} +
\frac{C_{{\rm P},2}}{\langle I_{\rm tot} \rangle^{2}}+...
\end{equation}
Rewriting the fractional contribution from source class $i$ to the total intensity $f_{i}=\langle I_{i} \rangle/\langle I_{\rm tot} \rangle$,
\begin{equation}
\frac{C_{\rm P, tot}}{\langle I_{\rm tot} \rangle^{2}} = 
f_{1}^{2}\frac{C_{{\rm P},1}}{\langle I_{1} \rangle^{2}} +
f_{2}^{2}\frac{C_{{\rm P},2}}{\langle I_{2} \rangle^{2}}+...
\end{equation}
If we allow a single source class $i$ to contribute all of the measured angular power, the source class is constrained such that
\begin{equation}
f_{i}^{2} \le \frac{C_{\rm P, tot}/\langle I_{\rm tot} \rangle^{2}}{C_{{\rm P},i}/\langle I_{i} \rangle^{2}}.
\end{equation}
Source classes whose predicted fluctuation angular power exceeds the measured fluctuation angular power therefore cannot contribute the entirety of the measured intensity.

It is important to note that the total intensity of the IGRB in this analysis is not equivalent to the intensity of the isotropic emission reported in~\citep{Abdo:2010nz:igrb}, since that analysis employed much more stringent selection cuts to remove charged particle contamination, and used a fitting procedure to remove contributions from non-isotropic components.  In this analysis the total intensity of the emission is simply the intensity that remains after the mask is applied, which may include some emission from non-isotropic components, as well as a non-negligible amount of charged particle contamination.  However, we emphasize that since the charged particle contamination is presumed to be nearly isotropic, with any potential fluctuations confined to large angular scales, it should not contribute to the intensity angular power at the multipoles considered here ($\ell \ge 155$), and so a more robust comparison of models with the data could be achieved by comparing the predicted intensity angular power to the measurement.

\begin{table*}[th]
\caption{\label{tab:constraints}
Maximum fractional contribution of various source populations to the IGRB intensity that is compatible with the best-fit constant value of the measured fluctuation angular power in all energy bins, $\langle C_{\rm P}/\langle I \rangle^{2}  \rangle = 9.05 \times 10^{-6}$~sr for the default data analysis or $\langle C_{\rm P}/\langle I \rangle^{2}  \rangle = 6.94 \times 10^{-6}$~sr for the Galactic-foreground--cleaned data analysis.  Indicative values for the fluctuation angular power $C_{\ell}/\langle I \rangle^{2}$ of each source class are taken from existing literature (see text for details) and evaluated at $\ell=100$.}
\begin{ruledtabular}
\begin{tabular}{lccc}
\multicolumn{1}{l}{Source class}&
\multicolumn{1}{c}{Predicted $C_{100}/\langle I \rangle^{2}$}&
\multicolumn{2}{c}{Maximum fraction of IGRB intensity}\\
\multicolumn{1}{c}{}&
\multicolumn{1}{c}{[sr]}&
\multicolumn{1}{c}{DATA}&
\multicolumn{1}{c}{DATA:CLEANED}\\
\colrule
Blazars & 					$2 \times10^{-4}$ 	& 	21\% 	&	19\%\\
Star-forming galaxies & 		$2 \times10^{-7}$ 	& 	100\%	&	100\%\\
Extragalactic dark matter annihilation & 	$1 \times 10^{-5}$ 	& 	95\% 	&	83\%\\
Galactic dark matter annihilation & 		$5 \times 10^{-5}$ 	& 	43\% 	&	37\%\\
Millisecond pulsars & 		$3 \times 10^{-2}$ 	& 	1.7\% 	&	1.5\%\\
\end{tabular}
\end{ruledtabular}
\end{table*}

We now compare our measurement to existing predictions from the literature for the angular power spectra of various gamma-ray source classes, and summarize these results in Table~\ref{tab:constraints}.  We caution that the predicted angular power can depend sensitively upon the adopted source model (in particular the shape of the flux distribution), the assumed source detection threshold, and, for cosmological source classes, assumptions regarding the effect on the observed energy spectrum of attenuation of high-energy photons by interactions with the EBL\@.  Consequently, the constraints derived in this section should be taken only as indicative values for these source populations.

Ref.~\citep{Ando:2006cr} predicted the fluctuation anisotropy from unresolved blazars $C_{\rm P}/\langle I \rangle^{2} \sim 2 \times 10^{-4}$~sr at $\ell \sim 100$ (see Fig.~4 of that work).  This is a factor of $\sim 20$ larger than the fluctuation angular power of $\sim 10^{-5}$~sr measured in the data, which suggests that emission from blazars, assuming the model adopted in that study, contributes less than $\sim 1/\sqrt{20} \sim 20\%$ of the total intensity.  Note, however, that the flux threshold for sources in the 1FGL catalog is between 0.5 and $1 \times 10^{-9}$ photons cm$^{-2}$ s$^{-1}$ for $|b|>30^\circ$,  higher than the threshold assumed in~\citep{Ando:2006cr}.  If the blazar luminosity function is identical to the one assumed in~\citep{Ando:2006cr}, this discrepancy in thresholds would imply that the prediction for the blazar anisotropy in~\citep{Ando:2006cr} is underestimated with respect to the one applicable to our analysis, since our masked maps include more bright unresolved blazars. As a result, the constraint on the fractional intensity contribution to the IGRB from blazars for this model from our measurement would, if anything, be stronger.

In contrast to the larger anisotropy expected from blazars, the fluctuation angular power at $\ell \sim 100$ predicted for star-forming galaxies by Ref.~\citep{Ando:2009nk} is $\sim 2 \times 10^{-7}$~sr at 1~GeV, far below the value measured in this analysis.  Since star-forming galaxies would thus provide a subdominant contribution to the measured angular power, this anisotropy measurement does not constrain their contribution to the total IGRB intensity.

The anisotropy from dark matter annihilation in extragalactic structures is predicted to be slightly smaller than that from unresolved blazars, although estimates can vary substantially due to differences in the adopted models.  Moreover, for extragalactic dark matter annihilation the amplitude of the expected anisotropy can be highly sensitive to the energy spectrum of the emission.  The source energy spectrum depends on the dark matter particle mass and dominant annihilation channels, while the observed energy spectrum is affected by redshifting and EBL attenuation.  These factors can introduce a non-trivial energy dependence into the amplitude of the anisotropy, particularly for high mass ($\sim 1$~TeV) dark matter candidates.  As a benchmark range, Refs.~\citep{Ando:2005xg, Ando:2006cr, Cuoco:2010jb} predict the anisotropy from annihilation of extragalactic dark matter to be $\sim 10^{-6}$--$10^{-5}$~sr at $\ell \sim 100$ at energies of a few GeV, comparable to the measured value.

The anisotropy from annihilation in Galactic dark matter substructure is expected to be much larger than that from extragalactic dark matter.  While variations in the assumed properties of Galactic substructure can lead to order-of-magnitude or larger variations in the predicted angular power, for typical assumptions the predicted fluctuation angular power is $\sim 5 \times 10^{-5}$~sr at $\ell \sim 100$ (e.g., Model A1 in Ref.~\citep{Ando:2009fp}), which implies that dark matter annihilation can contribute less than $\sim 43\%$ of the total intensity.  However, adopting alternative models for the substructure properties can increase or decrease the predicted angular power by as much as $\sim 2$ orders of magnitude~\citep{SiegalGaskins:2008ge,Fornasa:2009qh,Ando:2009fp}, so the measured angular power represents a strong constraint on some substructure models.

Galactic gamma-ray millisecond pulsars (MSPs) have also been considered as possible contributors to the intensity and anisotropy of the IGRB due to their extended latitude distribution~\citep{FaucherGiguere:2009df,SiegalGaskins:2010mp}.  The emission from Galactic MSPs is expected to feature very large fluctuation anisotropy due to the relatively low number density of this source class compared to dark matter substructure or extragalactic source populations.  Ref.~\citep{SiegalGaskins:2010mp} predicts fluctuation angular power at high Galactic latitudes of $\sim 0.03$~sr at $\ell \sim 100$ for this Galactic source class, which implies a contribution to the total IGRB intensity of no more than a few percent.

In addition to the specific source populations considered in this section, other Galactic source populations for which anisotropy predictions do not yet exist in the literature may also contribute to the anisotropy as well as the intensity of the high-latitude diffuse emission.  These include normal pulsars, as well as populations currently too faint to have had individual members detected by Fermi.  The properties of these populations can be constrained by both low-latitude and high-latitude source count analysis (in the case that individual members have been detected)~\citep{Strong:2006hf}, and also by the anisotropy analysis described in this study.  We leave the detailed study of this to future work.

We note that constraints derived in this section have not taken into account information about the likely energy spectrum of the dominant contributing population, discussed in \S\ref{sec:energydep}, which is incompatible with sources known or expected to feature spectral peaks at the energies we consider (for example, Galactic and extragalactic dark matter and MSPs).  A careful study combining all observables obtained in this work would almost certainly yield stronger constraints on contributing populations.
Furthermore, we have discussed the constraints obtainable on specific source populations by requiring that the total anisotropy from each population does not exceed the measured value.  We emphasize, however, that stronger bounds could be derived if some fraction of the total anisotropy could be robustly attributed to one or more confirmed source classes, thereby reducing the anisotropy available to additional contributors.

\section{Conclusions}
\label{sec:conclusions}

The statistical properties of the IGRB encode detailed information about the origin of this emission.  The advanced capabilities of the Fermi LAT, most notably its improved angular resolution and large effective area, have enabled a sensitive measurement of small angular-scale anisotropies in the IGRB\@.  Using $\sim 22$~months of data, we performed an angular power spectrum analysis of the high-latitude diffuse emission measured by the Fermi LAT\@.  Significant angular power above the photon noise level is detected in the data at multipoles $155 \le \ell \le 504$ in three energy bins spanning 1--10~GeV, and is measured at lower significance in the 10--50~GeV energy bin.  The primary limitation of the measurement at high energies is low event statistics, which results in the measurement uncertainties being dominated by the photon noise.  In this regime the measurement uncertainties scale roughly inversely to the number of events, and hence increasing the statistics by a factor of 2 or 3 could lead to a large enough improvement in the sensitivity of the analysis to allow a confident detection of angular power in this energy range and greater sensitivity to energy-dependent anisotropy.

The angular power measured in the data at $155 \le \ell \le 504$ is consistent with a constant value within each energy bin, and the scale independence of the signal suggests that it originates from one or more unclustered populations of point sources.  Comparing the measured angular power with predictions for known and proposed gamma-ray source classes, constraints can be obtained on the collective intensity and properties of source populations that contribute to the IGRB\@.
The fluctuation angular power detected in this analysis falls below the level predicted for many source classes, including blazars, MSPs, and some scenarios for dark matter annihilation in Galactic and extragalactic structures.  In these cases the measured amplitude of the fluctuation angular power limits the contribution to the total IGRB intensity of each source class.

The measured fluctuation angular power is consistent with a constant value over the energy range considered, however, due to the relatively large measurement uncertainties and limited number of energy bins, a mild energy dependence in this quantity cannot be excluded.  The absence of a strong energy dependence in the fluctuation anisotropy energy spectrum suggests that a single source class may provide the dominant contribution to the anisotropy while providing a constant fractional contribution to the intensity of the IGRB over the energy range considered.  We caution, however, that this analysis is not sensitive to structure in the anisotropy energy spectrum that is confined to small energy ranges, since the requirement of large event statistics to detect anisotropies at the measured level precludes fine energy binning of the data.  We anticipate that future analyses that draw on larger data sets will be more sensitive to localized features in the anisotropy energy spectrum.

The energy dependence of the intensity angular power of the data is well-described by that arising from a single source class with a power-law photon spectrum with index $\Gamma_{\rm s}=2.40 \pm 0.07$.  Interestingly, this value closely matches the mean intrinsic spectral index for blazars as determined from recent Fermi LAT measurements.  While alternative scenarios invoking contributions from more than one source class to explain the energy dependence of the angular power are in principle possible, the interpretation of the measured power as originating from a single source class with a power-law energy spectrum is an excellent fit to the data.  To identify a specific population or populations as the source of the measured IGRB anisotropy, detailed analysis of population models for plausible source classes will be essential in order to verify that both the predicted intensity energy spectrum of the IGRB and the corresponding anisotropy signal provides a consistent explanation of the data.

\begin{acknowledgments}

The \textit{Fermi} LAT Collaboration acknowledges generous ongoing support
from a number of agencies and institutes that have supported both the
development and the operation of the LAT as well as scientific data analysis.
These include the National Aeronautics and Space Administration and the
Department of Energy in the United States, the Commissariat \`a l'Energie Atomique
and the Centre National de la Recherche Scientifique / Institut National de Physique
Nucl\'eaire et de Physique des Particules in France, the Agenzia Spaziale Italiana
and the Istituto Nazionale di Fisica Nucleare in Italy, the Ministry of Education,
Culture, Sports, Science and Technology (MEXT), High Energy Accelerator Research
Organization (KEK) and Japan Aerospace Exploration Agency (JAXA) in Japan, and
the K.~A.~Wallenberg Foundation, the Swedish Research Council and the
Swedish National Space Board in Sweden.

Additional support for science analysis during the operations phase is gratefully
acknowledged from the Istituto Nazionale di Astrofisica in Italy and the Centre National d'\'Etudes Spatiales in France.

Some of the results in this paper have been derived using the HEALPix package.

EK is supported in part by NSF grants AST-0807649 and PHY-0758153 and
NASA grant NNX08AL43G.
JSG thanks the Galileo Galilei Institute for Theoretical Physics for hospitality, and acknowledges support from NASA through Einstein Postdoctoral Fellowship grant PF1-120089 awarded by the Chandra X-ray Center, which is operated by the Smithsonian Astrophysical Observatory for NASA under contract NAS8-03060.

\end{acknowledgments}

\bibliography{fermi_aniso_analysis_v2}

\begin{thebibliography}{55}
\expandafter\ifx\csname natexlab\endcsname\relax\def\natexlab#1{#1}\fi
\expandafter\ifx\csname bibnamefont\endcsname\relax
  \def\bibnamefont#1{#1}\fi
\expandafter\ifx\csname bibfnamefont\endcsname\relax
  \def\bibfnamefont#1{#1}\fi
\expandafter\ifx\csname citenamefont\endcsname\relax
  \def\citenamefont#1{#1}\fi
\expandafter\ifx\csname url\endcsname\relax
  \def\url#1{\texttt{#1}}\fi
\expandafter\ifx\csname urlprefix\endcsname\relax\def\urlprefix{URL }\fi
\providecommand{\bibinfo}[2]{#2}
\providecommand{\eprint}[2][]{\url{#2}}

\bibitem[{\citenamefont{{Kraushaar} et~al.}(1972)\citenamefont{{Kraushaar},
  {Clark}, {Garmire}, {Borken}, {Higbie}, {Leong}, and
  {Thorsos}}}]{Kraushaar:1972}
\bibinfo{author}{\bibfnamefont{W.~L.} \bibnamefont{{Kraushaar}}},
  \bibinfo{author}{\bibfnamefont{G.~W.} \bibnamefont{{Clark}}},
  \bibinfo{author}{\bibfnamefont{G.~P.} \bibnamefont{{Garmire}}},
  \bibinfo{author}{\bibfnamefont{R.}~\bibnamefont{{Borken}}},
  \bibinfo{author}{\bibfnamefont{P.}~\bibnamefont{{Higbie}}},
  \bibinfo{author}{\bibfnamefont{V.}~\bibnamefont{{Leong}}}, \bibnamefont{and}
  \bibinfo{author}{\bibfnamefont{T.}~\bibnamefont{{Thorsos}}},
  \bibinfo{journal}{Astrophys.J.} \textbf{\bibinfo{volume}{177}},
  \bibinfo{pages}{341} (\bibinfo{year}{1972}).

\bibitem[{\citenamefont{{Fichtel} et~al.}(1975)\citenamefont{{Fichtel},
  {Hartman}, {Kniffen}, {Thompson}, {Ogelman}, {Ozel}, {Tumer}, and
  {Bignami}}}]{Fichtel:1975}
\bibinfo{author}{\bibfnamefont{C.~E.} \bibnamefont{{Fichtel}}},
  \bibinfo{author}{\bibfnamefont{R.~C.} \bibnamefont{{Hartman}}},
  \bibinfo{author}{\bibfnamefont{D.~A.} \bibnamefont{{Kniffen}}},
  \bibinfo{author}{\bibfnamefont{D.~J.} \bibnamefont{{Thompson}}},
  \bibinfo{author}{\bibfnamefont{H.}~\bibnamefont{{Ogelman}}},
  \bibinfo{author}{\bibfnamefont{M.~E.} \bibnamefont{{Ozel}}},
  \bibinfo{author}{\bibfnamefont{T.}~\bibnamefont{{Tumer}}}, \bibnamefont{and}
  \bibinfo{author}{\bibfnamefont{G.~F.} \bibnamefont{{Bignami}}},
  \bibinfo{journal}{Astrophys.J.} \textbf{\bibinfo{volume}{198}},
  \bibinfo{pages}{163} (\bibinfo{year}{1975}).

\bibitem[{\citenamefont{Sreekumar et~al.}(1998)}]{Sreekumar:1997un}
\bibinfo{author}{\bibfnamefont{P.}~\bibnamefont{Sreekumar}}
  \bibnamefont{et~al.} (\bibinfo{collaboration}{EGRET Collaboration}),
  \bibinfo{journal}{Astrophys.J.} \textbf{\bibinfo{volume}{494}},
  \bibinfo{pages}{523} (\bibinfo{year}{1998}), \eprint{astro-ph/9709257}.

\bibitem[{\citenamefont{Strong et~al.}(2004)\citenamefont{Strong, Moskalenko,
  and Reimer}}]{Strong:2004ry}
\bibinfo{author}{\bibfnamefont{A.}~\bibnamefont{Strong}},
  \bibinfo{author}{\bibfnamefont{I.}~\bibnamefont{Moskalenko}},
  \bibnamefont{and} \bibinfo{author}{\bibfnamefont{O.}~\bibnamefont{Reimer}},
  \bibinfo{journal}{Astrophys.J.} \textbf{\bibinfo{volume}{613}},
  \bibinfo{pages}{956} (\bibinfo{year}{2004}), \eprint{astro-ph/0405441}.

\bibitem[{\citenamefont{Abdo et~al.}(2010{\natexlab{a}})}]{Abdo:2010nz:igrb}
\bibinfo{author}{\bibfnamefont{A.}~\bibnamefont{Abdo}} \bibnamefont{et~al.}
  (\bibinfo{collaboration}{Fermi LAT Collaboration}),
  \bibinfo{journal}{Phys.Rev.Lett.} \textbf{\bibinfo{volume}{104}},
  \bibinfo{pages}{101101} (\bibinfo{year}{2010}{\natexlab{a}}),
  \eprint{1002.3603}.

\bibitem[{\citenamefont{Kalashev et~al.}(2009)\citenamefont{Kalashev, Semikoz,
  and Sigl}}]{Kalashev:2007sn}
\bibinfo{author}{\bibfnamefont{O.~E.} \bibnamefont{Kalashev}},
  \bibinfo{author}{\bibfnamefont{D.~V.} \bibnamefont{Semikoz}},
  \bibnamefont{and} \bibinfo{author}{\bibfnamefont{G.}~\bibnamefont{Sigl}},
  \bibinfo{journal}{Phys.Rev.} \textbf{\bibinfo{volume}{D79}},
  \bibinfo{pages}{063005} (\bibinfo{year}{2009}), \eprint{0704.2463}.

\bibitem[{\citenamefont{Stecker and Salamon}(1996)}]{Stecker:1996ma}
\bibinfo{author}{\bibfnamefont{F.}~\bibnamefont{Stecker}} \bibnamefont{and}
  \bibinfo{author}{\bibfnamefont{M.}~\bibnamefont{Salamon}},
  \bibinfo{journal}{Astrophys.J.} \textbf{\bibinfo{volume}{464}},
  \bibinfo{pages}{600} (\bibinfo{year}{1996}), \eprint{astro-ph/9601120}.

\bibitem[{\citenamefont{Narumoto and Totani}(2006)}]{Narumoto:2006qg}
\bibinfo{author}{\bibfnamefont{T.}~\bibnamefont{Narumoto}} \bibnamefont{and}
  \bibinfo{author}{\bibfnamefont{T.}~\bibnamefont{Totani}},
  \bibinfo{journal}{Astrophys.J.} \textbf{\bibinfo{volume}{643}},
  \bibinfo{pages}{81} (\bibinfo{year}{2006}), \eprint{astro-ph/0602178}.

\bibitem[{\citenamefont{Inoue and Totani}(2009)}]{Inoue:2008pk}
\bibinfo{author}{\bibfnamefont{Y.}~\bibnamefont{Inoue}} \bibnamefont{and}
  \bibinfo{author}{\bibfnamefont{T.}~\bibnamefont{Totani}},
  \bibinfo{journal}{Astrophys.J.} \textbf{\bibinfo{volume}{702}},
  \bibinfo{pages}{523} (\bibinfo{year}{2009}), \eprint{0810.3580}.

\bibitem[{\citenamefont{Abdo
  et~al.}(2010{\natexlab{b}})}]{Collaboration:2010gqa:srccounts}
\bibinfo{author}{\bibfnamefont{A.}~\bibnamefont{Abdo}} \bibnamefont{et~al.},
  \bibinfo{journal}{Astrophys.J.} \textbf{\bibinfo{volume}{720}},
  \bibinfo{pages}{435} (\bibinfo{year}{2010}{\natexlab{b}}),
  \eprint{1003.0895}.

\bibitem[{\citenamefont{Abazajian et~al.}(2011)\citenamefont{Abazajian,
  Blanchet, and Harding}}]{Abazajian:2010pc}
\bibinfo{author}{\bibfnamefont{K.~N.} \bibnamefont{Abazajian}},
  \bibinfo{author}{\bibfnamefont{S.}~\bibnamefont{Blanchet}}, \bibnamefont{and}
  \bibinfo{author}{\bibfnamefont{J.}~\bibnamefont{Harding}},
  \bibinfo{journal}{Phys.Rev.} \textbf{\bibinfo{volume}{D84}},
  \bibinfo{pages}{103007} (\bibinfo{year}{2011}), \eprint{1012.1247}.

\bibitem[{\citenamefont{Stecker and Venters}(2011)}]{Stecker:2010di}
\bibinfo{author}{\bibfnamefont{F.~W.} \bibnamefont{Stecker}} \bibnamefont{and}
  \bibinfo{author}{\bibfnamefont{T.~M.} \bibnamefont{Venters}},
  \bibinfo{journal}{Astrophys.J.} \textbf{\bibinfo{volume}{736}},
  \bibinfo{pages}{40} (\bibinfo{year}{2011}), \eprint{1012.3678}.

\bibitem[{\citenamefont{Malyshev and Hogg}(2011)}]{Malyshev:2011zi}
\bibinfo{author}{\bibfnamefont{D.}~\bibnamefont{Malyshev}} \bibnamefont{and}
  \bibinfo{author}{\bibfnamefont{D.~W.} \bibnamefont{Hogg}},
  \bibinfo{journal}{Astrophys.J.} \textbf{\bibinfo{volume}{738}},
  \bibinfo{pages}{181} (\bibinfo{year}{2011}), \eprint{1104.0010}.

\bibitem[{\citenamefont{Fields et~al.}(2010)\citenamefont{Fields, Pavlidou, and
  Prodanovic}}]{Fields:2010bw}
\bibinfo{author}{\bibfnamefont{B.~D.} \bibnamefont{Fields}},
  \bibinfo{author}{\bibfnamefont{V.}~\bibnamefont{Pavlidou}}, \bibnamefont{and}
  \bibinfo{author}{\bibfnamefont{T.}~\bibnamefont{Prodanovic}},
  \bibinfo{journal}{Astrophys.J.} \textbf{\bibinfo{volume}{722}},
  \bibinfo{pages}{L199} (\bibinfo{year}{2010}), \eprint{1003.3647}.

\bibitem[{\citenamefont{Faucher-Giguere and
  Loeb}(2010)}]{FaucherGiguere:2009df}
\bibinfo{author}{\bibfnamefont{C.-A.} \bibnamefont{Faucher-Giguere}}
  \bibnamefont{and} \bibinfo{author}{\bibfnamefont{A.}~\bibnamefont{Loeb}},
  \bibinfo{journal}{JCAP} \textbf{\bibinfo{volume}{1001}}, \bibinfo{pages}{005}
  (\bibinfo{year}{2010}), \eprint{0904.3102}.

\bibitem[{\citenamefont{Ullio et~al.}(2002)\citenamefont{Ullio, Bergstrom,
  Edsjo, and Lacey}}]{Ullio:2002pj}
\bibinfo{author}{\bibfnamefont{P.}~\bibnamefont{Ullio}},
  \bibinfo{author}{\bibfnamefont{L.}~\bibnamefont{Bergstrom}},
  \bibinfo{author}{\bibfnamefont{J.}~\bibnamefont{Edsjo}}, \bibnamefont{and}
  \bibinfo{author}{\bibfnamefont{C.~G.} \bibnamefont{Lacey}},
  \bibinfo{journal}{Phys.Rev.} \textbf{\bibinfo{volume}{D66}},
  \bibinfo{pages}{123502} (\bibinfo{year}{2002}), \eprint{astro-ph/0207125}.

\bibitem[{\citenamefont{Elsaesser and Mannheim}(2005)}]{Elsaesser:2004ap}
\bibinfo{author}{\bibfnamefont{D.}~\bibnamefont{Elsaesser}} \bibnamefont{and}
  \bibinfo{author}{\bibfnamefont{K.}~\bibnamefont{Mannheim}},
  \bibinfo{journal}{Phys.Rev.Lett.} \textbf{\bibinfo{volume}{94}},
  \bibinfo{pages}{171302} (\bibinfo{year}{2005}), \eprint{astro-ph/0405235}.

\bibitem[{\citenamefont{Oda et~al.}(2005)\citenamefont{Oda, Totani, and
  Nagashima}}]{Oda:2005nv}
\bibinfo{author}{\bibfnamefont{T.}~\bibnamefont{Oda}},
  \bibinfo{author}{\bibfnamefont{T.}~\bibnamefont{Totani}}, \bibnamefont{and}
  \bibinfo{author}{\bibfnamefont{M.}~\bibnamefont{Nagashima}},
  \bibinfo{journal}{Astrophys.J.} \textbf{\bibinfo{volume}{633}},
  \bibinfo{pages}{L65} (\bibinfo{year}{2005}), \eprint{astro-ph/0504096}.

\bibitem[{\citenamefont{Kuhlen et~al.}(2008)\citenamefont{Kuhlen, Diemand, and
  Madau}}]{Kuhlen:2008aw}
\bibinfo{author}{\bibfnamefont{M.}~\bibnamefont{Kuhlen}},
  \bibinfo{author}{\bibfnamefont{J.}~\bibnamefont{Diemand}}, \bibnamefont{and}
  \bibinfo{author}{\bibfnamefont{P.}~\bibnamefont{Madau}},
  \bibinfo{journal}{Astrophys.J.} \textbf{\bibinfo{volume}{686}},
  \bibinfo{pages}{262} (\bibinfo{year}{2008}), \eprint{0805.4416}.

\bibitem[{\citenamefont{Springel et~al.}(2008)\citenamefont{Springel, White,
  Frenk, Navarro, Jenkins et~al.}}]{Springel:2008zz}
\bibinfo{author}{\bibfnamefont{V.}~\bibnamefont{Springel}},
  \bibinfo{author}{\bibfnamefont{S.}~\bibnamefont{White}},
  \bibinfo{author}{\bibfnamefont{C.}~\bibnamefont{Frenk}},
  \bibinfo{author}{\bibfnamefont{J.}~\bibnamefont{Navarro}},
  \bibinfo{author}{\bibfnamefont{A.}~\bibnamefont{Jenkins}},
  \bibnamefont{et~al.}, \bibinfo{journal}{Nature}
  \textbf{\bibinfo{volume}{456N7218}}, \bibinfo{pages}{73}
  (\bibinfo{year}{2008}).

\bibitem[{\citenamefont{Zavala et~al.}(2010)\citenamefont{Zavala, Springel, and
  Boylan-Kolchin}}]{Zavala:2009zr}
\bibinfo{author}{\bibfnamefont{J.}~\bibnamefont{Zavala}},
  \bibinfo{author}{\bibfnamefont{V.}~\bibnamefont{Springel}}, \bibnamefont{and}
  \bibinfo{author}{\bibfnamefont{M.}~\bibnamefont{Boylan-Kolchin}},
  \bibinfo{journal}{Mon.Not.Roy.Astron.Soc.} \textbf{\bibinfo{volume}{405}},
  \bibinfo{pages}{593} (\bibinfo{year}{2010}), \eprint{0908.2428}.

\bibitem[{\citenamefont{Abdo
  et~al.}(2010{\natexlab{c}})}]{Abdo:2010dk:cosmowimp}
\bibinfo{author}{\bibfnamefont{A.}~\bibnamefont{Abdo}} \bibnamefont{et~al.}
  (\bibinfo{collaboration}{Fermi LAT Collaboration}), \bibinfo{journal}{JCAP}
  \textbf{\bibinfo{volume}{1004}}, \bibinfo{pages}{014}
  (\bibinfo{year}{2010}{\natexlab{c}}), \eprint{1002.4415}.

\bibitem[{\citenamefont{Ando and Komatsu}(2006)}]{Ando:2005xg}
\bibinfo{author}{\bibfnamefont{S.}~\bibnamefont{Ando}} \bibnamefont{and}
  \bibinfo{author}{\bibfnamefont{E.}~\bibnamefont{Komatsu}},
  \bibinfo{journal}{Phys.Rev.} \textbf{\bibinfo{volume}{D73}},
  \bibinfo{pages}{023521} (\bibinfo{year}{2006}), \eprint{astro-ph/0512217}.

\bibitem[{\citenamefont{Lee et~al.}(2009)\citenamefont{Lee, Ando, and
  Kamionkowski}}]{Lee:2008fm}
\bibinfo{author}{\bibfnamefont{S.~K.} \bibnamefont{Lee}},
  \bibinfo{author}{\bibfnamefont{S.}~\bibnamefont{Ando}}, \bibnamefont{and}
  \bibinfo{author}{\bibfnamefont{M.}~\bibnamefont{Kamionkowski}},
  \bibinfo{journal}{JCAP} \textbf{\bibinfo{volume}{0907}}, \bibinfo{pages}{007}
  (\bibinfo{year}{2009}), \eprint{0810.1284}.

\bibitem[{\citenamefont{Dodelson et~al.}(2009)\citenamefont{Dodelson, Belikov,
  Hooper, and Serpico}}]{Dodelson:2009ih}
\bibinfo{author}{\bibfnamefont{S.}~\bibnamefont{Dodelson}},
  \bibinfo{author}{\bibfnamefont{A.~V.} \bibnamefont{Belikov}},
  \bibinfo{author}{\bibfnamefont{D.}~\bibnamefont{Hooper}}, \bibnamefont{and}
  \bibinfo{author}{\bibfnamefont{P.}~\bibnamefont{Serpico}},
  \bibinfo{journal}{Phys.Rev.} \textbf{\bibinfo{volume}{D80}},
  \bibinfo{pages}{083504} (\bibinfo{year}{2009}), \eprint{0903.2829}.

\bibitem[{\citenamefont{Xia et~al.}(2011)\citenamefont{Xia, Cuoco, Branchini,
  Fornasa, and Viel}}]{Xia:2011ax}
\bibinfo{author}{\bibfnamefont{J.-Q.} \bibnamefont{Xia}},
  \bibinfo{author}{\bibfnamefont{A.}~\bibnamefont{Cuoco}},
  \bibinfo{author}{\bibfnamefont{E.}~\bibnamefont{Branchini}},
  \bibinfo{author}{\bibfnamefont{M.}~\bibnamefont{Fornasa}}, \bibnamefont{and}
  \bibinfo{author}{\bibfnamefont{M.}~\bibnamefont{Viel}},
  \bibinfo{journal}{Mon.Not.Roy.Astron.Soc.} \textbf{\bibinfo{volume}{416}},
  \bibinfo{pages}{2247} (\bibinfo{year}{2011}), \eprint{1103.4861}.

\bibitem[{\citenamefont{Ando et~al.}(2007{\natexlab{a}})\citenamefont{Ando,
  Komatsu, Narumoto, and Totani}}]{Ando:2006cr}
\bibinfo{author}{\bibfnamefont{S.}~\bibnamefont{Ando}},
  \bibinfo{author}{\bibfnamefont{E.}~\bibnamefont{Komatsu}},
  \bibinfo{author}{\bibfnamefont{T.}~\bibnamefont{Narumoto}}, \bibnamefont{and}
  \bibinfo{author}{\bibfnamefont{T.}~\bibnamefont{Totani}},
  \bibinfo{journal}{Phys.Rev.} \textbf{\bibinfo{volume}{D75}},
  \bibinfo{pages}{063519} (\bibinfo{year}{2007}{\natexlab{a}}),
  \eprint{astro-ph/0612467}.

\bibitem[{\citenamefont{Ando et~al.}(2007{\natexlab{b}})\citenamefont{Ando,
  Komatsu, Narumoto, and Totani}}]{Ando:2006mt}
\bibinfo{author}{\bibfnamefont{S.}~\bibnamefont{Ando}},
  \bibinfo{author}{\bibfnamefont{E.}~\bibnamefont{Komatsu}},
  \bibinfo{author}{\bibfnamefont{T.}~\bibnamefont{Narumoto}}, \bibnamefont{and}
  \bibinfo{author}{\bibfnamefont{T.}~\bibnamefont{Totani}},
  \bibinfo{journal}{Mon.Not.Roy.Astron.Soc.} \textbf{\bibinfo{volume}{376}},
  \bibinfo{pages}{1635} (\bibinfo{year}{2007}{\natexlab{b}}),
  \eprint{astro-ph/0610155}.

\bibitem[{\citenamefont{Miniati et~al.}(2007)\citenamefont{Miniati,
  Koushiappas, and Di~Matteo}}]{Miniati:2007ke}
\bibinfo{author}{\bibfnamefont{F.}~\bibnamefont{Miniati}},
  \bibinfo{author}{\bibfnamefont{S.~M.} \bibnamefont{Koushiappas}},
  \bibnamefont{and}
  \bibinfo{author}{\bibfnamefont{T.}~\bibnamefont{Di~Matteo}},
  \bibinfo{journal}{Astrophys.J.} \textbf{\bibinfo{volume}{667}},
  \bibinfo{pages}{L1} (\bibinfo{year}{2007}), \eprint{astro-ph/0702083}.

\bibitem[{\citenamefont{Ando and Pavlidou}(2009)}]{Ando:2009nk}
\bibinfo{author}{\bibfnamefont{S.}~\bibnamefont{Ando}} \bibnamefont{and}
  \bibinfo{author}{\bibfnamefont{V.}~\bibnamefont{Pavlidou}},
  \bibinfo{journal}{Mon.Not.Roy.Astron.Soc.} \textbf{\bibinfo{volume}{400}},
  \bibinfo{pages}{2122} (\bibinfo{year}{2009}), \eprint{0908.3890}.

\bibitem[{\citenamefont{Siegal-Gaskins
  et~al.}(2011)\citenamefont{Siegal-Gaskins, Reesman, Pavlidou, Profumo, and
  Walker}}]{SiegalGaskins:2010mp}
\bibinfo{author}{\bibfnamefont{J.~M.} \bibnamefont{Siegal-Gaskins}},
  \bibinfo{author}{\bibfnamefont{R.}~\bibnamefont{Reesman}},
  \bibinfo{author}{\bibfnamefont{V.}~\bibnamefont{Pavlidou}},
  \bibinfo{author}{\bibfnamefont{S.}~\bibnamefont{Profumo}}, \bibnamefont{and}
  \bibinfo{author}{\bibfnamefont{T.~P.} \bibnamefont{Walker}},
  \bibinfo{journal}{Mon.Not.Roy.Astron.Soc.} \textbf{\bibinfo{volume}{415}},
  \bibinfo{pages}{1074S} (\bibinfo{year}{2011}), \eprint{1011.5501}.

\bibitem[{\citenamefont{Siegal-Gaskins}(2008)}]{SiegalGaskins:2008ge}
\bibinfo{author}{\bibfnamefont{J.~M.} \bibnamefont{Siegal-Gaskins}},
  \bibinfo{journal}{JCAP} \textbf{\bibinfo{volume}{0810}}, \bibinfo{pages}{040}
  (\bibinfo{year}{2008}), \eprint{0807.1328}.

\bibitem[{\citenamefont{Ando}(2009)}]{Ando:2009fp}
\bibinfo{author}{\bibfnamefont{S.}~\bibnamefont{Ando}},
  \bibinfo{journal}{Phys.Rev.} \textbf{\bibinfo{volume}{D80}},
  \bibinfo{pages}{023520} (\bibinfo{year}{2009}), \eprint{0903.4685}.

\bibitem[{\citenamefont{Fornasa et~al.}(2009)\citenamefont{Fornasa, Pieri,
  Bertone, and Branchini}}]{Fornasa:2009qh}
\bibinfo{author}{\bibfnamefont{M.}~\bibnamefont{Fornasa}},
  \bibinfo{author}{\bibfnamefont{L.}~\bibnamefont{Pieri}},
  \bibinfo{author}{\bibfnamefont{G.}~\bibnamefont{Bertone}}, \bibnamefont{and}
  \bibinfo{author}{\bibfnamefont{E.}~\bibnamefont{Branchini}},
  \bibinfo{journal}{Phys.Rev.} \textbf{\bibinfo{volume}{D80}},
  \bibinfo{pages}{023518} (\bibinfo{year}{2009}), \eprint{0901.2921}.

\bibitem[{\citenamefont{Cuoco et~al.}(2008)\citenamefont{Cuoco, Brandbyge,
  Hannestad, Haugboelle, and Miele}}]{Cuoco:2007sh}
\bibinfo{author}{\bibfnamefont{A.}~\bibnamefont{Cuoco}},
  \bibinfo{author}{\bibfnamefont{J.}~\bibnamefont{Brandbyge}},
  \bibinfo{author}{\bibfnamefont{S.}~\bibnamefont{Hannestad}},
  \bibinfo{author}{\bibfnamefont{T.}~\bibnamefont{Haugboelle}},
  \bibnamefont{and} \bibinfo{author}{\bibfnamefont{G.}~\bibnamefont{Miele}},
  \bibinfo{journal}{Phys.Rev.} \textbf{\bibinfo{volume}{D77}},
  \bibinfo{pages}{123518} (\bibinfo{year}{2008}), \eprint{0710.4136}.

\bibitem[{\citenamefont{Zhang and Sigl}(2008)}]{Zhang:2008rs}
\bibinfo{author}{\bibfnamefont{L.}~\bibnamefont{Zhang}} \bibnamefont{and}
  \bibinfo{author}{\bibfnamefont{G.}~\bibnamefont{Sigl}},
  \bibinfo{journal}{JCAP} \textbf{\bibinfo{volume}{0809}}, \bibinfo{pages}{027}
  (\bibinfo{year}{2008}), \eprint{0807.3429}.

\bibitem[{\citenamefont{Taoso et~al.}(2009)\citenamefont{Taoso, Ando, Bertone,
  and Profumo}}]{Taoso:2008qz}
\bibinfo{author}{\bibfnamefont{M.}~\bibnamefont{Taoso}},
  \bibinfo{author}{\bibfnamefont{S.}~\bibnamefont{Ando}},
  \bibinfo{author}{\bibfnamefont{G.}~\bibnamefont{Bertone}}, \bibnamefont{and}
  \bibinfo{author}{\bibfnamefont{S.}~\bibnamefont{Profumo}},
  \bibinfo{journal}{Phys.Rev.} \textbf{\bibinfo{volume}{D79}},
  \bibinfo{pages}{043521} (\bibinfo{year}{2009}), \eprint{0811.4493}.

\bibitem[{\citenamefont{Ibarra et~al.}(2010)\citenamefont{Ibarra, Tran, and
  Weniger}}]{Ibarra:2009nw}
\bibinfo{author}{\bibfnamefont{A.}~\bibnamefont{Ibarra}},
  \bibinfo{author}{\bibfnamefont{D.}~\bibnamefont{Tran}}, \bibnamefont{and}
  \bibinfo{author}{\bibfnamefont{C.}~\bibnamefont{Weniger}},
  \bibinfo{journal}{Phys.Rev.} \textbf{\bibinfo{volume}{D81}},
  \bibinfo{pages}{023529} (\bibinfo{year}{2010}), \eprint{0909.3514}.

\bibitem[{\citenamefont{Cuoco et~al.}(2011)\citenamefont{Cuoco, Sellerholm,
  Conrad, and Hannestad}}]{Cuoco:2010jb}
\bibinfo{author}{\bibfnamefont{A.}~\bibnamefont{Cuoco}},
  \bibinfo{author}{\bibfnamefont{A.}~\bibnamefont{Sellerholm}},
  \bibinfo{author}{\bibfnamefont{J.}~\bibnamefont{Conrad}}, \bibnamefont{and}
  \bibinfo{author}{\bibfnamefont{S.}~\bibnamefont{Hannestad}},
  \bibinfo{journal}{Mon.Not.Roy.Astron.Soc.} \textbf{\bibinfo{volume}{414}},
  \bibinfo{pages}{2040} (\bibinfo{year}{2011}), \eprint{1005.0843}.

\bibitem[{\citenamefont{Gorski et~al.}(2005)\citenamefont{Gorski, Hivon,
  Banday, Wandelt, Hansen et~al.}}]{Gorski:2004by}
\bibinfo{author}{\bibfnamefont{K.}~\bibnamefont{Gorski}},
  \bibinfo{author}{\bibfnamefont{E.}~\bibnamefont{Hivon}},
  \bibinfo{author}{\bibfnamefont{A.}~\bibnamefont{Banday}},
  \bibinfo{author}{\bibfnamefont{B.}~\bibnamefont{Wandelt}},
  \bibinfo{author}{\bibfnamefont{F.}~\bibnamefont{Hansen}},
  \bibnamefont{et~al.}, \bibinfo{journal}{Astrophys.J.}
  \textbf{\bibinfo{volume}{622}}, \bibinfo{pages}{759} (\bibinfo{year}{2005}),
  \eprint{astro-ph/0409513}.

\bibitem[{\citenamefont{Ackermann et~al.}(2010)}]{Ackermann:2010ip:CREaniso}
\bibinfo{author}{\bibfnamefont{M.}~\bibnamefont{Ackermann}}
  \bibnamefont{et~al.} (\bibinfo{collaboration}{Fermi LAT Collaboration}),
  \bibinfo{journal}{Phys.Rev.} \textbf{\bibinfo{volume}{D82}},
  \bibinfo{pages}{092003} (\bibinfo{year}{2010}), \eprint{1008.5119}.

\bibitem[{\citenamefont{Atwood et~al.}(2009)}]{Atwood:2009ez}
\bibinfo{author}{\bibfnamefont{W.}~\bibnamefont{Atwood}} \bibnamefont{et~al.}
  (\bibinfo{collaboration}{Fermi LAT Collaboration}),
  \bibinfo{journal}{Astrophys.J.} \textbf{\bibinfo{volume}{697}},
  \bibinfo{pages}{1071} (\bibinfo{year}{2009}), \eprint{0902.1089}.

\bibitem[{\citenamefont{Ackermann et~al.}(2009)}]{Ackermann:2009zz}
\bibinfo{author}{\bibfnamefont{M.}~\bibnamefont{Ackermann}}
  \bibnamefont{et~al.} (\bibinfo{collaboration}{Fermi LAT Collaboration}),
  \bibinfo{journal}{AIP Conf.Proc.} \textbf{\bibinfo{volume}{1085}},
  \bibinfo{pages}{763} (\bibinfo{year}{2009}).

\bibitem[{\citenamefont{Hensley et~al.}(2010)\citenamefont{Hensley,
  Siegal-Gaskins, and Pavlidou}}]{Hensley:2009gh}
\bibinfo{author}{\bibfnamefont{B.~S.} \bibnamefont{Hensley}},
  \bibinfo{author}{\bibfnamefont{J.~M.} \bibnamefont{Siegal-Gaskins}},
  \bibnamefont{and} \bibinfo{author}{\bibfnamefont{V.}~\bibnamefont{Pavlidou}},
  \bibinfo{journal}{Astrophys.J.} \textbf{\bibinfo{volume}{723}},
  \bibinfo{pages}{277} (\bibinfo{year}{2010}), \eprint{0912.1854}.

\bibitem[{\citenamefont{Siegal-Gaskins and
  Pavlidou}(2009)}]{SiegalGaskins:2009ux}
\bibinfo{author}{\bibfnamefont{J.~M.} \bibnamefont{Siegal-Gaskins}}
  \bibnamefont{and} \bibinfo{author}{\bibfnamefont{V.}~\bibnamefont{Pavlidou}},
  \bibinfo{journal}{Phys.Rev.Lett.} \textbf{\bibinfo{volume}{102}},
  \bibinfo{pages}{241301} (\bibinfo{year}{2009}), \eprint{0901.3776}.

\bibitem[{\citenamefont{Abdo
  et~al.}(2010{\natexlab{d}})}]{Collaboration:2010ru:1fgl}
\bibinfo{author}{\bibfnamefont{A.}~\bibnamefont{Abdo}} \bibnamefont{et~al.}
  (\bibinfo{collaboration}{Fermi LAT Collaboration}),
  \bibinfo{journal}{Astrophys.J.Suppl.} \textbf{\bibinfo{volume}{188}},
  \bibinfo{pages}{405} (\bibinfo{year}{2010}{\natexlab{d}}),
  \eprint{1002.2280}.

\bibitem[{\citenamefont{Hivon et~al.}(2001)\citenamefont{Hivon, Gorski,
  Netterfield, Crill, Prunet et~al.}}]{Hivon:2001jp}
\bibinfo{author}{\bibfnamefont{E.}~\bibnamefont{Hivon}},
  \bibinfo{author}{\bibfnamefont{K.}~\bibnamefont{Gorski}},
  \bibinfo{author}{\bibfnamefont{C.}~\bibnamefont{Netterfield}},
  \bibinfo{author}{\bibfnamefont{B.}~\bibnamefont{Crill}},
  \bibinfo{author}{\bibfnamefont{S.}~\bibnamefont{Prunet}},
  \bibnamefont{et~al.} (\bibinfo{year}{2001}), \eprint{astro-ph/0105302}.

\bibitem[{\citenamefont{Komatsu et~al.}(2002)\citenamefont{Komatsu, Wandelt,
  Spergel, Banday, and Gorski}}]{Komatsu:2001wu}
\bibinfo{author}{\bibfnamefont{E.}~\bibnamefont{Komatsu}},
  \bibinfo{author}{\bibfnamefont{B.~D.} \bibnamefont{Wandelt}},
  \bibinfo{author}{\bibfnamefont{D.~N.} \bibnamefont{Spergel}},
  \bibinfo{author}{\bibfnamefont{A.~J.} \bibnamefont{Banday}},
  \bibnamefont{and} \bibinfo{author}{\bibfnamefont{K.~M.}
  \bibnamefont{Gorski}}, \bibinfo{journal}{Astrophys.J.}
  \textbf{\bibinfo{volume}{566}}, \bibinfo{pages}{19} (\bibinfo{year}{2002}),
  \eprint{astro-ph/0107605}.

\bibitem[{\citenamefont{Jing}(2005)}]{Jing:2004fq}
\bibinfo{author}{\bibfnamefont{Y.}~\bibnamefont{Jing}},
  \bibinfo{journal}{Astrophys.J.} \textbf{\bibinfo{volume}{620}},
  \bibinfo{pages}{559} (\bibinfo{year}{2005}), \eprint{astro-ph/0409240}.

\bibitem[{\citenamefont{Knox}(1995)}]{Knox:1995dq}
\bibinfo{author}{\bibfnamefont{L.}~\bibnamefont{Knox}},
  \bibinfo{journal}{Phys.Rev.} \textbf{\bibinfo{volume}{D52}},
  \bibinfo{pages}{4307} (\bibinfo{year}{1995}), \eprint{astro-ph/9504054}.

\bibitem[{\citenamefont{Su et~al.}(2010)\citenamefont{Su, Slatyer, and
  Finkbeiner}}]{Su:2010qj}
\bibinfo{author}{\bibfnamefont{M.}~\bibnamefont{Su}},
  \bibinfo{author}{\bibfnamefont{T.~R.} \bibnamefont{Slatyer}},
  \bibnamefont{and} \bibinfo{author}{\bibfnamefont{D.~P.}
  \bibnamefont{Finkbeiner}}, \bibinfo{journal}{Astrophys.J.}
  \textbf{\bibinfo{volume}{724}}, \bibinfo{pages}{1044} (\bibinfo{year}{2010}),
  \eprint{1005.5480}.

\bibitem[{\citenamefont{Casandjian and Grenier}(2009)}]{Casandjian:2009wq}
\bibinfo{author}{\bibfnamefont{J.-M.} \bibnamefont{Casandjian}}
  \bibnamefont{and} \bibinfo{author}{\bibfnamefont{I.}~\bibnamefont{Grenier}}
  (\bibinfo{collaboration}{Fermi LAT Collaboration}) (\bibinfo{year}{2009}),
  \eprint{0912.3478}.

\bibitem[{\citenamefont{Schlegel et~al.}(1998)\citenamefont{Schlegel,
  Finkbeiner, and Davis}}]{Schlegel:1997yv}
\bibinfo{author}{\bibfnamefont{D.~J.} \bibnamefont{Schlegel}},
  \bibinfo{author}{\bibfnamefont{D.~P.} \bibnamefont{Finkbeiner}},
  \bibnamefont{and} \bibinfo{author}{\bibfnamefont{M.}~\bibnamefont{Davis}},
  \bibinfo{journal}{Astrophys.J.} \textbf{\bibinfo{volume}{500}},
  \bibinfo{pages}{525} (\bibinfo{year}{1998}), \eprint{astro-ph/9710327}.

\bibitem[{\citenamefont{Abdo et~al.}(2011)}]{Collaboration:2011bm}
\bibinfo{author}{\bibfnamefont{A.}~\bibnamefont{Abdo}} \bibnamefont{et~al.}
  (\bibinfo{collaboration}{Fermi LAT Collaboration}) (\bibinfo{year}{2011}),
  \eprint{1108.1435}.

\bibitem[{\citenamefont{Strong}(2007)}]{Strong:2006hf}
\bibinfo{author}{\bibfnamefont{A.~W.} \bibnamefont{Strong}},
  \bibinfo{journal}{Astrophys.Space Sci.} \textbf{\bibinfo{volume}{309}},
  \bibinfo{pages}{35} (\bibinfo{year}{2007}), \eprint{astro-ph/0609359}.

\end{thebibliography}

\end{document}